\documentclass[fleqn,usenatbib]{mnras}

\usepackage{newtxtext,newtxmath}

\usepackage[T1]{fontenc}
\usepackage{ae,aecompl}
\usepackage{times}
\usepackage{enumitem}
\usepackage{times}
\usepackage{textcomp}
\usepackage{verbatim}
\usepackage{esdiff}

\newcommand{\msun}{{\,\rm M_\odot}}

\newcommand{\kms}{\,{\rm km}\,{\rm s}^{-1}}

\newcommand{\Gyr}{\,{\rm Gyr}}

\newcommand{\pc}{\,{\rm pc}}
\newcommand{\kpc}{\,{\rm kpc}}

\newcommand{\cpm}{\,{\rm cm}^2\,{\rm g}^{-1}}

\def\jcap{J. Cosmol.  Astropart. Phys.}
\def\aap{A\&A}
\def\apj{ApJ}

\def\apjl{ApJ}
\def\mnras{MNRAS}
\def\araa{ARA\&A}
\def\aj{AJ}

\def\physrep{Phys. Rep.}
\def\nat{Nature}

\def\apjs{ApJS}
\def\prd{Phys. Rev. D}

\usepackage{graphicx}	
\usepackage{amsmath}	

\usepackage{tikz}

\makeatletter
\newcommand{\rmnum}[1]{\romannumeral #1}
\newcommand{\Rmnum}[1]{\expandafter\@slowromancap\romannumeral #1@}

\renewcommand\paragraph{\@startsection{paragraph}{4}{\z@}{3.25ex\@plus1ex\@minus.2ex}{-1em}{\normalfont\it\normalsize}}

\def\checkmark{\tikz\fill[scale=0.4](0,.35) -- (.25,0) -- (1,.7) -- (.25,.15) -- cycle;}

\title[Dissipative Dark Matter on FIRE]{Dissipative Dark Matter on FIRE: \Rmnum{1}. Structural and kinematic properties of dwarf galaxies}

\author[Shen et al.]{\parbox{18.0cm}{
Xuejian Shen$^{1}$\thanks{Contact e-mail: \href{mailto:xshen@caltech.edu}{xshen@caltech.edu}},
Philip F. Hopkins$^{1}$,
Lina Necib$^{2,3,4}$,
Fangzhou Jiang$^{1,4}$,
Michael Boylan-Kolchin$^{5}$,
Andrew Wetzel$^{6}$
\\
}\vspace{0.3cm}\\
$^{1}$TAPIR, California Institute of Technology, Pasadena, CA 91125, USA\\
$^{2}$Walter Burke Institute for Theoretical Physics, California Institute of Technology, Pasadena, CA 91125, USA\\
$^{3}$Center for Cosmology, Department of Physics and Astronomy,
University of California, Irvine, CA 92697, USA\\
$^{4}$Carnegie Observatories, 813 Santa Barbara Street, Pasadena, CA 91101, USA\\
$^{5}$Department of Astronomy, The University of Texas at Austin, 2515 Speedway Stop C1400, Austin, TX 78712, USA\\
$^{6}$Department of Physics \& Astronomy, University of California, Davis, CA 95616, USA
}

\date{Accepted XXX. Received YYY; in original form ZZZ}

\pubyear{2021}

\begin{document}

\label{firstpage}
\pagerange{\pageref{firstpage}--\pageref{lastpage}}
\maketitle

\begin{abstract}
We present the first set of cosmological baryonic zoom-in simulations of galaxies including dissipative self-interacting dark matter (dSIDM). These simulations utilize the Feedback In Realistic Environments (FIRE-2) galaxy formation physics, but allow the dark matter to have dissipative self-interactions analogous to Standard Model forces, parameterized by the self-interaction cross-section per unit mass, $(\sigma/m)$, and the dimensionless degree of dissipation, $0<f_{\rm diss}<1$. We survey this parameter space, including constant and velocity-dependent cross-sections, and focus on structural and kinematic properties of dwarf galaxies with $M_{\rm halo} \sim 10^{10-11}\msun$ and $M_{\ast} \sim 10^{5-8}\msun$. Central density profiles (parameterized as $\rho \propto r^{\alpha}$) of simulated dwarfs become cuspy when $(\sigma/m)_{\rm eff} \gtrsim 0.1\,{\rm cm^{2}\,g^{-1}}$ (and $f_{\rm diss}=0.5$ as fiducial). The power-law slopes asymptote to $\alpha \approx -1.5$ in low-mass dwarfs independent of cross-section, which arises from a dark matter ``cooling flow''. Through comparisons with dark matter only simulations, we find the profile in this regime is insensitive to the inclusion of baryons. However, when $(\sigma/m)_{\rm eff} \ll 0.1\,{\rm cm^{2}\,g^{-1}}$, baryonic effects can produce cored density profiles comparable to non-dissipative cold dark matter (CDM) runs but at smaller radii. Simulated galaxies with $(\sigma/m) \gtrsim 10\,{\rm cm^{2}\,g^{-1}}$ and the fiducial $f_{\rm diss}$ develop significant coherent rotation of dark matter, accompanied by halo deformation, but this is unlike the well-defined thin ``dark disks'' often attributed to baryon-like dSIDM. The density profiles in this high cross-section model exhibit lower normalizations given the onset of halo deformation. For our surveyed dSIDM parameters, halo masses and galaxy stellar masses do not show appreciable difference from CDM, but dark matter kinematics and halo concentrations/shapes can differ. 
\end{abstract}

\begin{keywords}
methods : numerical -- galaxies : dwarf -- galaxies : haloes -- cosmology : dark matter -- cosmology : theory
\end{keywords}


\section{Introduction}

Despite its veiled nature, dark matter is considered the main driver of structure formation in the Universe. The current paradigm --- the cosmological constant plus cold dark matter ($\Lambda {\rm CDM}$) cosmological model --- has been successful in describing the large scale structures in the Universe~\citep{Blumenthal1984,Davis1985}. This model assumes that dark matter is non-relativistic and is effectively collisionless, apart from its gravitational interactions with itself and Standard Model particles. However, in recent decades, evidence from astrophysical observations and absence of signal from particle physics experiments have motivated conjectures on alternative dark matter models. On the astrophysics side, the $\Lambda {\rm CDM}$ model faces significant challenges in matching observations at small scales~\citep[see a recent review][]{Bullock2017}. For example, the {\it core-cusp} problem states that the central profiles of dark matter dominated systems, e.g. dwarf spheroidal galaxies (dSphs) and low surface brightness galaxies (LSBs), are cored~\citep[e.g.,][]{Flores1994,Moore1994,deBlok2001,KDN2006,Gentile2004,Simon2005,Spano2008,KDN2011a,KDN2011b,Oh2011,Walker2011,Oh2015,Chan2015,Zhu2016}, in contrast to the universal cuspy central density profile found in dark matter only (DMO) simulations~\citep{Navarro1996,Navarro1997,Moore1999,Klypin2001,Navarro2004,Diemand2005}. The {\it too-big-to-fail} (TBTF) problem states that a substantial population of massive concentrated subhaloes appears in DMO simulations, which is incompatible with the stellar kinematics of observed satellite galaxies around the Milky Way or M31~\citep{MBK2011,MBK2012,Tollerud2014}. This mismatch has been extended to field dwarf galaxies in the Local Group~\citep{SGK2014,Kirby2014} and beyond~\citep{Papastergis2015}. Although the inclusion of bursty star formation and feedback processes has been shown to alleviate the tensions~\citep[e.g.,][]{Governato2010,Pontzen2012,Madau2014,Brooks2014,Wetzel2016,Sawala2016,SGK2019}, a population of compact dwarf galaxies in the local Universe are missing in cosmological simulations of CDM (plus baryons) that can produce dark matter cores~\citep[e.g.,][]{Santos2018,Jiang2019,SGK2019}. Relate to this, the rotation curves of dwarf galaxies appear to be more diverse than CDM predictions in the field~\citep{Oman2015} and Milky Way satellites~\citep{Kaplinghat2019}. Therefore, it is important to explore how non-standard dark matter models -- in conjunction with baryonic physics -- could help solve the small-scale anomalies. On the particle physics side, one of the most popular candidates for CDM (the class of Weakly Interacting Massive Particles, WIMPs) has not been discovered despite decades of efforts and a significant proportion of its parameter space being ruled out~\citep[e.g.,][]{Bertone2005,Bertone2010,Aprile2018}. The null results in collider production and direct/indirect detection experiments of classical CDM candidates have motivated ideas about alternative dark matter models~\citep[e.g.,][]{Hogan2000,Spergel2000,Dalcanton2001,Buckley2018} and explorations of the rich phenomenology from potential non-gravitational dark matter interactions. Many of these alternative dark matter models could behave dramatically differently from CDM at astrophysical scales and could potentially solve the small-scale problems mentioned above. 

\begin{table*}
    \centering
    \begin{tabular}{p{0.06\textwidth}|p{0.05\textwidth}|p{0.05\textwidth}|p{0.05\textwidth}|p{0.05\textwidth}|p{0.05\textwidth}|p{0.04\textwidth}|p{0.04\textwidth}|p{0.04\textwidth}|p{0.04\textwidth}|p{0.04\textwidth}|p{0.20\textwidth}}
        \hline
        Simulation & $M^{\rm cdm}_{\rm halo}$ & $R^{\rm cdm}_{\rm vir}$ & $M^{\rm cdm}_{\ast}$ & $r^{\rm cdm}_{1/2}$ & $r^{\rm conv}_{\rm dm}$ & $\sigma 1$ & $\sigma 10$ & $\sigma 1$ & $\sigma 0.1$ & $\sigma(v)$ & Notes\\
        name & $[\msun]$ & $[\kpc]$ & $[\msun]$ & $[\kpc]$ & $[\pc]$ & elastic & $f_{\rm diss} 0.5$ & $f_{\rm diss} 0.5$ & $f_{\rm diss} 0.5$ & $f_{\rm diss} 0.5$ & \\
        \hline
        \hline
    \end{tabular}
    
    \textbf{Ultra faint dwarf\\}
    \begin{tabular}{p{0.06\textwidth}|p{0.05\textwidth}|p{0.05\textwidth}|p{0.05\textwidth}|p{0.05\textwidth}|p{0.05\textwidth}|p{0.04\textwidth}|p{0.04\textwidth}|p{0.04\textwidth}|p{0.04\textwidth}|p{0.03\textwidth}|p{0.21\textwidth}}
        m09	  & 2.5e9 & 35.6 & 7.0e4 & 0.46 & 65 & \checkmark & \checkmark & \checkmark & \checkmark & \checkmark & other parameter choices explored \\
        \hline
    \end{tabular}
    
    \textbf{Classical dwarfs\\}
    \begin{tabular}{p{0.06\textwidth}|p{0.05\textwidth}|p{0.05\textwidth}|p{0.05\textwidth}|p{0.05\textwidth}|p{0.05\textwidth}|p{0.04\textwidth}|p{0.04\textwidth}|p{0.04\textwidth}|p{0.04\textwidth}|p{0.04\textwidth}|p{0.20\textwidth}}
        m10b& 9.4e9 & 55.2 & 5.8e5 & 0.36 & 77 &            &\checkmark & \checkmark & \checkmark & \checkmark & late-forming\\
        m10q& 7.5e9 & 51.1 & 1.7e6 & 0.72 & 73 & \checkmark &\checkmark & \checkmark & \checkmark & \checkmark & isolated, early-forming\\
        m10v& 8.5e9 & 53.5 & 1.4e5 & 0.32 & 65 &            &\checkmark & \checkmark & \checkmark & \checkmark & isolated, late-forming\\
        \hline
    \end{tabular}
    
    \textbf{Bright dwarfs\\}
    \begin{tabular}{p{0.06\textwidth}|p{0.05\textwidth}|p{0.05\textwidth}|p{0.05\textwidth}|p{0.05\textwidth}|p{0.05\textwidth}|p{0.04\textwidth}|p{0.04\textwidth}|p{0.04\textwidth}|p{0.04\textwidth}|p{0.04\textwidth}|p{0.20\textwidth}}
        m11a & 3.6e10 & 86.7 & 3.7e7 & 1.2 & 310 & \checkmark & \checkmark & \checkmark & \checkmark  & \checkmark  & diffuse, cored \\
        m11b & 4.2e10 & 90.7 & 4.2e7 & 1.7 & 250 &            & \checkmark & \checkmark & \checkmark  & \checkmark  & intermediate-forming\\
        m11q & 1.5e11 & 138.7 & 2.9e8 & 3.1 & 240 & \checkmark &            & \checkmark & \checkmark  & \checkmark  & early-forming, cored\\
        \hline
    \end{tabular}

   \textbf{ Milky Way-mass galaxies\\}
    \begin{tabular}{p{0.065\textwidth}|p{0.05\textwidth}|p{0.05\textwidth}|p{0.05\textwidth}|p{0.05\textwidth}|p{0.05\textwidth}|p{0.04\textwidth}|p{0.04\textwidth}|p{0.04\textwidth}|p{0.04\textwidth}|p{0.04\textwidth}|p{0.20\textwidth}}
        m11f      & 4.5e11 & 200.2 & 1.0e10 & 2.9 & 280 &       &       &            & \checkmark & \checkmark  & quiescent late history\\
        m12i l.r. & 1.1e12 & 272.3 & 1.1e11 & 2.0 & 290 &       &       & \checkmark & \checkmark & \checkmark  & Milky Way like\\
        m12f l.r. & 1.5e12 & 302.8 & 1.3e11 & 4.1 & 310 &       &       & \checkmark & \checkmark & \checkmark  & Milky Way like\\
        m12m l.r. & 1.5e12 & 299.3 & 1.4e11 & 6.1 & 360 &       &       & \checkmark & \checkmark & \checkmark  & early-forming, boxy bulge\\
        m12i h.r. & 9.8e11 & 259.9 & 2.4e10 & 3.7 & 150 &       &       &           & \checkmark & \checkmark  & Milky Way like\\
        \hline
    \end{tabular}
    
    \caption{ \textbf{Simulations of the FIRE-2 dSIDM suite.} The simulated galaxies are labelled and grouped by their halo masses. They are classified into four categories: ultra faint dwarfs; classical dwarfs, with typical halo mass $\lesssim 10^{10} \msun$; bright dwarfs, with typical halo mass $\sim 10^{10-11} \msun$; Milky Way-mass galaxies, with typical halo mass $\sim 10^{12}\msun$. These haloes are randomly picked from the standard FIRE-2 simulation suite~\citep{Hopkins2018}, sampling various star formation and merger histories. All units are physical. \newline \hspace{\textwidth}
    (\textbf{1}) Name of the simulation. ``l.r.'' (``h.r.'') indicates low (high)-resolution version of the simulation. \newline \hspace{\textwidth}
    (\textbf{2}) $M^{\rm cdm}_{\rm halo}$: Virial mass of the halo (definition given in Section~\ref{sec:halo_prop}) in the CDM simulation with baryons at $z=0$. \newline \hspace{\textwidth}
    (\textbf{3}) $R^{\rm cdm}_{\rm vir}$: Virial radius of the halo (definition given in Section~\ref{sec:halo_prop}) in the CDM simulation with baryons at $z=0$. \newline \hspace{\textwidth}
    (\textbf{4}) $M^{\rm cdm}_{\ast}$: Galaxy Stellar mass (see Section~\ref{sec:halo_prop}) in the CDM simulation at $z=0$. \newline \hspace{\textwidth}
    (\textbf{5}) $r^{\rm cdm}_{\rm 1/2}$: Galaxy stellar half mass radius (see Section~\ref{sec:halo_prop}) in the CDM simulation at $z=0$. \newline \hspace{\textwidth}
    (\textbf{6}) $r^{\rm conv}_{\rm dm}$: Radius of convergence in dark matter properties at $z=0$ (calculated for the CDM DMO simulations in the standard FIRE-2 series~\citep{Hopkins2018} based on the \citet{Power2003} criterion). As shown in \citet{Hopkins2018}, the convergence radii in simulations with baryons can in fact extend to much smaller radii. In Appendix~\ref{appsec:conv}, we show that these are rather conservative estimates of the true convergence radii in dSIDM runs. \newline \hspace{\textwidth}
    (\textbf{7-11}) Parameters of the dark matter models. $\sigma$ (with the number after it) indicates the self-interaction cross-section, $\sigma/m$, in unit of $\cpm$. $\sigma(v)$ denotes the velocity-dependent cross-section, introduced in Section~\ref{sec:sim}. $f_{\rm diss}$ indicates the dimensionless degree of dissipation. \newline \hspace{\textwidth}
    (\textbf{12}) Notes: Additional information of each simulation. 
    }
    \label{tab:sims}
\end{table*}

Self-interacting dark matter (SIDM) is an important category of alternative dark matter models that has been proposed and discussed in the literature for about three decades~\citep[e.g.,][]{Carlson1992, deLaix1995, Firmani2000,Spergel2000}. It is well motivated by hidden dark sectors as extensions to the Standard Model~\citep[e.g.,][]{Ackerman2009,Arkani-Hamed2009,Feng2009,Feng2010,Loeb2011,vandenAarssen2012,CyrRacine2013,Tulin2013,Cline2014}. The introduction of SIDM could potentially solve some small-scale problems~\citep[see the review of][and references therein]{Tulin2018}. Dark matter self-interactions enable effective heat conduction and could result in an isothermal distribution of dark matter with cores at halo centers, which alleviates the {\it core-cusp} problem. Meanwhile, it could also make dark matter haloes (subhaloes) less dense and alleviate the TBTF problem. Previous DMO simulations have found that a self-interaction cross-section of $\sim 1\cpm$ could solve the {\it core-cusp} and TBTF problems in dwarf galaxies simultaneously~\citep[e.g.,][]{Vogelsberger2012, Rocha2013, Zavala2013, Elbert2015}. In addition, SIDM with comparable cross-sections also have the potential to explain~\citep[e.g.,][]{Kamada2017,Creasey2017,Omid2020} the diversity of rotation curves of dwarf galaxies~\citep{Oman2015,Kaplinghat2019}. Following studies of galaxy clusters in SIDM suggested a cross-section of $\sim 0.1\cpm$~\citep[e.g.,][]{Kaplinghat2016,Elbert2018}, which motivates the velocity-dependence of self-interaction cross-section. These previous studies on SIDM focused on elastic dark matter self-interactions. However, in many particle physics realizations of SIDM, dark matter have inelastic (or specifically dissipative) self-interactions~\citep[e.g.,][]{Arkani-Hamed2009,Alves2010,Kaplan2010,Loeb2011,CyrRacine2013,Cline2014,Boddy2014,Wise2014,Foot2015b,Schutz2015,Boddy2016,Finkbeiner2016,Zhang2017,Blennow2017,Gresham2018}. The impact of dissipative processes of dark matter has not yet been explored in the context of cosmological structure formation. 

In addition, the focus on purely elastic SIDM (eSIDM) in previous studies has been motivated by solving some small-scale problems~(making galaxy centers less dense). Since dissipative dark matter self-interactions tend to make centers of haloes denser to first order consideration, dSIDM was largely omitted in previous studies of SIDM. However, apart from dark matter physics, some baryonic physics processes, including bursty star formation and stellar/supernovae feedback and tidal disruption, have also been shown to strongly impact the structure of dark matter haloes and help alleviate some small-scale problems. Specifically, gas outflows driven by stellar/supernovae feedback could create fluctuations in the central potential, which irreversibly transfer energy to CDM particles and generate dark matter cores~\citep{Governato2010,Governato2012,Pontzen2012,Madau2014}. Some more recent CDM simulations could resolve the small-scale problems by more realistic modeling of gas cooling, star formation and stellar/supernovae feedback~\citep[e.g.,][]{Brooks2014,Dutton2016,Fattahi2016,Sawala2016,Wetzel2016,SGK2019,Buck2019}. The interplay between baryons and SIDM in galaxy formation has been more carefully considered in subsequent SIDM simulations that include baryonic physics~\citep[e.g.,][]{Vogelsberger2014,Elbert2015,Fry2015,Robles2017,Despali2019,Fitts2019,Robles2019}. The inclusion of baryons 
substantially reduces the distinct signatures in dwarf galaxies caused by elastic dark matter self-interactions, especially in bright dwarfs with $r_{\rm 1/2}\gtrsim 400\pc$~\citep{Fitts2019}. This could hide dark matter physics that lead to enhanced central density originally, other than those proposed specifically to lower the central density. The parameter space for dSIDM, as an example of such models, reopens due to these recent developments. The contraction of the halo driven by dSIDM interactions could help produce the compact dwarf galaxies found in the local Universe that are missing in CDM simulations plus baryons~\citep[e.g.,][]{Santos2018,Jiang2019,SGK2019} and increase the diversity of dwarf galaxy rotation curves.

A finite self-gravitating system has negative heat capacity and the heat conduction will eventually result in the ``gravothermal catastrophe'' of the system~\citep[e.g.,][]{Bell1968,Bell1980}. In the eSIDM case, effective heat conduction is realized by dark matter self-interactions and the inner cores of isolated eSIDM haloes will ultimately experience gravothermal collapse and cuspy density profiles will reappear~\citep[e.g.,][]{Burkert2000,Kochanek2000,Balberg2002,Colin2002,Koda2011,Vogelsberger2012,Elbert2015,Correa2021}. However, for the most favored elastic self-interaction cross-sections $\sim \operatorname{0.1\,-\,1}\cpm$ (assuming velocity-independent), the ``gravothermal catastrophe'' would not have enough time to happen in haloes within their typical lifetime. In the presence of dissipative self-interactions, the gravothermal evolution of a halo can be accelerated significantly, which affects the structure of dwarf galaxies within a Hubble time. \citet{Essig2019} recently used an semi-analytical fluid model to investigate the structure of isolated spherically symmetric haloes in dissipative SIDM (dSIDM) and presented the first constraint on the energy loss and cross-section of dSIDM. This work was followed by \citet{Huo2019} with non-cosmological N-body simulations of isolated dark matter haloes with the NFW profile~\citep{NFW1996} initially. Moreover, when dissipation of dark matter self-interaction is strong enough, a patch of dark matter could lose its kinetic energy faster than rebuilding hydrostatic equilibrium with surrounding matter. Substructures of dissipative dark matter, e.g. dark disks and dark stars, could be generated under this circumstance. For example, dark matter scenarios with a highly dissipative component~(sourced by an $U(1)\,$-like hidden sector) have been studied by \citet{Fan2013,Fan2013b,Fan2014,Randall2015,Foot2013,Foot2015b,Foot2016,Chang2019}. \citet{Randall2015} claimed that a dark disk composed of highly dissipative dark matter could appear and help explain the exotic mass-to-light ratios of some Milky Way satellites. However, the analytical or semi-analytical studies discussed above were limited to isolated DMO haloes with various geometrical simplifications. The influences of baryonic physics, hierarchical halo mergers, deviations from simple fluid approximations in dark matter haloes were not properly captured in these previous studies. In addition, multi-component dark matter with inelastic interactions have been considered in simulations in \citet{Todoroki2019,Vogelsberger2019}, but the dominant process is exothermic in these studies.

In this paper, we perform the first study of dSIDM models using cosmological baryonic (hydrodynamical) zoom-in simulations of galaxies. We aim at studying the evolution tracks of dSIDM haloes and looking for properties of dSIDM haloes that distinguish them from their CDM counterparts. These simulations have incorporated the FIRE-2 model~\citep{Hopkins2018} for hydrodynamics and galaxy formation physics that could produce galaxies consistent with various local and high redshift observables in collisionless CDM simulations~\citep[e.g.,][]{Ma2018,SGK2018,Hafen2019,SGK2019}. The setup also enables predictions in the regime where hierarchical mergers and strong non-linear gravitational effects could drive systems away from the idealized analytical solutions. All these factors allow more robust constraints on dSIDM models. The paper is arranged as follows: In Section~\ref{sec:sim}, we discuss the details of the simulations and briefly introduce the dSIDM models we study. We derive relevant time scales for dSIDM haloes analytically in Section~\ref{sec:timescale} and study the stellar masses and host halo masses of simulated dwarf galaxies in Section~\ref{sec:halo_prop}. Then we present the mass density profiles of simulated dwarf galaxies and quantitatively study the impact of dissipation on galaxy structure in Section~\ref{sec:denpro}. We study the kinematic properties of dark matter and the shapes of haloes in simulations in Section~\ref{sec:kin} and Section~\ref{sec:shape}. Subsequently, in Section~\ref{sec:discussion}, we use analytical methods to explain the phenomena in dSIDM simulations and summarize the evolution pattern of dSIDM haloes in different regimes. In Section~\ref{sec:vary}, we explore the results of simulations with other choices of $f_{\rm diss}$ as well as the DMO simulations and compare their differences from the fiducial simulations. The summary and conclusion of the paper are presented in Section~\ref{sec:conclusion}.

\section{Simulations}
\label{sec:sim}

\begin{figure}
    \centering
    \includegraphics[width=0.49\textwidth]{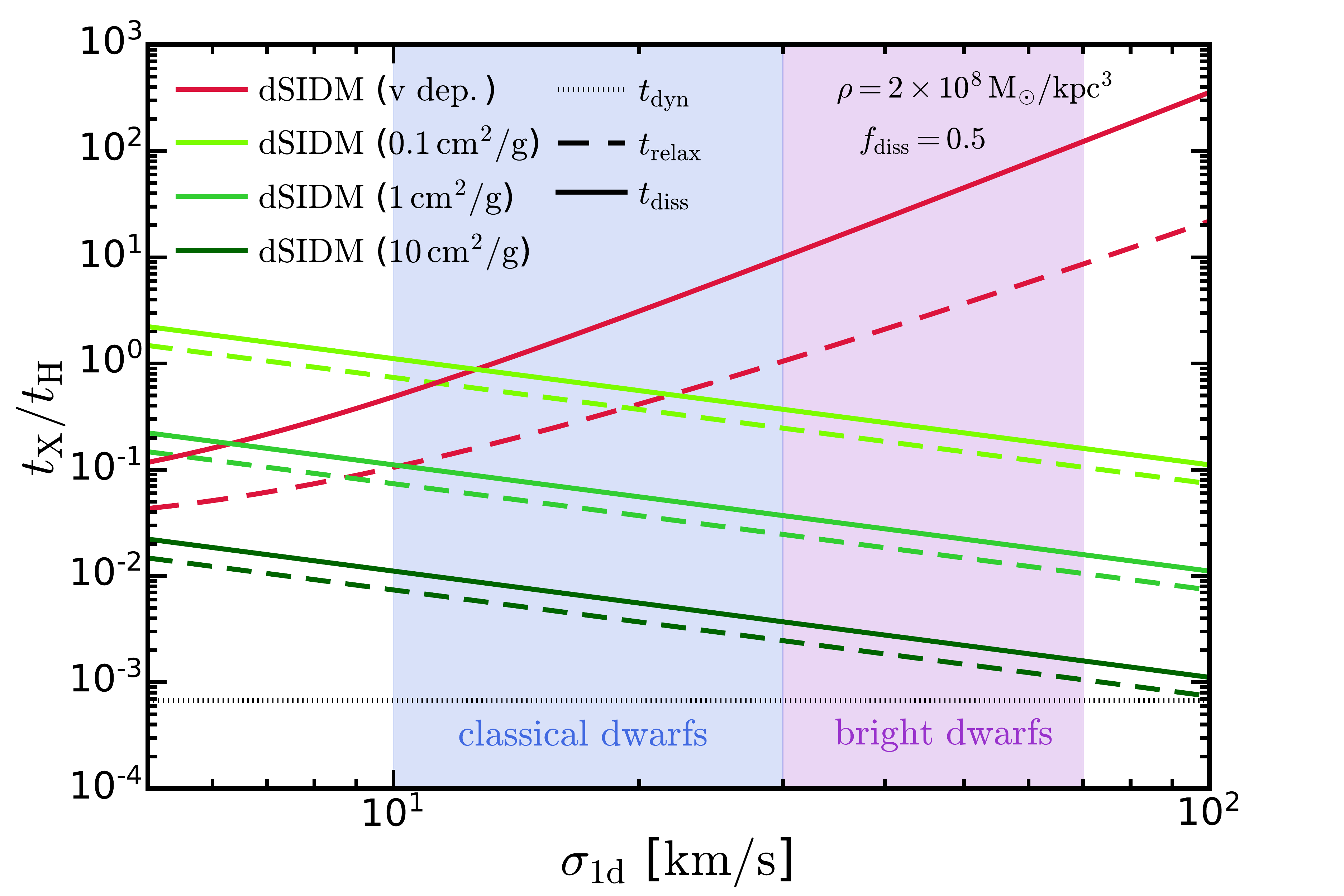}
    \includegraphics[width=0.49\textwidth]{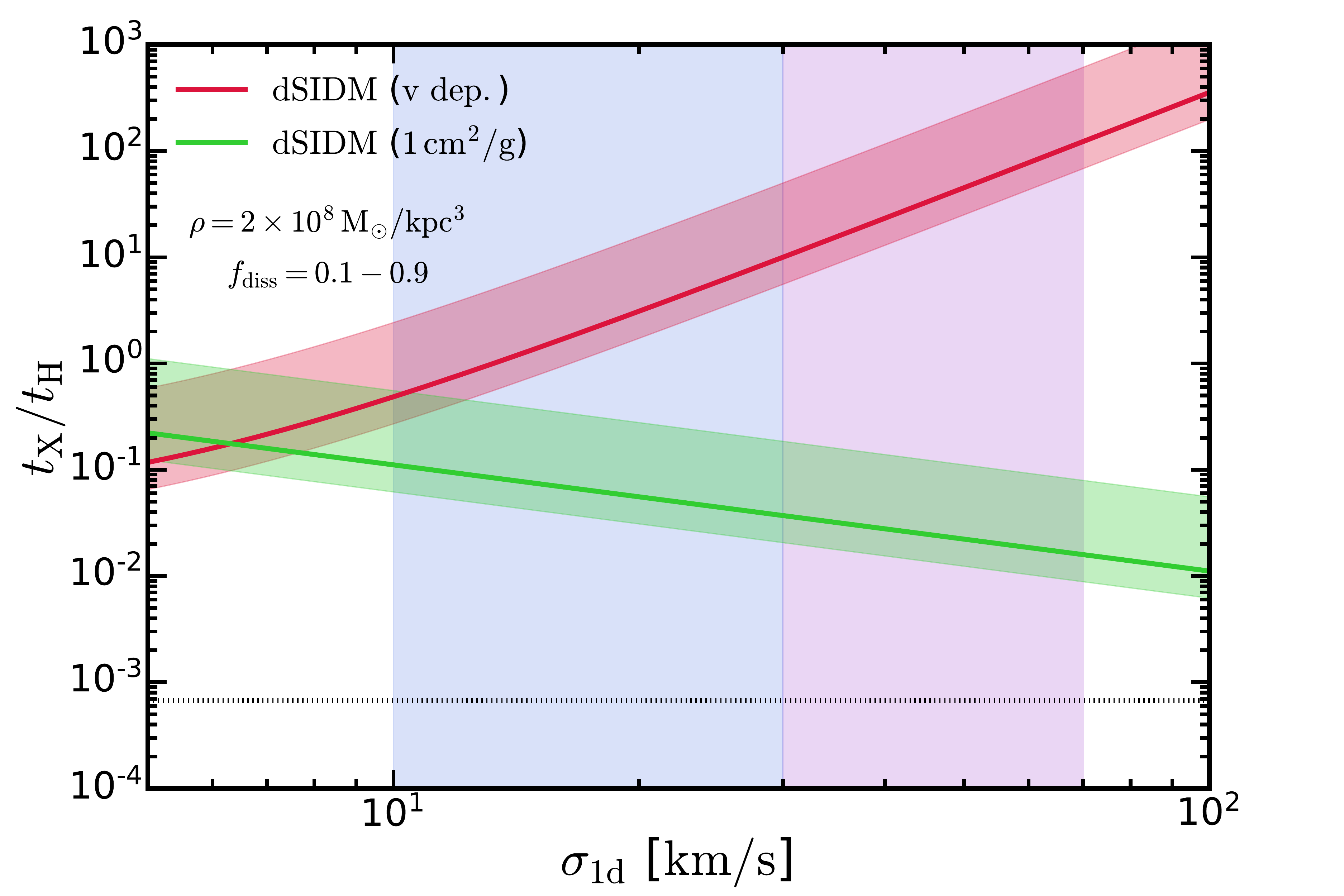}
    \caption{{\it Top}: \textbf{Relevant time scales of the physical processes involved in dSIDM haloes versus one-dimensional velocity dispersion of the system.} We have assumed that the local dark matter density is $\rho_{\rm dm} = 2\times 10^{8}\,\msun/{\kpc}^{3}$, a typical value at dwarf galaxy centers. We show the collision time scale ($t_{\rm coll}$) and dissipation time scales ($t_{\rm diss}$) of all the dSIDM models studied in this paper as well as the dynamical time scale ($t_{\rm dyn}$). All the time scales are normalized by the Hubble time scale at $z=0$ ($t_{\rm H}\equiv 1/{\rm H}_{0}$). The dissipation time scales are calculated assuming $f_{\rm diss}=0.5$. The shaded regions show the typical one-dimensional velocity dispersions in the classical (e.g. Milky Way satellites) and bright dwarf galaxies (e.g. LSB galaxies). In dwarf galaxies, dissipation and collision time scales are much larger than the dynamical time scale, but can become considerably shorter than the Hubble time scale. The velocity-dependent model becomes less dissipative ($t_{\rm diss}/t_{\rm H}$ becomes larger) in more massive galaxies (with larger velocity dispersion) while models with constant cross-sections become more dissipative. {\it Bottom}: \textbf{Dissipation time scales versus one-dimensional velocity dispersion of the system with $f_{\rm diss}$ varying from $0.1$ to $0.9$.} The symbols are the same as the top panel. For each model, the upper boundary of the shaded region corresponds to the case $f_{\rm diss}=0.1$ and the lower boundary corresponds to the case $f_{\rm diss}=0.9$.
    }
    \label{fig:timescale}
\end{figure}

\subsection{Overview of the simulation suite}
We present the new FIRE-2 dSIDM simulation suite, which consists of $\sim\,$45 cosmological hydrodynamical zoom-in simulations of galaxies chosen at representative mass scales with CDM, eSIDM and dSIDM models. The simulations here are part of the Feedback In Realistic Environments project~\citep[FIRE, ][]{Hopkins2014}, specifically the ``FIRE-2'' version of the code with details described in \citet{Hopkins2018}. The simulations adopt the code {\sc Gizmo}~\citep{Hopkins2015}, with hydrodynamics solved using the mesh-free Lagrangian Godunov ``MFM'' method. The simulations include heating and cooling from a meta-galactic radiation background and stellar sources in the galaxies, star formation in self-gravitating molecular, Jeans-unstable gas and stellar/supernovae/radiation feedback. The FIRE physics, source code, and numerical parameters are identical to those described in \citet{Hopkins2018,SGK2019b}. For dwarf galaxies, the baryonic particle masses of simulations are $m_{\rm b} \simeq 250 \operatorname{-} 2000\msun$. For Milky Way-mass galaxies, the high-resolution `latte' runs have $m_{\rm b} = 7000\msun$ while the low-resolution runs have $m_{\rm b} = 56000\msun$. In all simulations, the dark matter particle masses are roughly five times larger, according to the universal baryon fraction. For dwarf galaxies, the minimum gravitational force softening length reached by gas in the simulations is $h_{\rm b} \simeq 0.5 \operatorname{-} 2\pc$. For Milky Way-mass galaxies, the value is $h_{\rm b} \simeq 0.3 \operatorname{-} 0.5\pc$ ($1.4\pc$) for high-resolution (low-resolution) runs. The physical dark matter force resolution of the simulations of dwarf (Milky Way-mass) galaxies is $\epsilon_{\rm dm} = 40 \pc$ ($30 \pc$). Force softening for gas uses the fully conservative adaptive algorithm from \citet{Price2007}, meaning that the gravitational force assumes the identical mass distribution as the hydrodynamic equations (resulting in identical hydrodynamic and gravitational resolution). The simulations are identified with the main ``target'' halo around which the high-resolution zoom-in region is centered. In post-processing, we identify subhaloes (of the main ``target'' halo) with the {\sc Rockstar}~\citep{Behroozi2013} halo finder and create merger trees of haloes (subhaloes) with the code {\sc Consistent Trees}~\citep{Behroozi2012,Behroozi2013b}. As shown in Table~\ref{tab:sims}, the simulation suite consists of one ultra faint dwarf (\textbf{m09}), three classical dwarf galaxies (\textbf{m10q, m10b, m10v}), three bright dwarf galaxies (\textbf{m11a, m11b, m11q}) and four Milky Way-mass galaxies (\textbf{m11f, m12i, m12f, m12m}). The analysis in this paper will primarily focus on the classical and bright dwarf galaxies and we defer analysis on Milky Way-mass galaxies to a follow-up work.

\subsection{Dissipative dark matter parameterization}

Dark matter self-interactions are simulated in a Monte-Carlo fashion following the implementation in \citet{Rocha2013} and the scattering process is assumed to be isotropic. In this paper, we study a simplified empirical dSIDM model: two dark matter particles lose a constant fraction, $f_{\rm diss}$, of their kinetic energy in the center of momentum frame when they collide with each other. The extreme version of this type of interaction is the fusion process ($f_{\rm diss}=1$) of dark matter composites. Such model has been discussed in the context of self-interacting asymmetric dark matter \citep[e.g.,][]{Wise2014,Wise2015,Detmold2014,Krnjaic2015,Gresham2018}. Self-interaction mediated by a scalar mediator can give rise to strong attractive forces, and large bound states of dark matter (``nuggets'') can form in the absence of competing repulsive forces \citep{Wise2014,Gresham2018}. These dark nuggets are the smoking gun signature of fermionic asymmetric dark matter \citep[see][for a review]{Zurek2014}. The residual self-interaction between nuggets is highly dissipative and mimics the fusion process of nucleons.

Beyond this, dissipative portals present in other SIDM models as well. For strongly-interacting dark composites in a hidden non-Abelian sector~\citep[e.g.,][]{Alves2010,Cline2014,Boddy2014}, dark matter will consist of dark baryons/mesons and glueballs (or glueballinos if incorporating super-symmetry). For example, inelastic scattering to excited state(s) and glueball emission will be possible when glueballinos have mass $m_{\chi} \gg \Lambda$  \citep{Boddy2014}. Hyperfine-transitions of dark mesons/baryons have been suggested in \citet{Alves2010, Alves2010b} and the late time up-scattering to excited states can induce dissipation. Excited states and dissipative (endothermic) processes are also ubiquitous in generic SIDM models \citep[e.g.,][]{Arkani-Hamed2009,Loeb2011}, models featuring a dark SU(2)-like sector \citep[e.g.,][]{Chen2009,Cirelli2010} or a dark U(1)-like sector \citep[e.g.,][]{Kaplan2010,Fan2013,Schutz2015,Foot2015}. However, the exact behaviour of dissipation is model-dependent and could be quite different from what we are modelling here.

For each galaxy, we run simulations with a default dissipation fraction $f_{\rm diss} = 0.5$~\footnote{Other choices of $f_{\rm diss}$ are explored with \textbf{m09} in Section~\ref{sec:m09}.} and with constant self-interaction cross-sections $(\sigma/m) = 0.1/1/10\,\cpm$ or a velocity-dependent cross-section model:
\begin{equation}
    \dfrac{\sigma(v)}{m} = \dfrac{(\sigma/m)_{0}}{1+(v/v_0)^4},
    \label{eq:cx_vel}
\end{equation}
where the fiducial choice of parameters is $(\sigma/m)_{0} = 10 \cpm$ and $v_0 = 10 \kms$. The velocity dependence of the self-interaction cross-section is empirically motivated by the relatively tight constraints on SIDM at galaxy cluster scale~\citep[e.g.,][]{Markevitch2004,Randall2008,Kaplinghat2016} and the relatively high cross-section needed to solve some small-scale problems~\citep[e.g.,][]{Vogelsberger2012,Rocha2013,Zavala2013,Elbert2015,Kaplinghat2016}. Meanwhile, the velocity dependence is a generic feature of many particle physics realizations of dark matter. The asymptotic $(v/v_0)^{-4}$ velocity dependence we adopt is motivated by particle physics models featuring dark matter self-interactions mediated by light gauge bosons~\citep[e.g.,][]{Feng2009,Kaplan2010,CyrRacine2013,Boddy2016,Zhang2017}. The sharp decline in cross-section could also appear in some models of strongly interacting composites. In these models, when the de Broglie wavelength of the particle become smaller than the characteristic length scale of the interaction, $\sim 1/\Lambda_{\rm dm}$, the self-interaction cross-section is expected to drop significantly~\citep[e.g.,][]{Boddy2014, Cline2014, Tulin2018}. 

\section{Relevant time scales}
\label{sec:timescale}

In this section, we derive analytical formulae for relevant time scales in dSIDM haloes, including the dynamical time scale, the collision time scale and the dissipation time scale. These analytical formulae can be used to understand the influence of dissipation on galaxy structures in different circumstances. We will present results for models with constant and velocity-dependent cross-sections, respectively.

\subsection{Dynamical time scale}
The local dynamical time scale in a system is defined as
\begin{align}
    t_{\rm dyn} & \equiv \sqrt{\dfrac{1}{4\pi G \rho}} \nonumber \\
    & = 0.0042 \Gyr \, \Big(\dfrac{\rho}{10^9 \msun/\kpc^3}\Big)^{-1/2},
    \label{eq:tdyn}
\end{align}
where $G$ is the gravitational constant and $\rho$ is the local matter density. At the centers of dwarf galaxies, the mass density is dominated by dark matter, so $\rho$ is simply the local dark matter mass density. 

\subsection{Collision time scale}
\label{sec:timescale_coll}
The collision time scale of dark matter self-interaction is
\begin{equation}
    t_{\rm coll} \equiv \dfrac{1}{\langle \rho v_{\rm rel} \dfrac{\sigma}{m} \rangle},
\end{equation}
where $\rho$ is local dark matter mass density, $v_{\rm rel}$ is the relative velocity between dark matter particles and $\langle ... \rangle$ denotes the average over all possible encounters. This measures the time scale that one dark matter particle is expected to have one self-interaction with any other dark matter particles. For simplicity, we assume that the velocities of dark matter particles locally obey the Maxwell-Boltzmann distribution. Therefore, the average can be treated as a thermal average
\begin{equation}
    \langle X \rangle = \dfrac{1}{2\sqrt{\pi} \sigma_{\rm 1d}^3}\int_0^{\infty}{\rm d}v_{\rm rel} v_{\rm rel}^{2} e^{-v_{\rm rel}^2/4\sigma_{\rm 1d}^2} X,
    \label{eq:thermal_average}
\end{equation}
where $\sigma_{\rm 1d}$ is the local one-dimensional velocity dispersion of dark matter. After taking the thermal average, the collision time scale is
\begin{align}
    t_{\rm coll} & = 0.206\Gyr\, \Big(\dfrac{\rho}{10^9 \msun/\kpc^3}\Big)^{-1} \Big(\dfrac{(\sigma/m)}{1\cpm}\Big)^{-1} \Big(\dfrac{\sigma_{\rm 1d}}{10\kms}\Big)^{-1} \nonumber \\ 
    &  \hspace{3.8cm} {\rm [constant\, \operatorname{cross-section}]}; \nonumber \\
    t_{\rm coll} & = 0.661\Gyr\, \Big(\dfrac{\rho}{10^9 \msun/\kpc^3}\Big)^{-1} \Big(\dfrac{(\sigma/m)_{0}}{10\cpm}\Big)^{-1} \Big(\dfrac{\sigma_{\rm 1d}}{10\kms}\Big)^{-1} \nonumber \\  
    & \Big(\dfrac{\sigma_{\rm 1d}}{v_0}\Big)^4 \, \Bigg[- 2 {\rm Ci}\Big(\dfrac{v_0^2}{4\sigma_{\rm 1d}^2}\Big) \cos{\Big(\dfrac{v_0^2}{4\sigma_{\rm 1d}^2}\Big)} + \sin{\Big(\dfrac{v_0^2}{4\sigma_{\rm 1d}^2}\Big)} \Big(\pi - 2{\rm Si}\Big(\dfrac{v_0^2}{4\sigma_{\rm 1d}^2}\Big)\Big) \Bigg]^{-1} \nonumber \\ 
    & \simeq 0.165\Gyr\, \Big(\dfrac{\rho}{10^9 \msun/\kpc^3}\Big)^{-1} \Big(\dfrac{(\sigma/m)_{0}}{10\cpm}\Big)^{-1} \Big(\dfrac{\sigma_{\rm 1d}}{10\kms}\Big)^{-1} \nonumber \\  
    & \Big(\dfrac{\sigma_{\rm 1d}}{v_0}\Big)^4 \, \ln{\Big(\dfrac{\sigma_{\rm 1d}}{v_0}\Big)}^{-1} \hspace{4.4cm} [\sigma_{\rm 1d}\gg v_{0}] \nonumber \\ 
    &  \hspace{3.8cm} {\rm [\operatorname{velocity-dependent}\, \operatorname{cross-section}]},
    \label{eq:tcoll}
\end{align}
where ${\rm Si}(x)=\int_{0}^{x} {\rm d}t\, {\rm sin}(t)/t$ and ${\rm Ci}(x)= - \int_{x}^{\infty} {\rm d}t\, {\rm cos}(t)/t$ are sine and cosine integrals, $(\sigma/m)_{0}$ and $v_{\rm 0}$ are parameters of the velocity-dependent cross-section. For our fiducial choice of $v_{0}=10\kms$, galaxies of masses $\gtrsim 10^{11}\msun$ (massive dwarfs/Milky Way-mass galaxies) will have velocity dispersions in the limit $\sigma_{\rm 1d} \gg v_0$. We can see that the collision time scale of the velocity-dependent model is usually much larger than the constant cross-section model after the thermal average. This is due to the velocity suppression of collisions between particles with high relative velocities, which contribute more to the total interaction rate. In addition, the collision time scale in different models scales with velocity dispersion in opposite ways. For the models with {\it \textbf{constant}} cross-sections, the collision time scale is {\it \textbf{shorter}} in systems with higher densities or higher velocity dispersions, which indicates that {\it \textbf{self-interaction has stronger impact in more massive systems}}. On the other hand, for the {\it \textbf{velocity-dependent}} model, the collision time scale sharply {\it \textbf{increases}} in systems with higher velocity dispersions, which indicates that {\it \textbf{self-interaction has weaker impact in more massive systems}}. 

\subsection{Dissipation time scale}
The dissipation time scale here is defined as the time scale for an order unity fraction of local dark matter kinetic energy to be dissipated away through dark matter self-interactions
\begin{equation}
    t_{\rm diss} \equiv \dfrac{3}{2}\rho \sigma_{\rm 1d}^{2}/C, 
\end{equation}
where $\sigma_{\rm 1d}$ is the one-dimensional velocity dispersion and $C$ is the effective cooling rate defined as
\begin{equation}
    C \equiv \Big\langle n (\rho v_{\rm rel} \dfrac{\sigma}{m}) E_{\rm loss}  \Big\rangle = \Big\langle \rho^{2} \,\dfrac{\sigma}{m}\,v_{\rm rel}\dfrac{E_{\rm loss}}{m} \Big\rangle,
    \label{eq:coolingrate}
\end{equation}
where $n$ is the local number density of dark matter particles, $E_{\rm loss}$ is the kinetic energy loss per collision in the center of momentum frame and $\langle ... \rangle$ again denotes the thermal average. For the fractional dissipation model we study in this paper, $E_{\rm loss}/m = (1/4)f_{\rm diss} v_{\rm rel}^2$. The dissipation time scale measures how fast the kinetic energy is dissipated away from the system and, after order one dissipation time scale, the local dark matter structure is expected to be dramatically affected. 

After taking the thermal average, the dissipation time scale is
\begin{align}
    t_{\rm diss} & = \dfrac{3}{4f_{\rm diss}} t_{\rm coll} \nonumber \\
      & = 0.310\Gyr\, \Big(\dfrac{f_{\rm diss}}{0.5}\Big)^{-1} \Big(\dfrac{\rho}{10^9 \msun/\kpc^3}\Big)^{-1} \Big(\dfrac{(\sigma/m)}{1\cpm}\Big)^{-1} \nonumber \\  
      &\Big(\dfrac{\sigma_{\rm 1d}}{10\kms}\Big)^{-1} \nonumber \\  
      &  \hspace{4.2cm} {\rm [constant\, \operatorname{cross-section}]}; \nonumber \\
    t_{\rm diss} & = 7.926\Gyr\, \Big(\dfrac{f_{\rm diss}}{0.5}\Big)^{-1} \Big(\dfrac{\rho}{10^9 \msun/\kpc^3}\Big)^{-1} \Big(\dfrac{(\sigma/m)_{0}}{10\cpm}\Big)^{-1} \nonumber \\  
      &\Big(\dfrac{\sigma_{\rm 1d}}{10\kms}\Big)^{-1} \Big(\dfrac{\sigma_{\rm 1d}}{v_0}\Big)^6 \, \Bigg[ 8\Big(\dfrac{\sigma_{\rm 1d}}{v_0}\Big)^2 - 2 {\rm Ci}\Big(\dfrac{v_0^2}{4\sigma_{\rm 1d}^2}\Big) \sin{\Big(\dfrac{v_0^2}{4\sigma_{\rm 1d}^2}\Big)} \nonumber \\
      & - \cos{\Big(\dfrac{v_0^2}{4\sigma_{\rm 1d}^2}\Big)} \Big(\pi - 2{\rm Si}\Big(\dfrac{v_0^2}{4\sigma_{\rm 1d}^2}\Big)\Big) \Bigg]^{-1} \nonumber \\
      & \simeq 0.991\Gyr\, \Big(\dfrac{f_{\rm diss}}{0.5}\Big)^{-1} \Big(\dfrac{\rho}{10^9 \msun/\kpc^3}\Big)^{-1} \Big(\dfrac{(\sigma/m)_{0}}{10\cpm}\Big)^{-1} \nonumber \\  
      &\Big(\dfrac{\sigma_{\rm 1d}}{10\kms}\Big)^{-1} \Big(\dfrac{\sigma_{\rm 1d}}{v_0}\Big)^4, \hspace{3.6cm} [\sigma_{\rm 1d} \gg v_{0}] \nonumber \\
      & \hspace{4.2cm} {\rm [velocity\, dependent\, model]}.
      \label{eq:tdiss}
\end{align}
In the model with a constant cross-section, the dissipation time scale has the same scaling behaviour as the collision time scale defined in Equation~\ref{eq:tcoll} and differs only by a factor of $0.75/f_{\rm diss}$. In the velocity-dependent model, the scaling behaviours of the dissipation and collision time scales are also quite similar when $\sigma_{\rm 1d} \gg v_{0}$. The dissipation time scale of the velocity-dependent model is usually much larger than the constant cross-section model after thermal average. This again can be attributed to the velocity suppression of collisions between particles with high relative velocities, which not only contribute more to the total collision rate but also induce higher energy loss per collision. Similar to what has been found for the collision time scale, dissipation is {\it \textbf{more significant in more massive systems}} in the models with {\it \textbf{constant}} cross-sections. Dissipation, however, is {\it \textbf{less significant in more massive systems}} in the {\it \textbf{velocity-dependent}} model. 

\begin{figure*}
    \centering
    \includegraphics[width=0.245\textwidth]{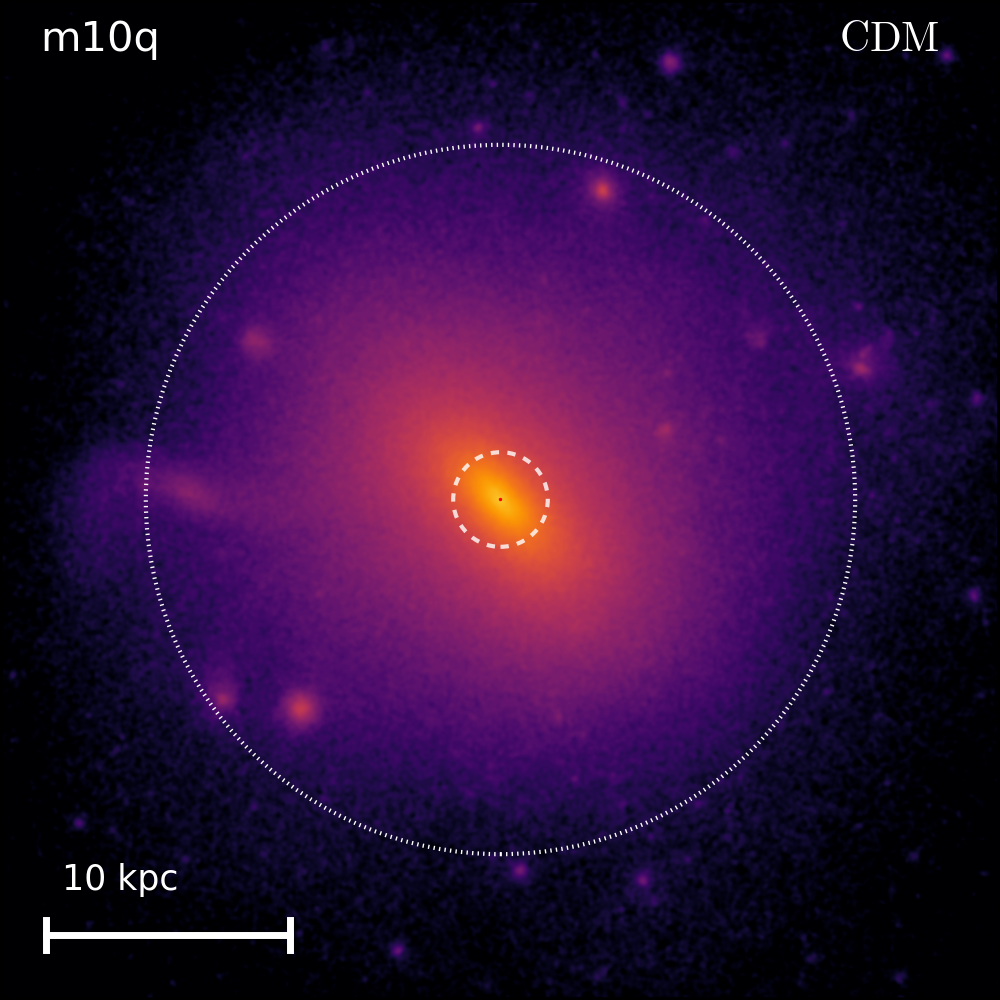}
    \includegraphics[width=0.245\textwidth]{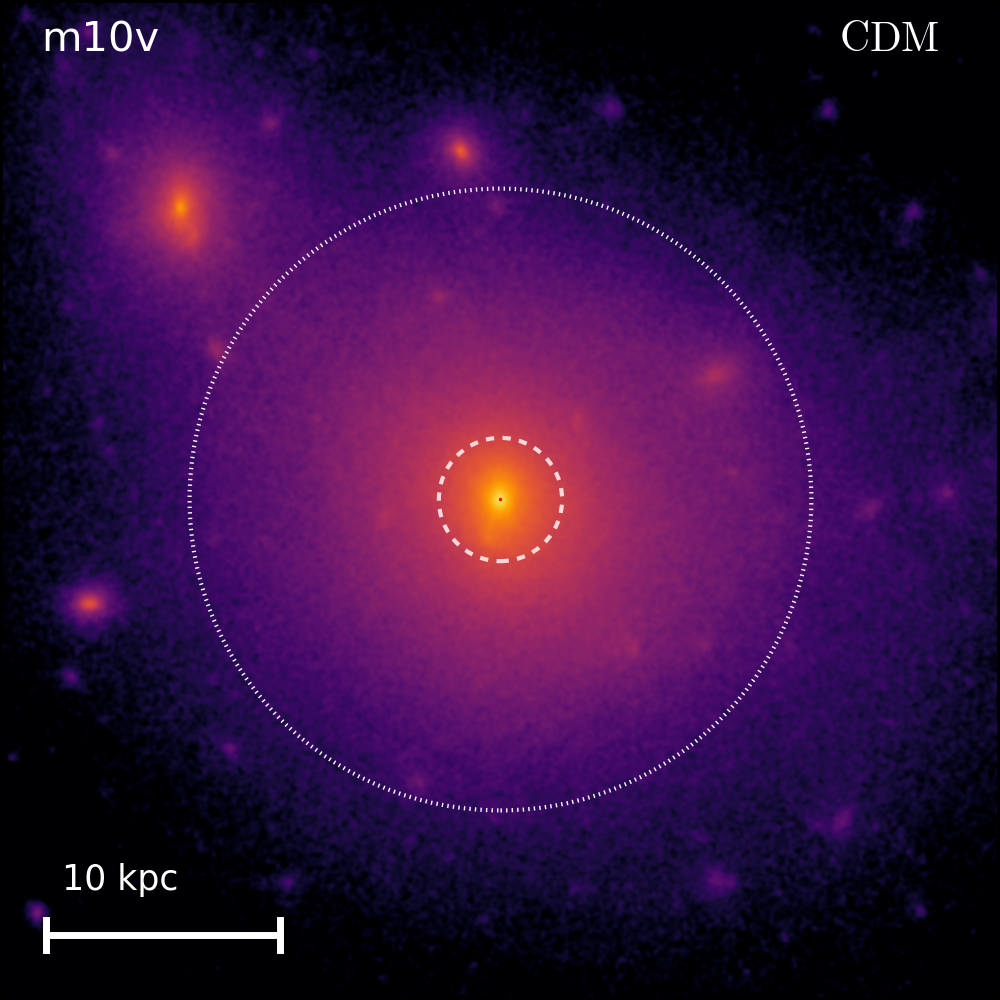}
    \includegraphics[width=0.245\textwidth]{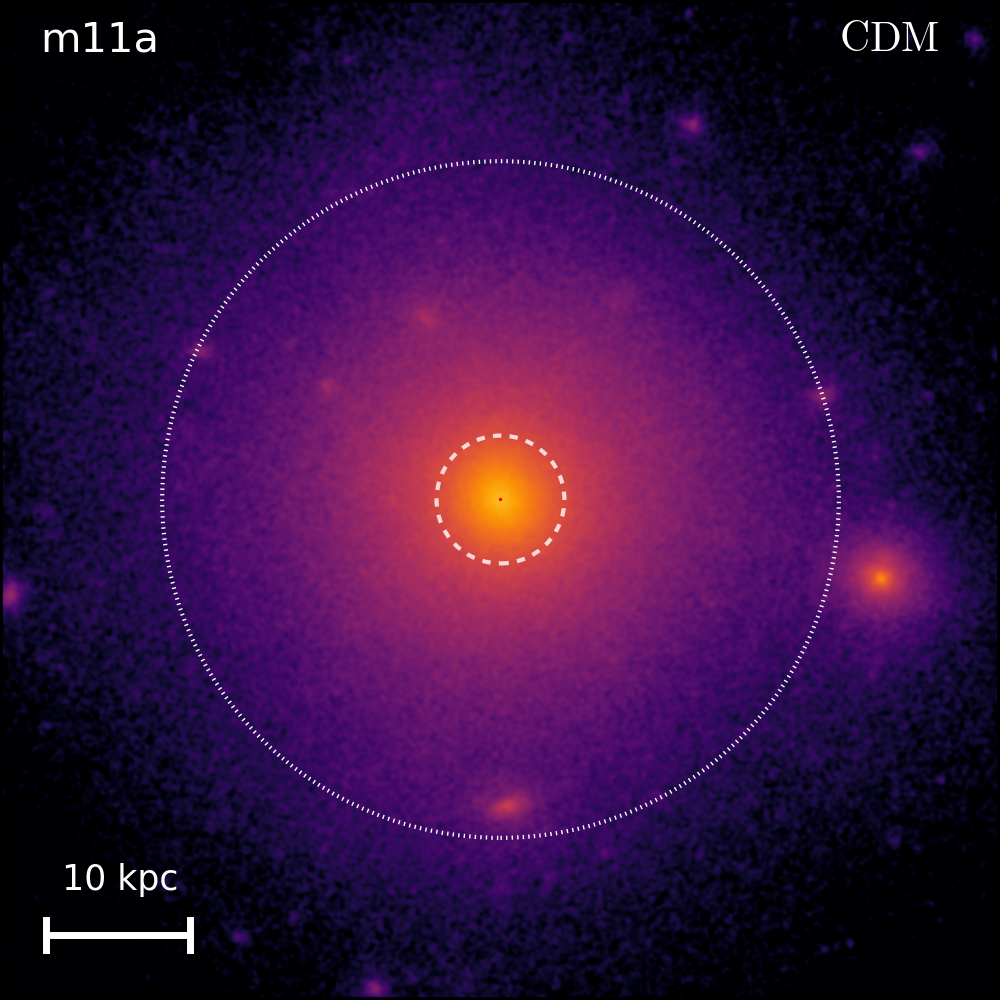}
    \includegraphics[width=0.245\textwidth]{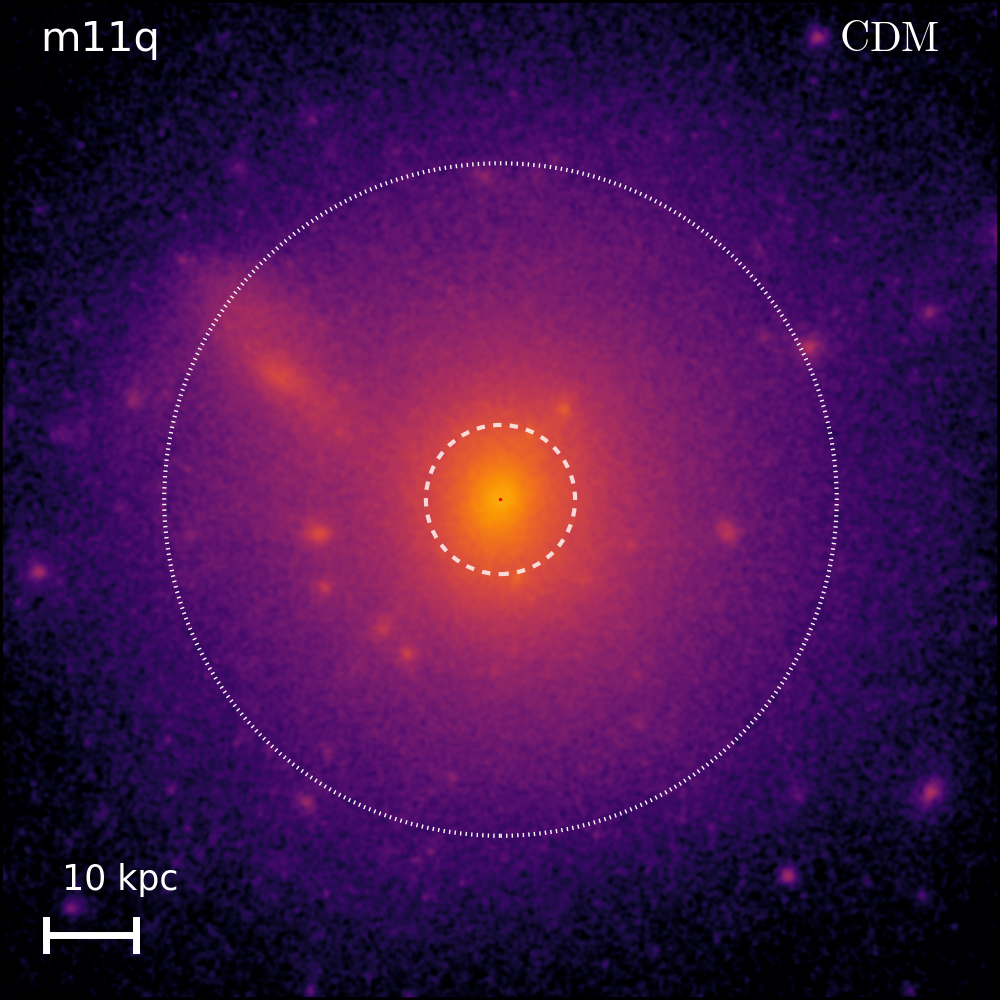}
    \includegraphics[width=0.245\textwidth]{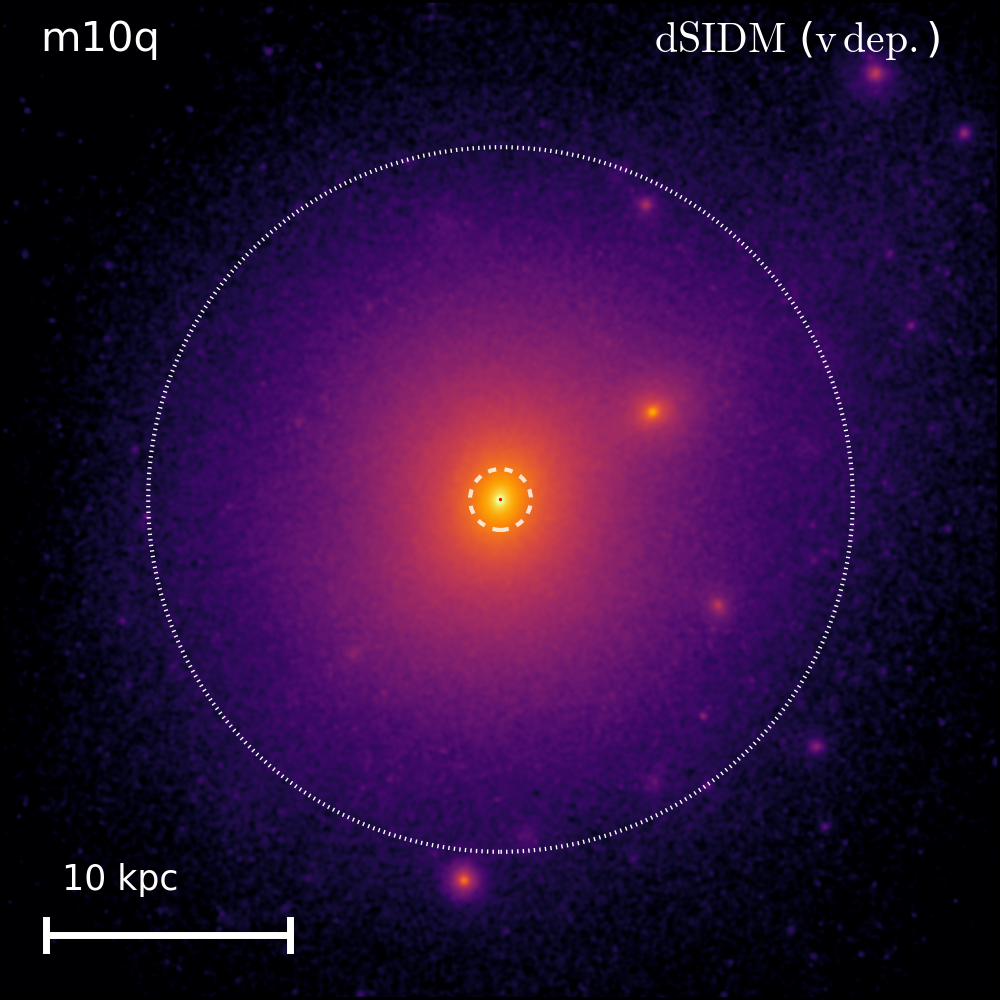}
    \includegraphics[width=0.245\textwidth]{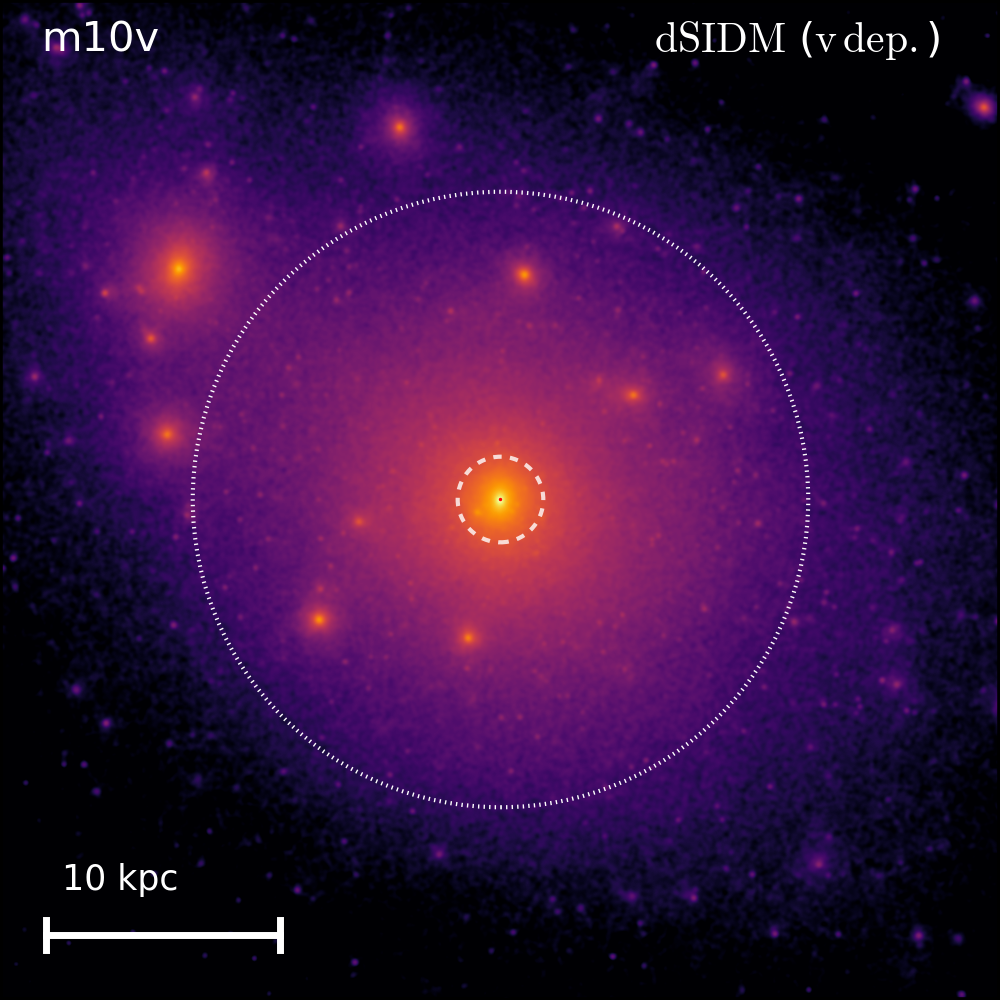}
    \includegraphics[width=0.245\textwidth]{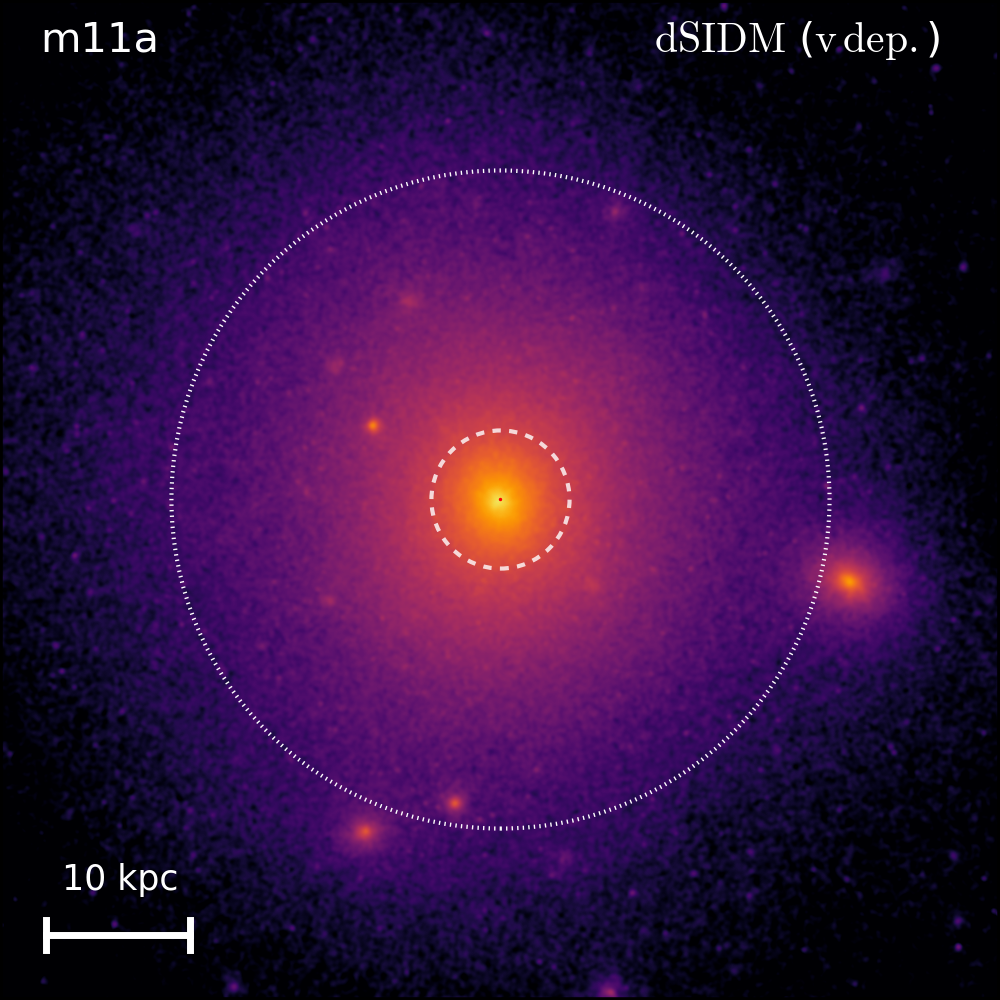}
    \includegraphics[width=0.245\textwidth]{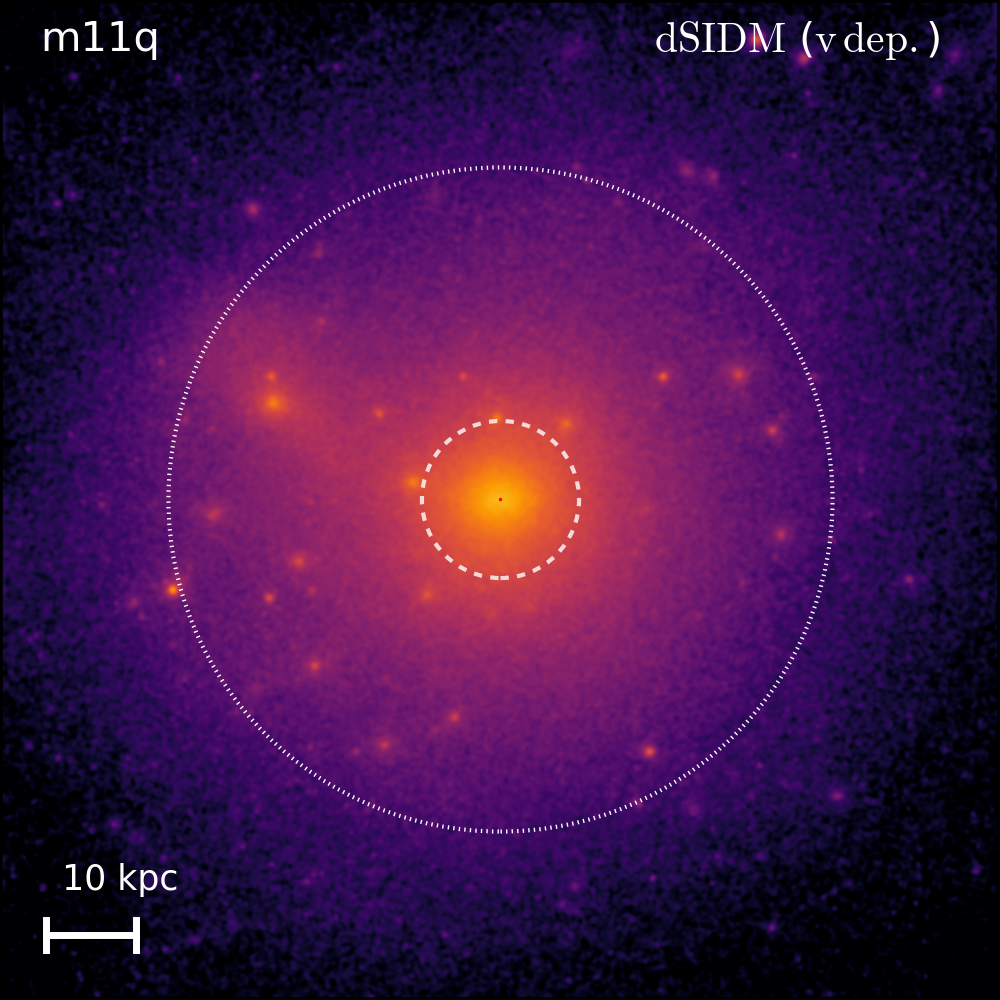}
    \includegraphics[width=0.245\textwidth]{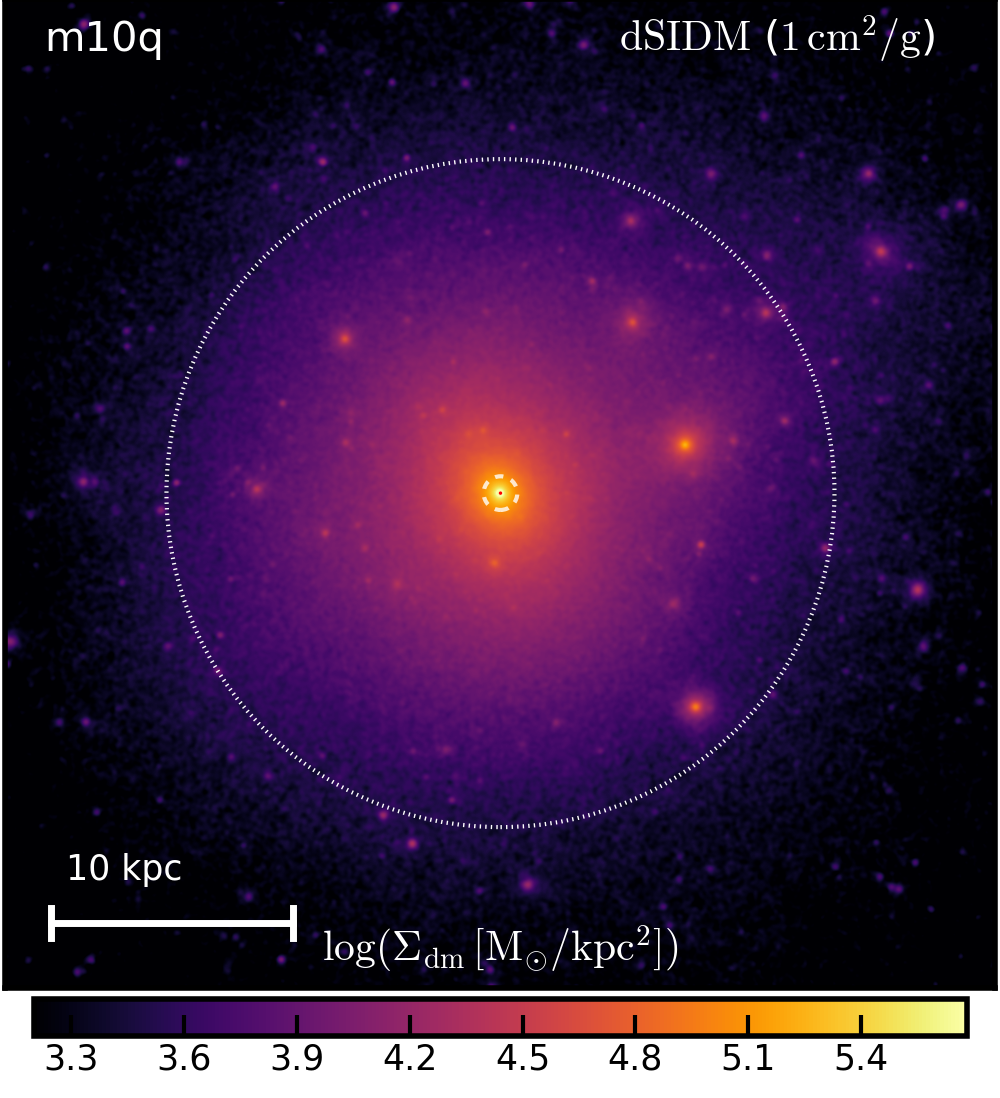}
    \includegraphics[width=0.245\textwidth]{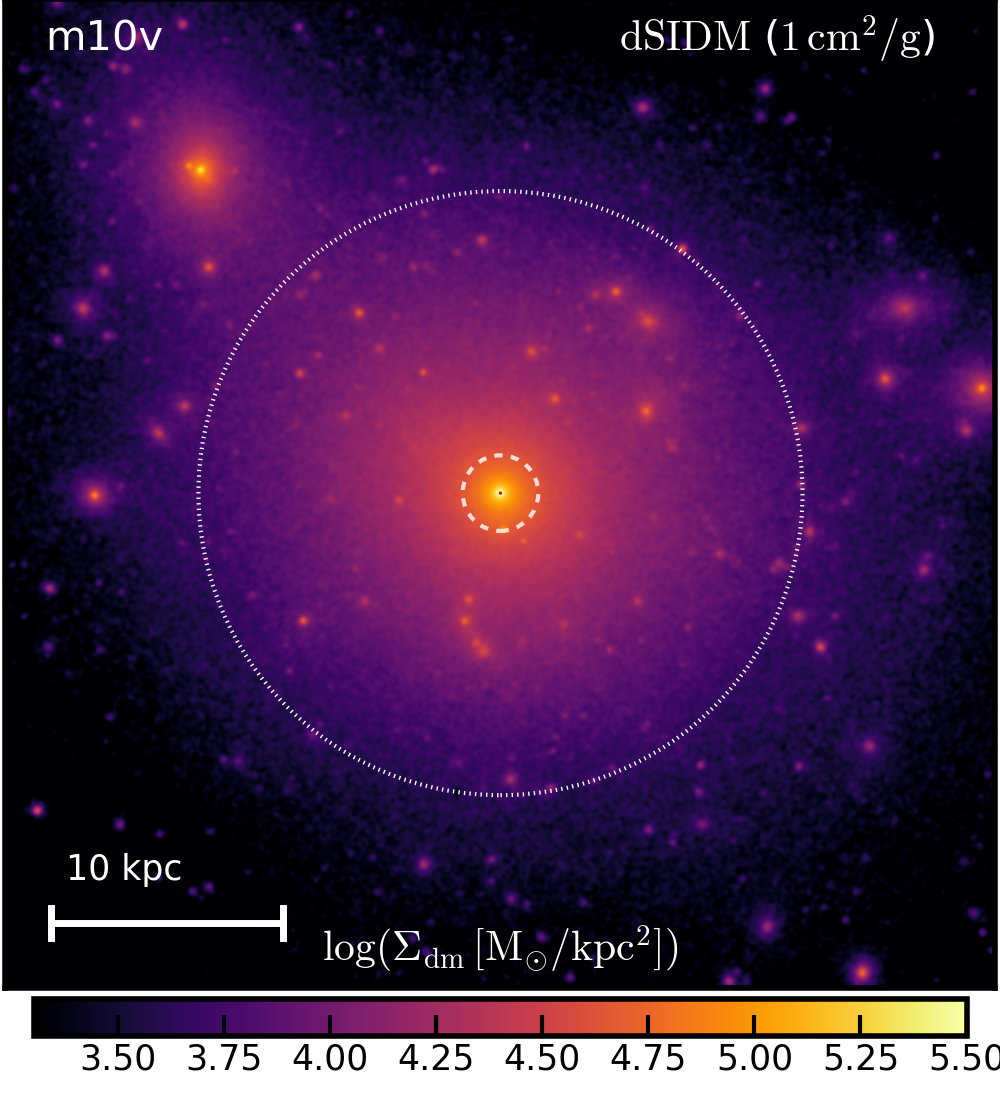}
    \includegraphics[width=0.245\textwidth]{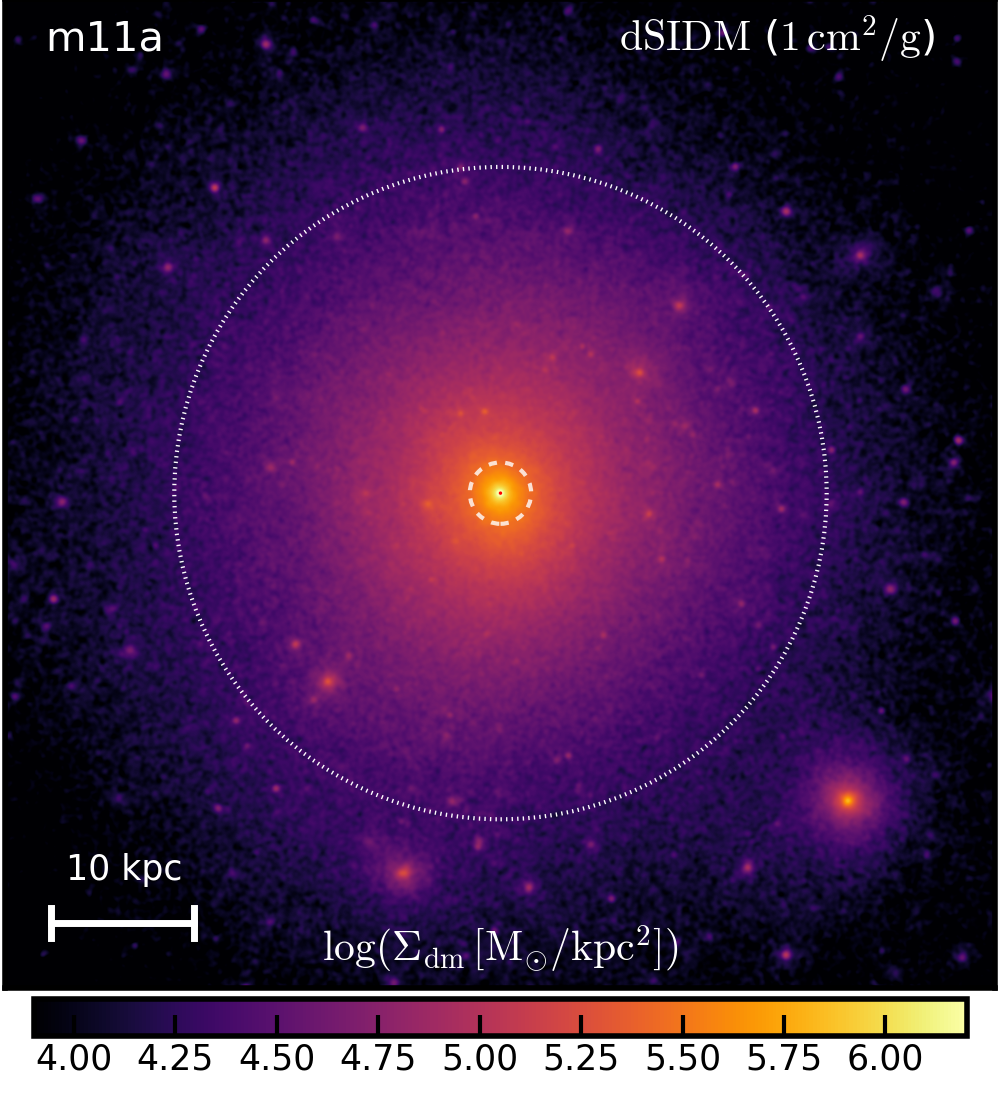}
    \includegraphics[width=0.245\textwidth]{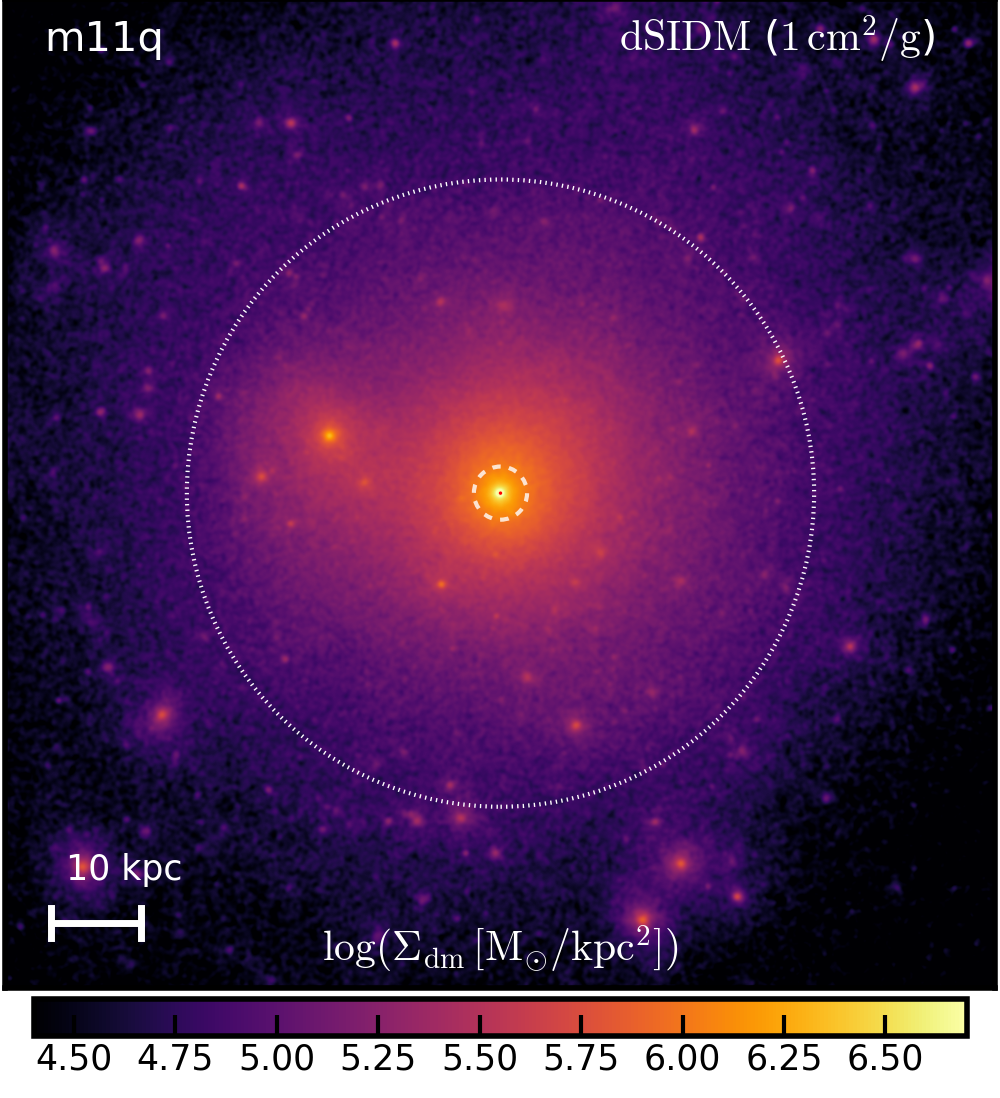}
    
    \caption{\textbf{Visualizations of four dark matter haloes in simulations with CDM versus dSIDM.} The images are dark matter surface density maps, projected along the z-direction of simulation coordinates, at $z=0$ with a logarithmic stretch. The dynamical ranges are adjusted based on the maximum/median intensities of the pixels (but remain the same for the same halo). The side lengths of the images are all chosen to be $0.8\times R_{\rm vir}$ of the CDM run. In the first row, we show the haloes in the CDM. In the second row, we show the haloes in the velocity-dependent dSIDM model. In the third row, we show haloes in the dSIDM model with constant cross-section $1\cpm$. The haloes are ordered from left to right by their virial masses. In each image, the outer dotted circle indicates the radius $R_{\rm 500}$ (the density enclosed is $500$ times the critical density at $z=0$) which represents the overall size of the halo. The inner dashed circle indicates the radius $R_{\rm core}\equiv 10\times R_{\rm 0.1\%}$ (the mass enclosed in a sphere of radius $R_{\rm 0.1\%}$ is $0.1\%$ the virial mass of the halo) which represents the core size of the halo. Comparing the core sizes, the haloes in the dSIDM model are visibly more concentrated than their CDM counterparts. For the velocity-dependent dSIDM model, since the self-interaction cross-section decreases in more massive haloes, the increased concentration of halo is less apparent in more massive haloes. For the dSIDM with constant cross-section, haloes of all masses are consistently more concentrated than their CDM counterparts.}
    \label{fig:image1}
\end{figure*}

In Figure~\ref{fig:timescale}, we show the relevant time scales discussed above as a function of the one-dimensional velocity dispersion of the system; in particular, we show the collision and dissipation time scales of the dSIDM models studied in this paper as well as the dynamical time scale, assuming that the local dark matter mass density is $\rho = 2\times 10^{8}\,\msun/{\kpc}^3$, which is a typical value at dwarf galaxy centers. The time scales are all normalized by the Hubble time scale at $z=0$, roughly representing the lifetime of the system. In the top panel, the dissipation time scales are calculated assuming $f_{\rm diss}=0.5$ while, in the bottom panel, the shaded regions indicate the variation of $t_{\rm diss}$ with $f_{\rm diss}=\operatorname{0.1\,-\,0.9}$. With the vertical shaded regions in both panels, we show the typical ranges of one-dimensional velocity dispersions of the classical (e.g., Milky Way satellites) and bright dwarf galaxies (e.g., LSB galaxies). For the dSIDM models with constant cross-sections, the collision time scales are always proportional to the dissipation time scales and, they are order of magnitude comparable to each other. Both of them are shorter than the Hubble time scale but larger than the dynamical time scale in dwarf galaxies. The dissipation time scale decreases in systems with higher velocity dispersions, so we expect these constant cross-section models to become more dissipative in more massive dwarfs. For the velocity-dependent dSIDM model, the collision and dissipation time scales are no longer proportional to each other, and they both increase as the velocity dispersion increases, opposite to the behaviour of models with constant cross-sections. The dissipation time scale of the velocity-dependent model is comparable to the Hubble time scale in the classical dwarfs but becomes at least an order of magnitude larger than the Hubble time scale in the bright dwarfs, suggesting negligible effects of dissipation in this case.

\begin{figure*}
    \centering
    \rotatebox{90}{\hspace{1.2cm} \textbf{density profile}}
    \includegraphics[trim=0.1cm 0.0cm 3cm 0.6cm, clip, width=0.32\textwidth]{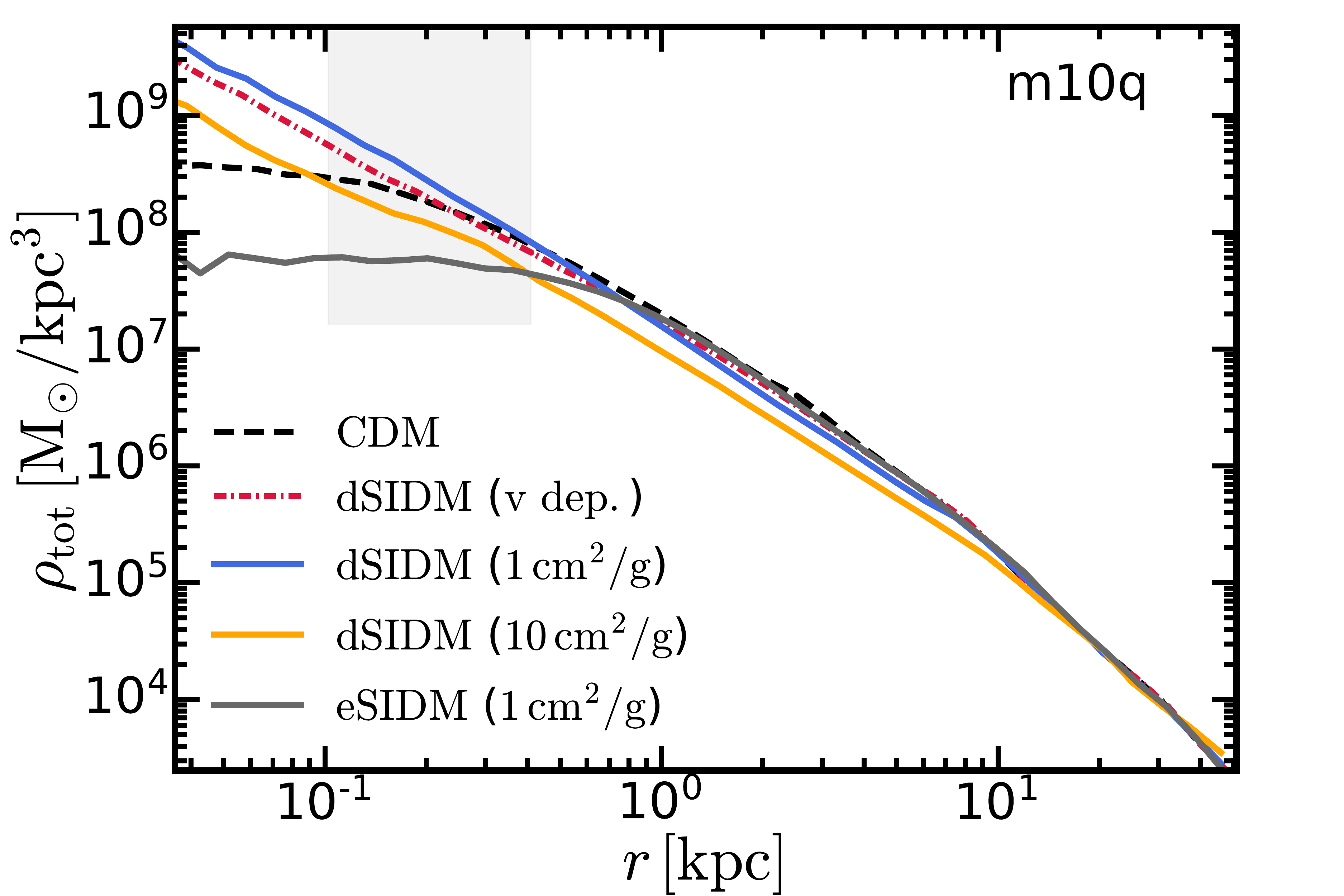}
    \includegraphics[trim=0.1cm 0.0cm 3cm 0.6cm, clip, width=0.32\textwidth]{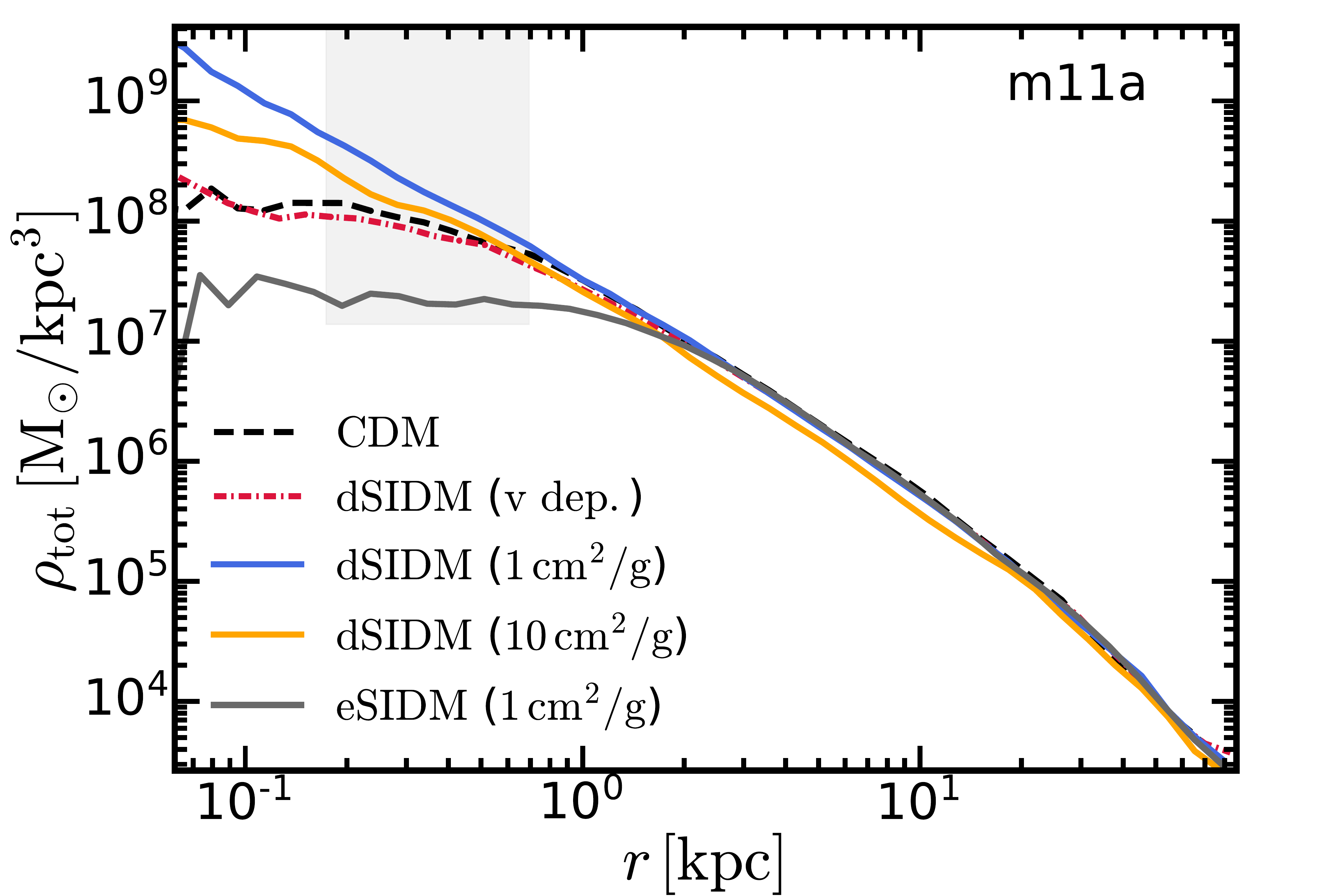}
    \includegraphics[trim=0.1cm 0.0cm 3cm 0.6cm, clip, width=0.32\textwidth]{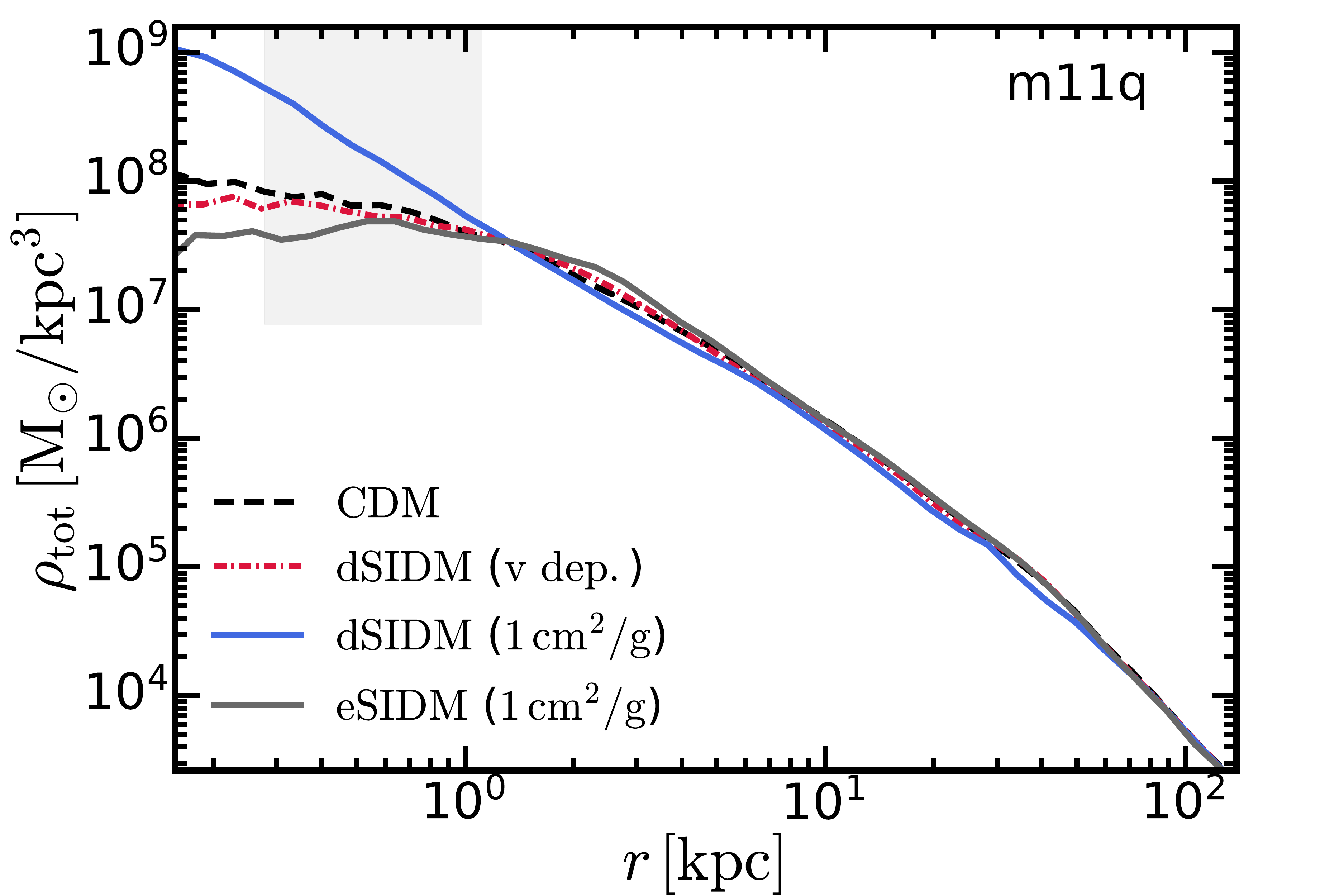}
    
    \rotatebox{90}{\hspace{1.2cm} \textbf{circular velocity}}
    \includegraphics[trim=0.1cm 0.0cm 3cm 0.6cm, clip, width=0.32\textwidth]{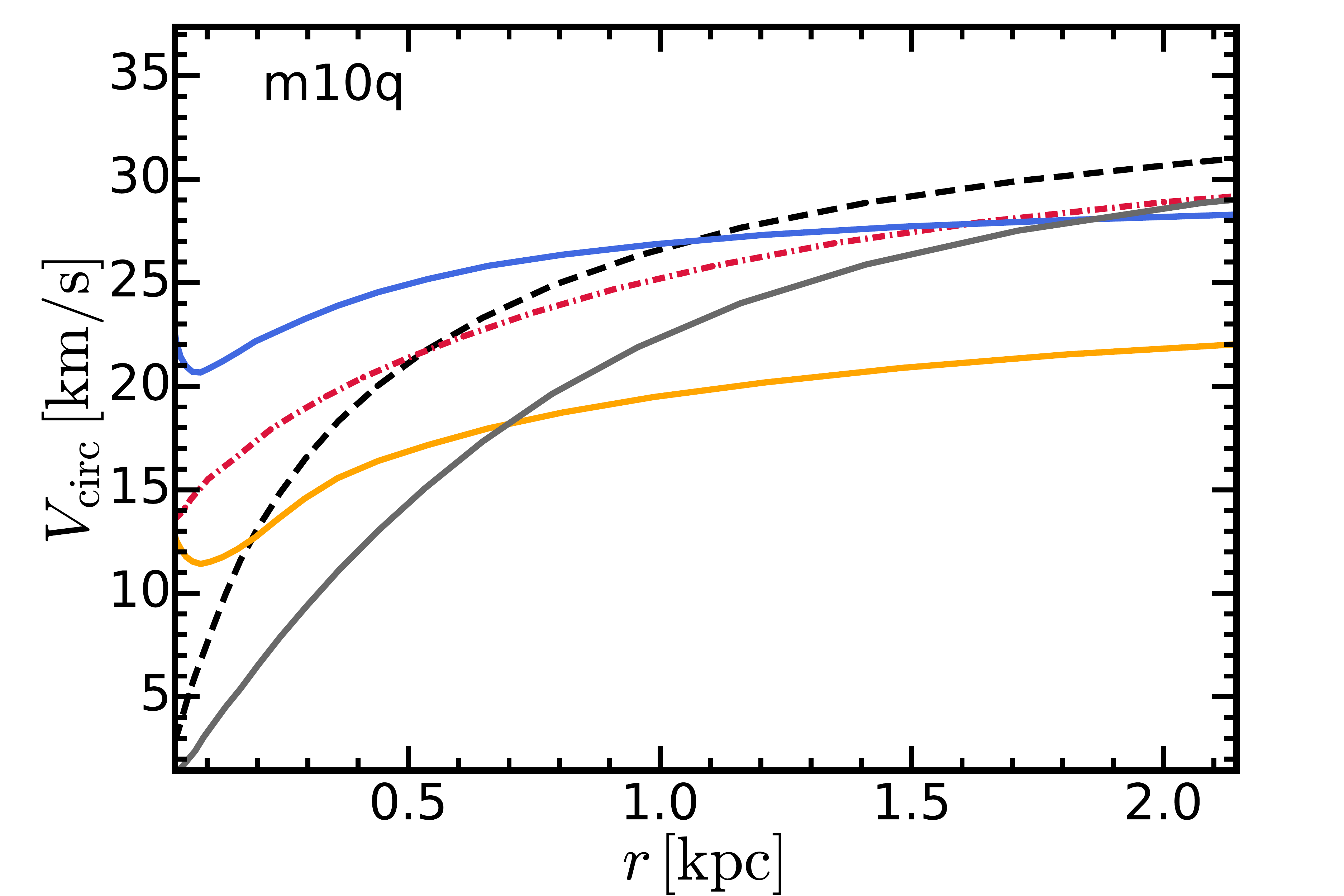}
    \includegraphics[trim=0.1cm 0.0cm 3cm 0.6cm, clip, width=0.32\textwidth]{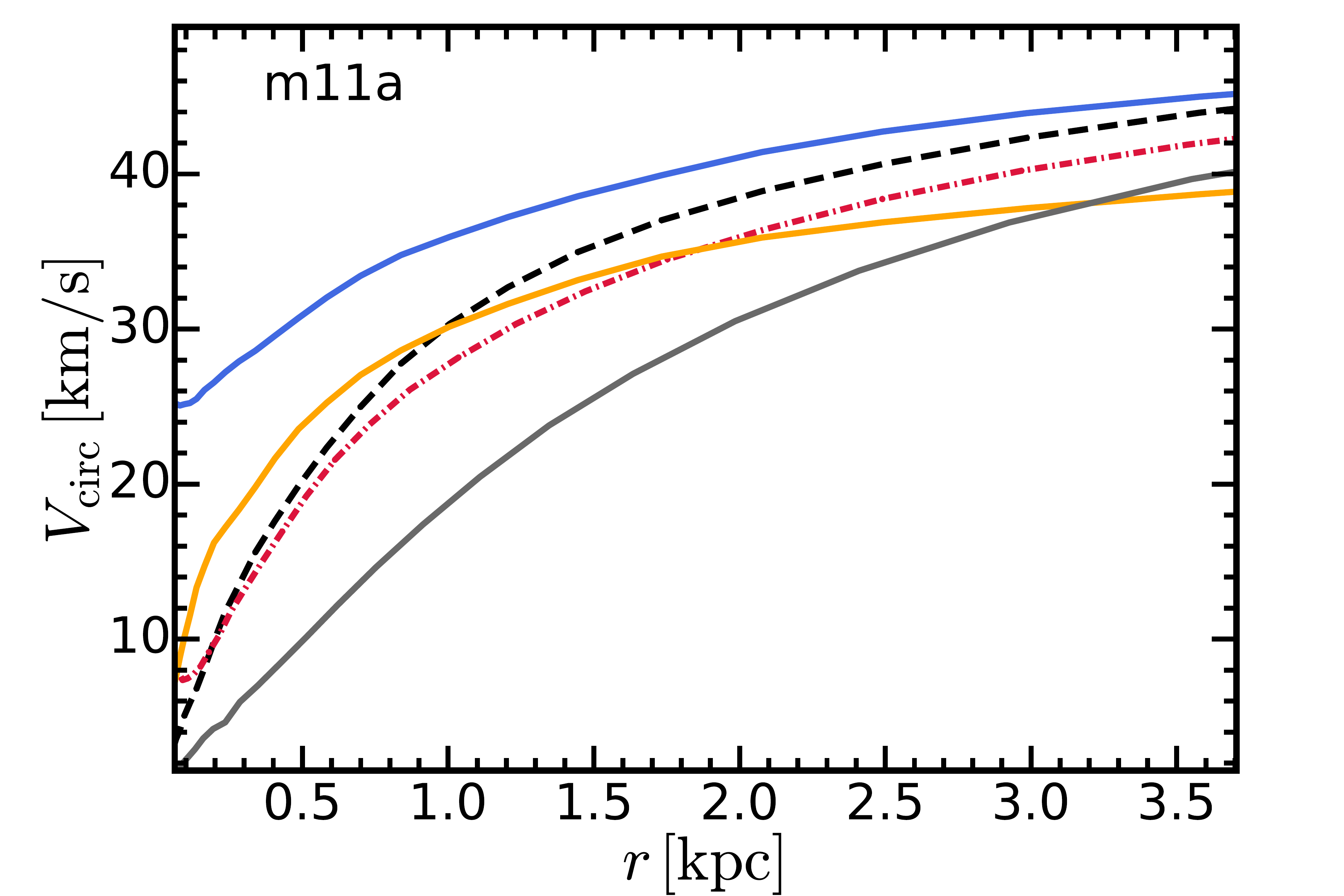}
    \includegraphics[trim=0.1cm 0.0cm 3cm 0.6cm, clip, width=0.32\textwidth]{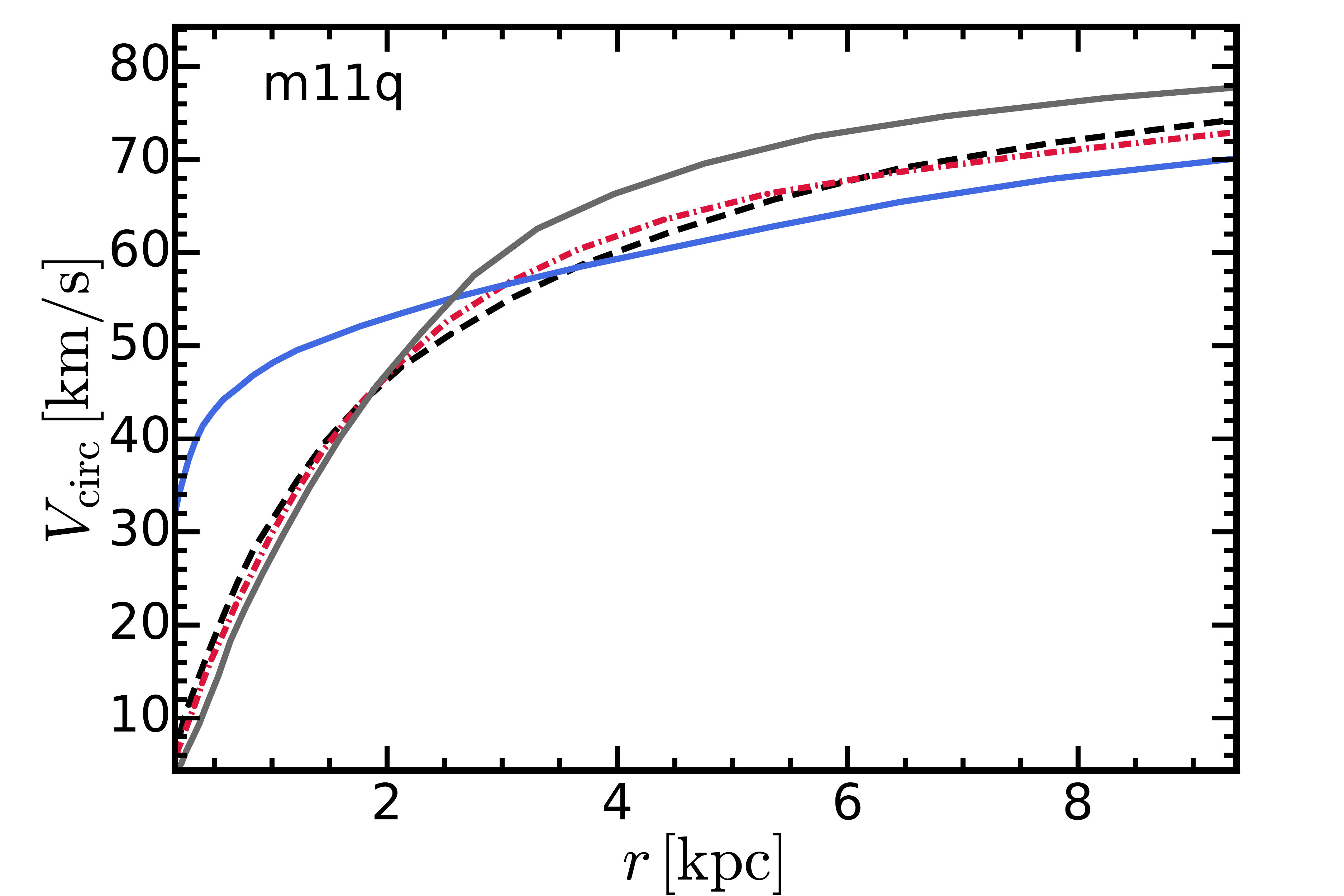}
    
    \rotatebox{90}{\hspace{0.8cm} \textbf{velocity dispersion}}
    \includegraphics[trim=0.1cm 0.0cm 3cm 0.3cm, clip, width=0.32\textwidth]{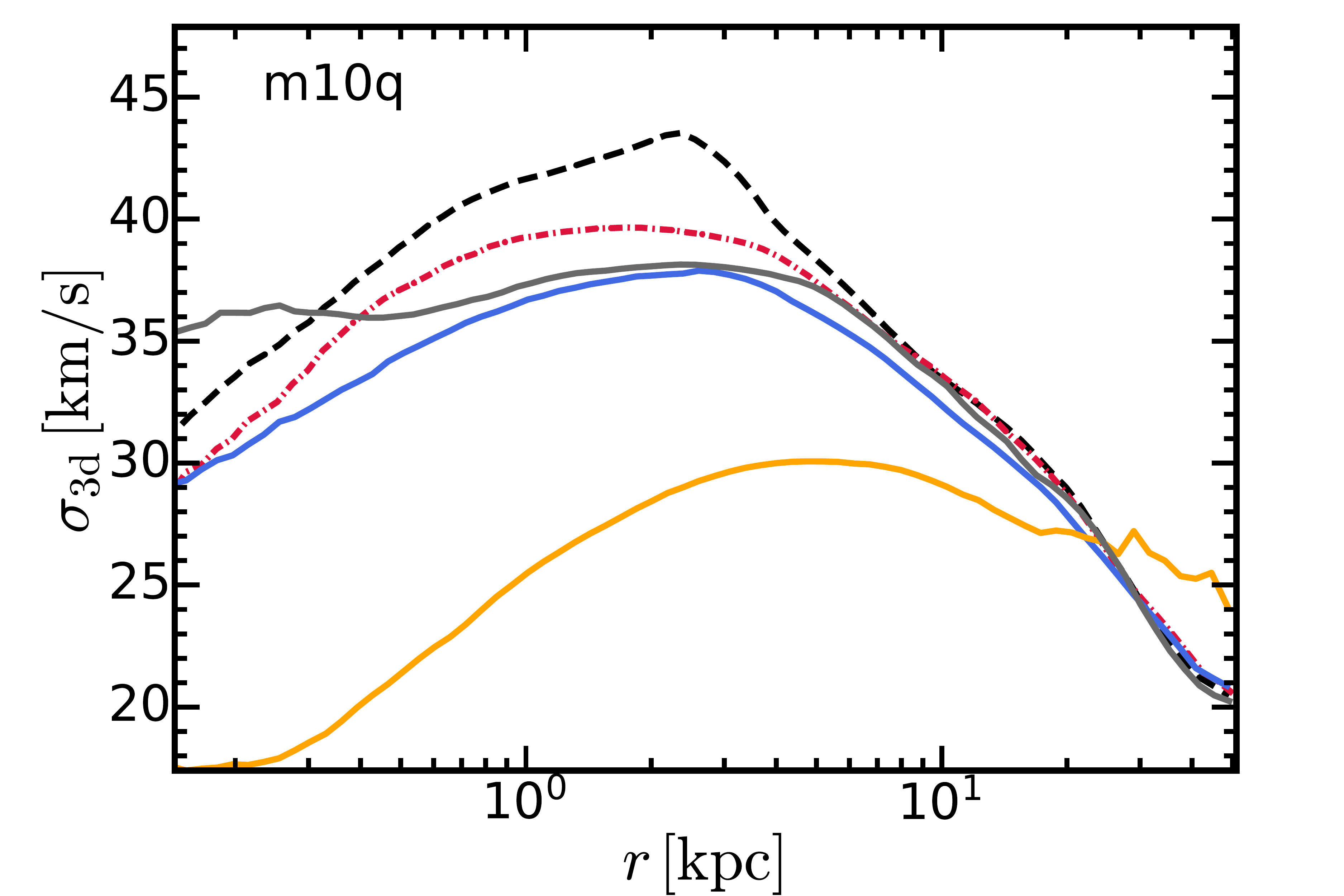}
    \includegraphics[trim=0.1cm 0.0cm 3cm 0.3cm, clip, width=0.32\textwidth]{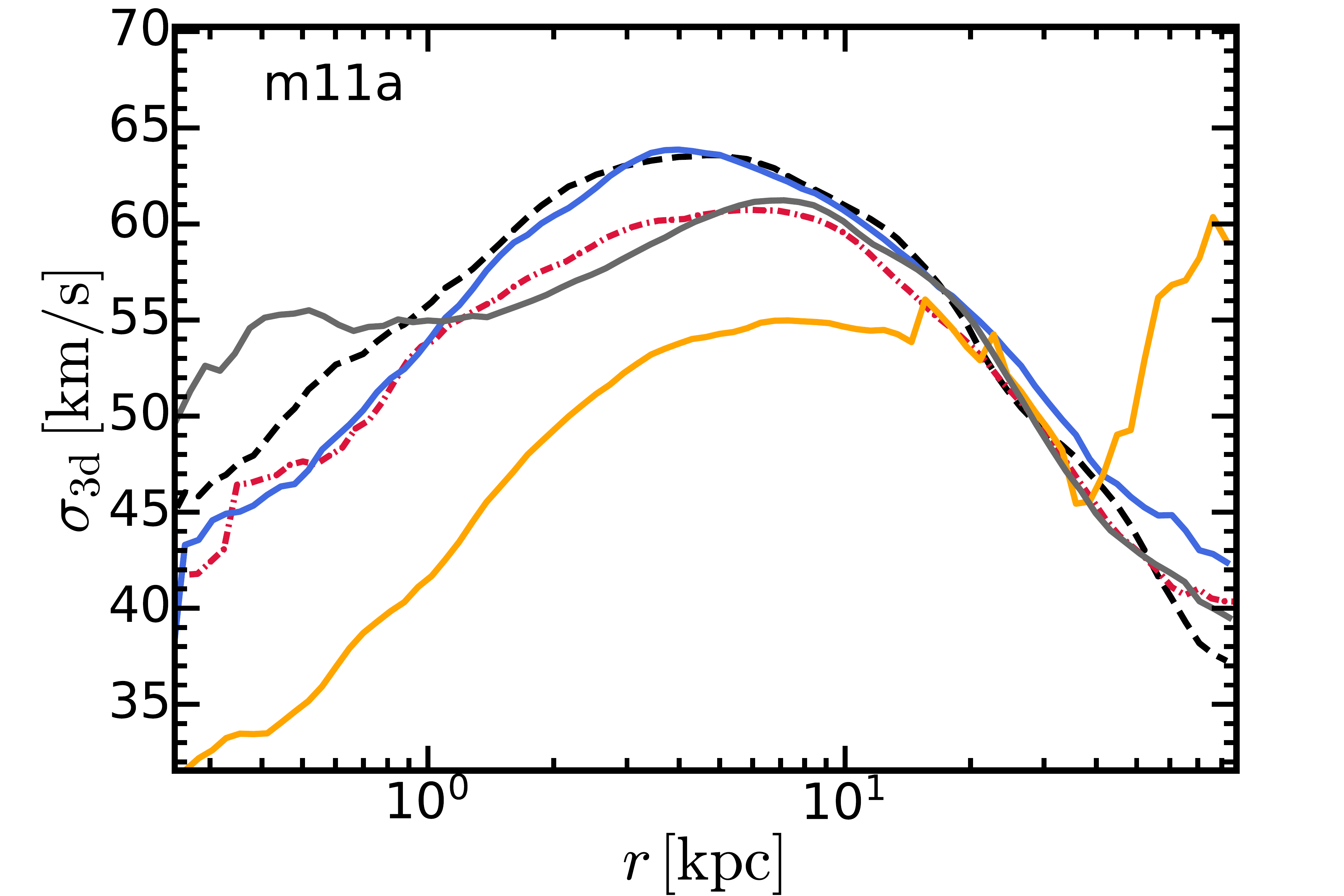}
    \includegraphics[trim=0.1cm 0.0cm 3cm 0.3cm, clip, width=0.32\textwidth]{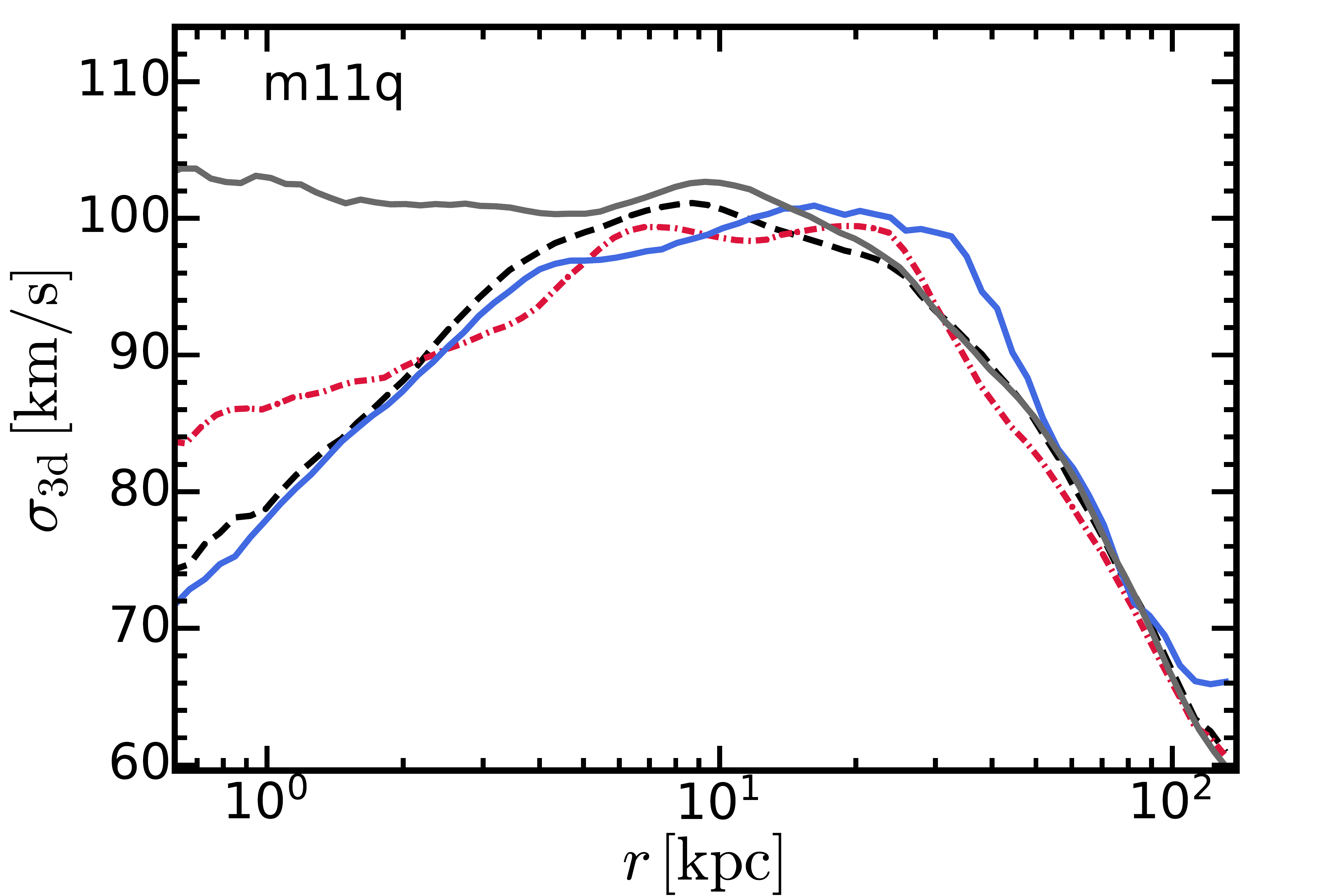}
    
    \rotatebox{90}{\hspace{1.0cm} \textbf{velocity anisotropy}}
    \includegraphics[trim=0.1cm 0.0cm 3cm 0.6cm, clip, width=0.32\textwidth]{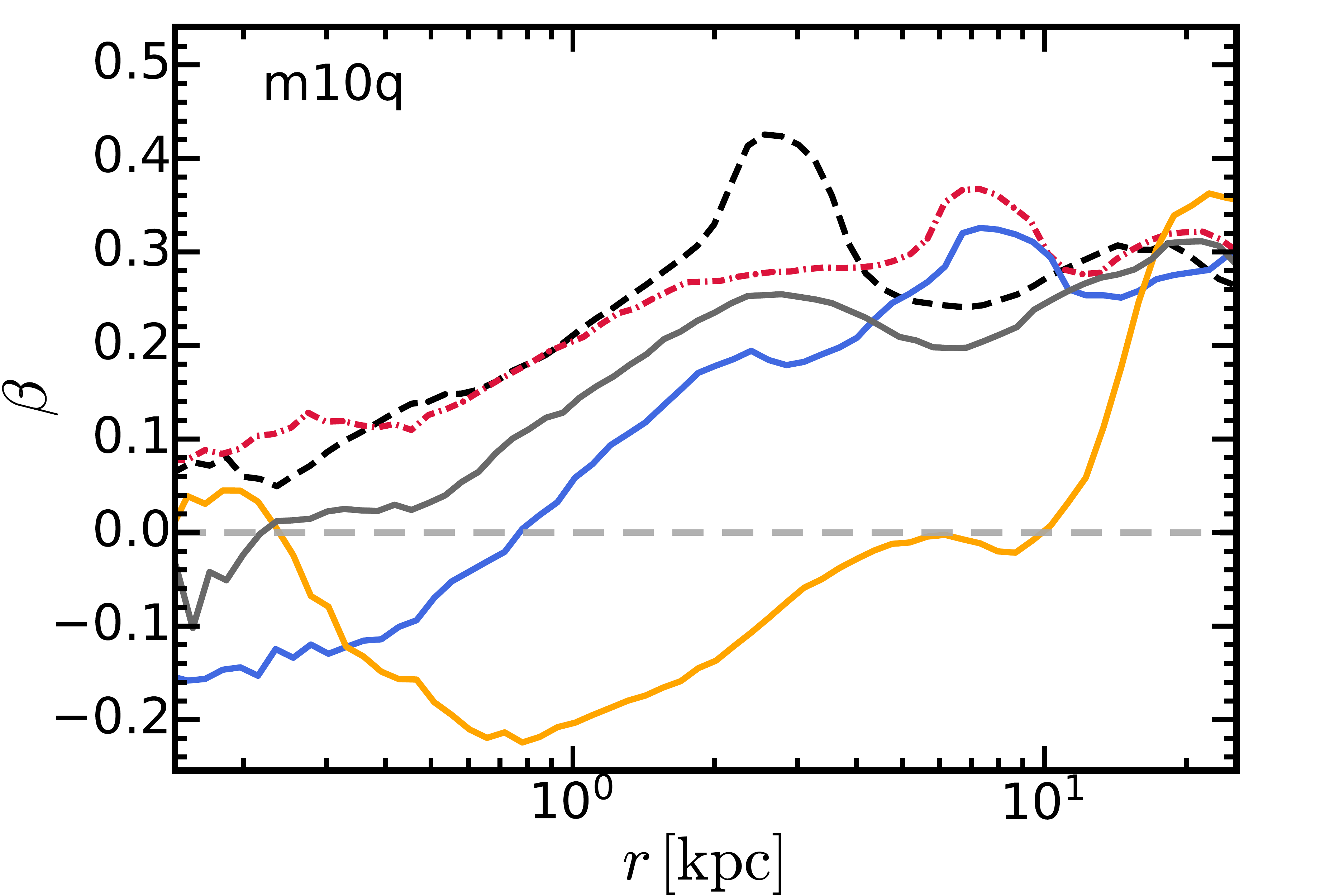}
    \includegraphics[trim=0.1cm 0.0cm 3cm 0.6cm, clip, width=0.32\textwidth]{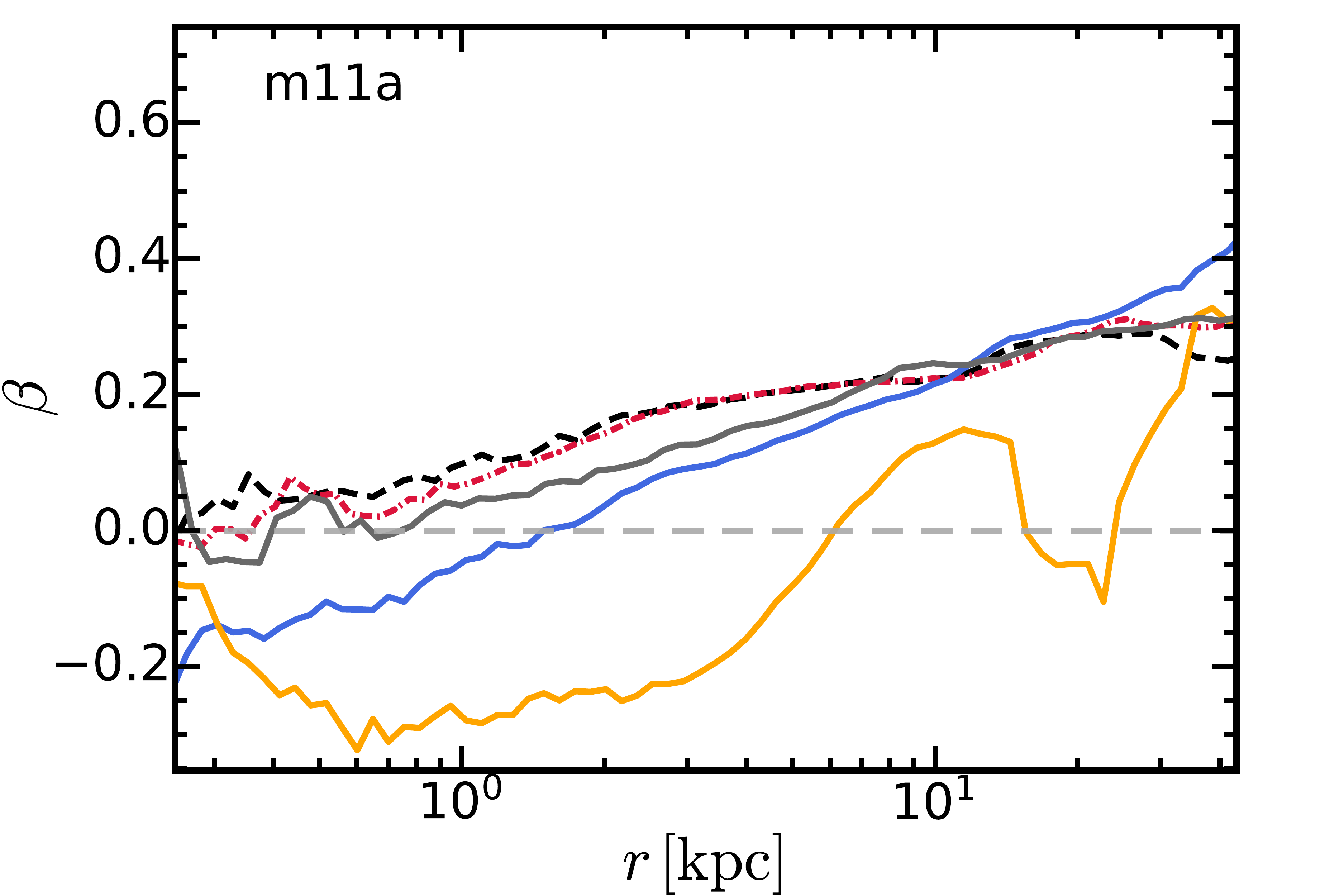}
    \includegraphics[trim=0.1cm 0.0cm 3cm 0.6cm, clip, width=0.32\textwidth]{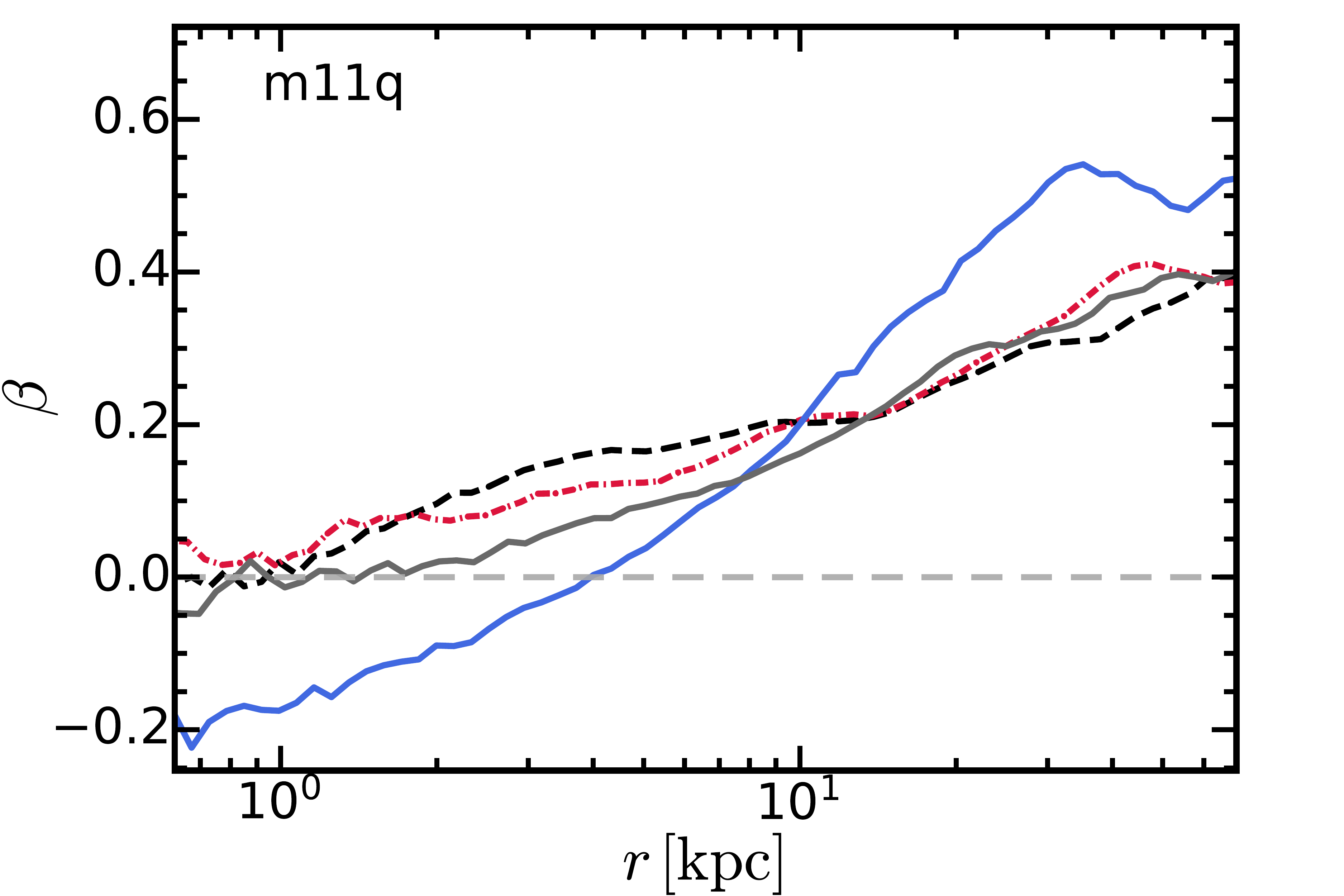}
    
    \rotatebox{90}{\hspace{1.0cm} \textbf{rotation versus dispersion}}
    \includegraphics[trim=0.1cm 0.0cm 3cm 0.6cm, clip, width=0.32\textwidth]{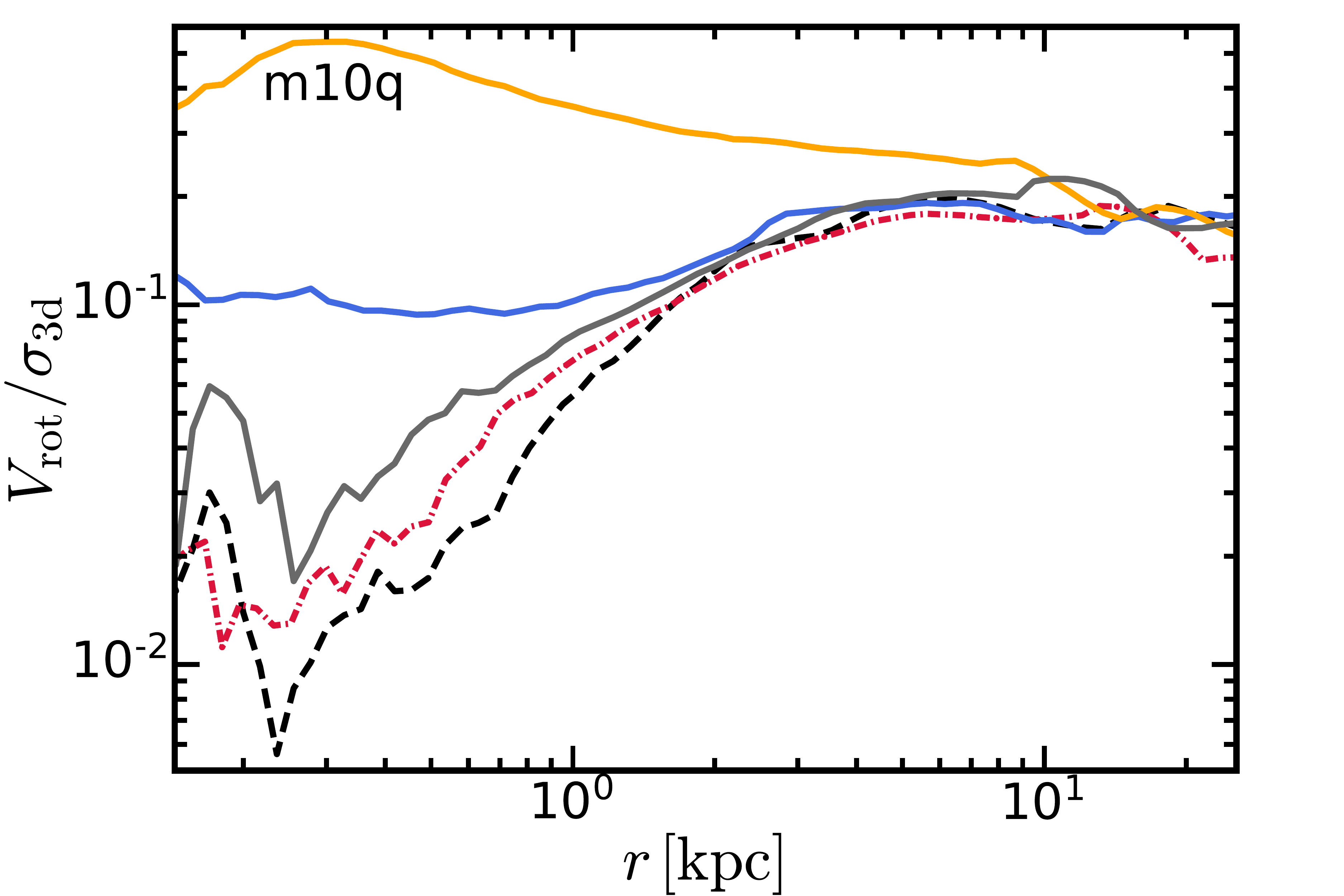}
    \includegraphics[trim=0.1cm 0.0cm 3cm 0.6cm, clip, width=0.32\textwidth]{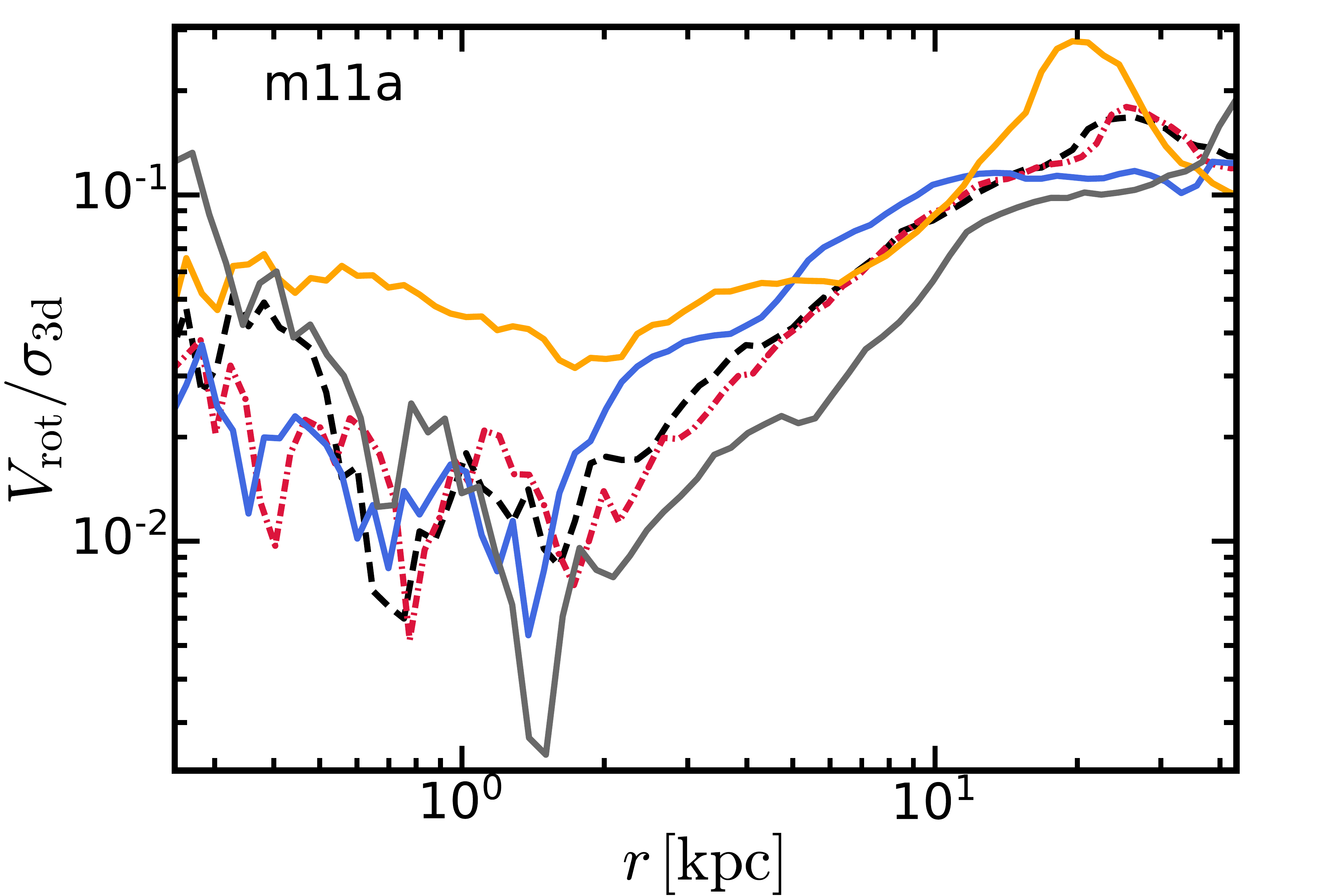}
    \includegraphics[trim=0.1cm 0.0cm 3cm 0.6cm, clip, width=0.32\textwidth]{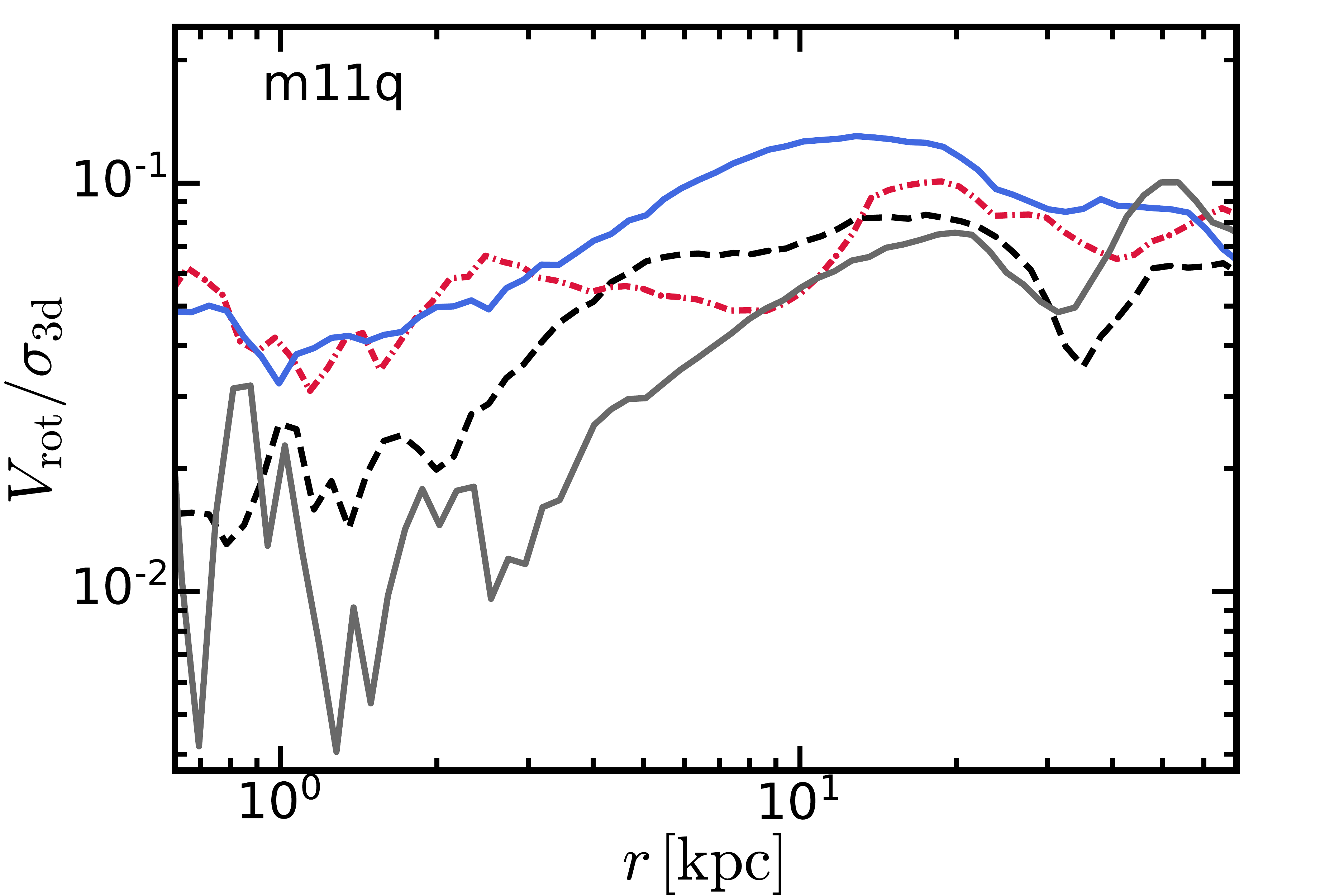}
    \caption{\textbf{A gallery view of the structural and kinematic properties of dwarf galaxies in simulations.} From top to bottom, in each row, we show the three-dimensional total mass density ($\rho_{\rm tot}=\rho_{\rm dm}+\rho_{\rm star}+\rho_{\rm gas}$), circular velocity ($V_{\rm circ}\equiv \sqrt{GM^{\rm tot}_{\rm enc}(r)/r}$), three-dimensional velocity dispersion of dark matter ($\sigma_{\rm 3d}\equiv \sqrt{\sigma^2_{\rm r}+\sigma^2_{\theta}+\sigma^2_{\phi}} $), velocity anisotropy of dark matter ($\beta \equiv 1- (\sigma^2_{\theta}+\sigma^2_{\phi})/2\sigma^2_{\rm r}$) and rotation velocity versus velocity dispersion of dark matter ($V_{\rm rot}/\sigma_{\rm 3d}$) averaged in spherical shells as a function of galactocentric distance for three simulated galaxies. We compare three categories of dark matter models: CDM; eSIDM (elastic SIDM model with a constant cross-section $1\cpm$); dSIDM (dissipative SIDM models with various cross-sections, as defined in Table~\ref{tab:sims}). The gray shaded regions in the first row of plots indicate $0.2\%-0.8\% \,R^{\rm cdm}_{\rm vir}$, which is the aperture we will later use to measure the slopes of the density profiles (see Section~\ref{sec:denpro} and Figure~\ref{fig:denpro_lowmass}-\ref{fig:slope}). The gray dashed horizontal line in the fourth row is a reference line, indicating isotropic velocity dispersion ($\beta$=0). In general, dSIDM models produce cuspy central density profiles in the simulated dwarf galaxies, opposed to the cored central density profile in CDM and eSIDM models. As a consequence, the circular velocities at the center of the galaxies increase. In dSIDM models with $(\sigma/m)\geq 1\cpm$, coherent rotation of dark matter becomes prominent and random velocity dispersion is suppressed. 
    }
    \label{fig:gallery}
\end{figure*}

\subsection{Comparison to the cooling of baryons}

The cooling induced by dissipative dark matter self-interactions can be compared to the cooling of baryons, which is usually described by the cooling function $\Lambda$. For dSIDM, the effective cooling function is
\begin{align}
    \Lambda_{\rm eff} \sim \dfrac{T}{n t_{\rm diss}} & \sim (\sigma/m)\, f_{\rm diss}\,\sigma_{\rm 1d}^{3} \nonumber \\
    & \sim
    \begin{cases}
       \sigma_{\rm 1d}^{3} \sim T^{3/2}   &  {[\rm constant\, \operatorname{cross-section}]}\\
       \sigma_{\rm 1d}^{-1} \sim T^{-1/2} &  {[\rm velocity\,dependent\,model]}
    \end{cases}
\end{align}
where $T$ is $m \sigma_{\rm 1d}^{2}/k_{\rm B}$ for weakly collisional dark matter. The cooling function in the constant cross-section model is similar to the cooling curve of gas below $\sim 10^{4}\,{\rm K}$ while the cooling function in the velocity dependent model is similar to the $10^{4}-10^{7}\,{\rm K}$ gas cooling curve. Other behaviours are possible if a velocity-dependence of $f_{\rm diss}$ is introduced, e.g. $\Lambda_{\rm eff}$ would be a constant if $f_{\rm diss} \sim T^{\rm 1/2}$ with the same velocity-dependent cross-section. However, the most important qualitative difference between the dSIDM studied here and baryons is not the behaviour of the cooling curve but the fact that baryons (gas) are effectively in the $f_{\rm diss} \rightarrow 0$ and $(\sigma/m) \rightarrow \infty$ regime. The effective interaction cross-section of gas is enormous compared to favored SIDM interaction cross-sections and the energy loss per ``collision'' is small. Gas cooling is the result of a large amount of particle interactions in a locally thermalized region. On contrary, dSIDM with $t_{\rm coll}$ order of magnitude comparable to $t_{\rm diss}$ cannot achieve local thermalization effectively when cools down.

\subsection{Effective cross-section}
It is useful to define an "effective cross-section" for the velocity-dependent dSIDM model
\begin{equation}
    \Big(\dfrac{\sigma}{m}\Big)_{\rm eff} = \Big\langle \dfrac{\sigma}{m} v_{\rm rel} \Big\rangle / \langle v_{\rm rel} \rangle,
\end{equation}
where $v_{\rm rel}$ is the relative velocity between encountering particles and $\langle ... \rangle$ is a thermal average as discussed in Section~\ref{sec:timescale_coll}. This definition ensures that a dSIDM model with a constant cross-section taking the value of this "effective cross-section" will result in the identical rate of dark matter self-interaction, assuming that dark matter particles are in thermal equilibrium. This definition allows a proper comparison between velocity-dependent and independent SIDM models. Using Equation~\ref{eq:thermal_average}, we find
\begin{align}
     \Big(\dfrac{\sigma}{m}\Big)_{\rm eff} = \dfrac{(\sigma/m)_{0}}{32}\, \Big(\dfrac{v_0}{\sigma_{\rm 1d}}\Big)^4 \, & \Bigg[- 2 {\rm Ci}\Big(\dfrac{v_0^2}{4\sigma_{\rm 1d}^2}\Big) \cos{\Big(\dfrac{v_0^2}{4\sigma_{\rm 1d}^2}\Big)} \nonumber \\
     & + \sin{\Big(\dfrac{v_0^2}{4\sigma_{\rm 1d}^2}\Big)} \Big(\pi - 2{\rm Si}\Big(\dfrac{v_0^2}{4\sigma_{\rm 1d}^2}\Big)\Big) \Bigg],
     \label{eq:sigma_eff}
\end{align}
where the notation is the same as Equation~\ref{eq:tcoll}. The asymptotic behaviour of $(\sigma/m)_{\rm eff}$ is dominated by the $\sigma_{\rm 1d}^{-4}$ term, which is similar to the velocity-dependent cross-section defined in Equation~\ref{eq:cx_vel}. The factor $32$ in the denominator comes from the thermal average and indicates that dSIDM models with velocity-dependent cross-section are not as efficient as those with constant cross-sections, owing again to the velocity suppression.

\section{Simulation Results}
\label{sec:results}

In this section, we present the structural and kinematic properties of simulated dwarf galaxies in different dark matter models and study the impact of dissipation on galaxy structures.

\subsection{Overview}
\label{sec:overview}

In Figure~\ref{fig:image1}, we show images of four dark matter haloes in our simulation suite at $z=0$. Each image is a two-dimensional surface density map of dark matter, projected along the z-direction of simulation coordinates, with a logarithmic stretch. The dynamical ranges are adjusted based on the maximum and median intensities of pixels. The haloes are ordered from left to right by their halo masses (see Section~\ref{sec:halo_prop} for the definition). We show the images in CDM, the dSIDM with constant cross-section $(\sigma/m)=1\cpm$ and the velocity-dependent dSIDM model for comparison. The haloes in dSIDM models are visibly more concentrated than their CDM counterparts when comparing their core sizes (dashed circles). For the velocity-dependent dSIDM model, since the self-interaction cross-section decreases in more massive haloes which typically have higher velocity dispersions, the increased concentration of the halo becomes less apparent. On contrary, in dSIDM models with constant cross-sections, haloes of all masses are consistently more concentrated than their CDM counterparts. Meanwhile, the substructures also appear to be more abundant and concentrated in dSIDM models, For example, in \textbf{m10q}, the number of subhaloes (within the virial radius) with $M>10^{6}\msun$ increases by about $20\%$, and the median concentration increases by about $25\%$ in the dSIDM model with $(\sigma/m)=1\cpm$. But we will focus on the main halo in this paper and defer the analysis on substructures to follow-up work. 

\begin{figure}
    \centering
    \includegraphics[width=0.49\textwidth]{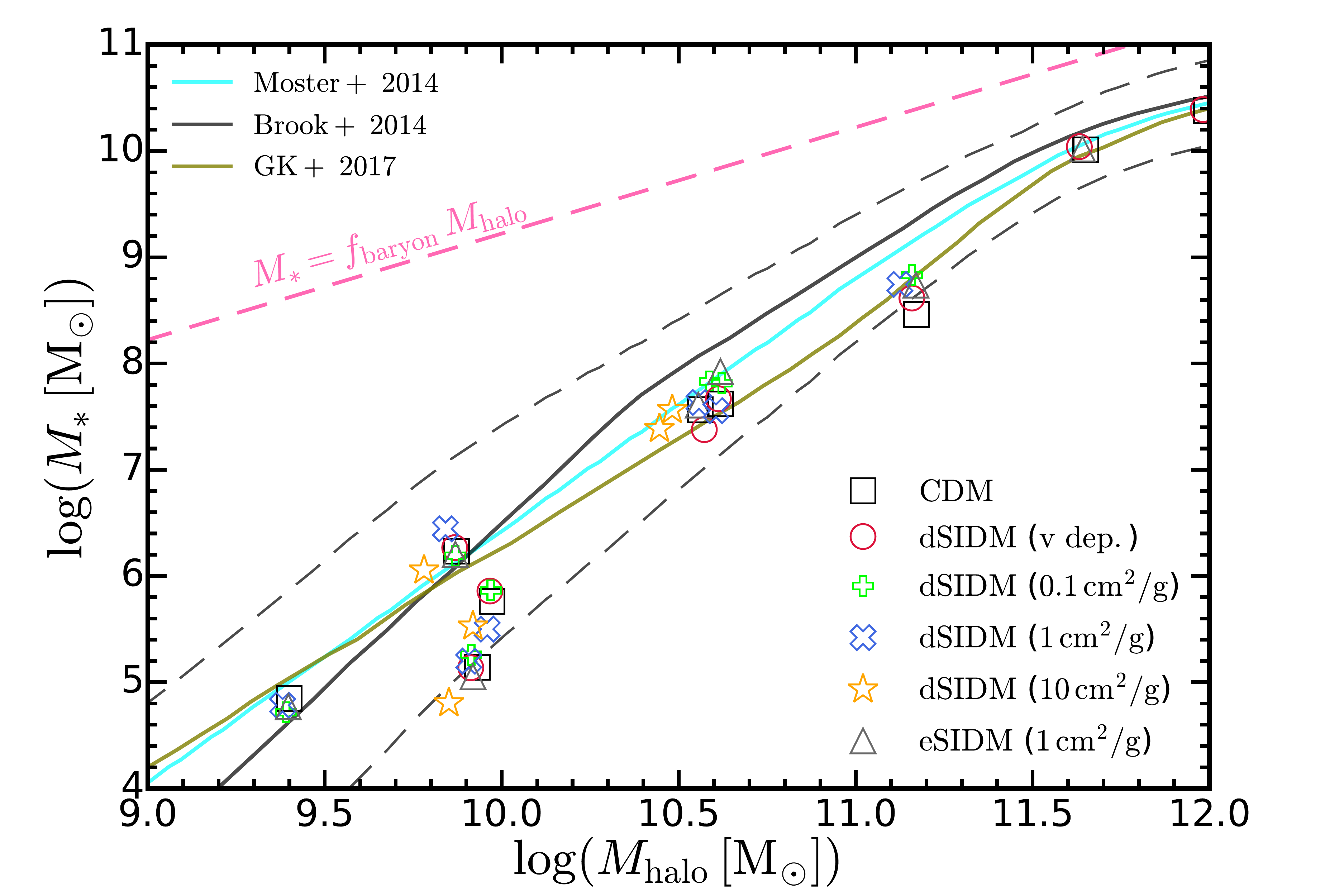}
    \caption{\textbf{Stellar mass versus halo mass relation of galaxies in simulations.} The stellar masses and halo masses of simulated dwarf galaxies are presented with open markers (as labelled). We compare them with the observational results derived through abundance matching from \citet{Moster2014,Brook2014,SGK2017}. The black dashed lines show $\sim 95\%$ inclusion contour assuming the scatter of the relation estimated in \citet{SGK2017}. Regardless of the dark matter model, the simulated galaxies are consistent with the observational relation. }
    \label{fig:mstar-mhalo}
\end{figure}

In Figure~\ref{fig:gallery}, we present a gallery view of the total mass density, circular velocity, three-dimensional velocity dispersion of dark matter, velocity anisotropy of dark matter, rotation velocity versus velocity dispersion of dark matter, averaged in spherical shells as a function galactocentric distance for three simulated galaxies. Details of the measurements of the kinematic properties and relevant definitions are introduced in Section~\ref{sec:kin}. Under the influence of baryonic feedback, the density profiles in CDM are generally shallower than the cuspy NFW profiles at galaxy centers, which is expected for these galaxies for their $M_{\ast}/M_{\rm halo}$ values~\citep[e.g.,][]{diCintio2014,Chan2015,Onorbe2015,Tollet2016,Lazar2020}. In the eSIDM model, due to effective heat conduction, the profiles are even flatter at galaxy centers compared to the CDM case, but the difference becomes less apparent in the bright dwarf (\textbf{m11q}) where thermal conduction through self-interactions is subdominant compared to baryonic feedback. In dSIDM models, when the effective self-interaction cross-section is large (and equivalently dissipation is efficient assuming a fixed $f_{\rm diss}$), the central density profiles are cuspy and power-law like. For the velocity-dependent dSIDM model, in the classical dwarf galaxies like \textbf{m10q}, the velocity-dependent cross-section is high and a cuspy central profile emerges. In more massive galaxies like \textbf{m11a} and \textbf{m11q}, the velocity-dependent cross-section there becomes much smaller, accompanied by stronger baryonic feedback. As a consequence, the profiles in these systems become cored again though the central mass density is still higher than the CDM case. An interesting outlier here is the dSIDM model with constant $(\sigma/m)=10\cpm$, exhibiting cuspy central density profile but with lower normalization, which is likely due to the deformed shape of the halo (see Section~\ref{sec:shape}). A more detailed discussion on the mass density profiles will be presented in Section~\ref{sec:denpro}.

\begin{figure*}
    \centering
    \includegraphics[trim=0.6cm 1cm 3.2cm 1.2cm, clip,width=0.49\textwidth]{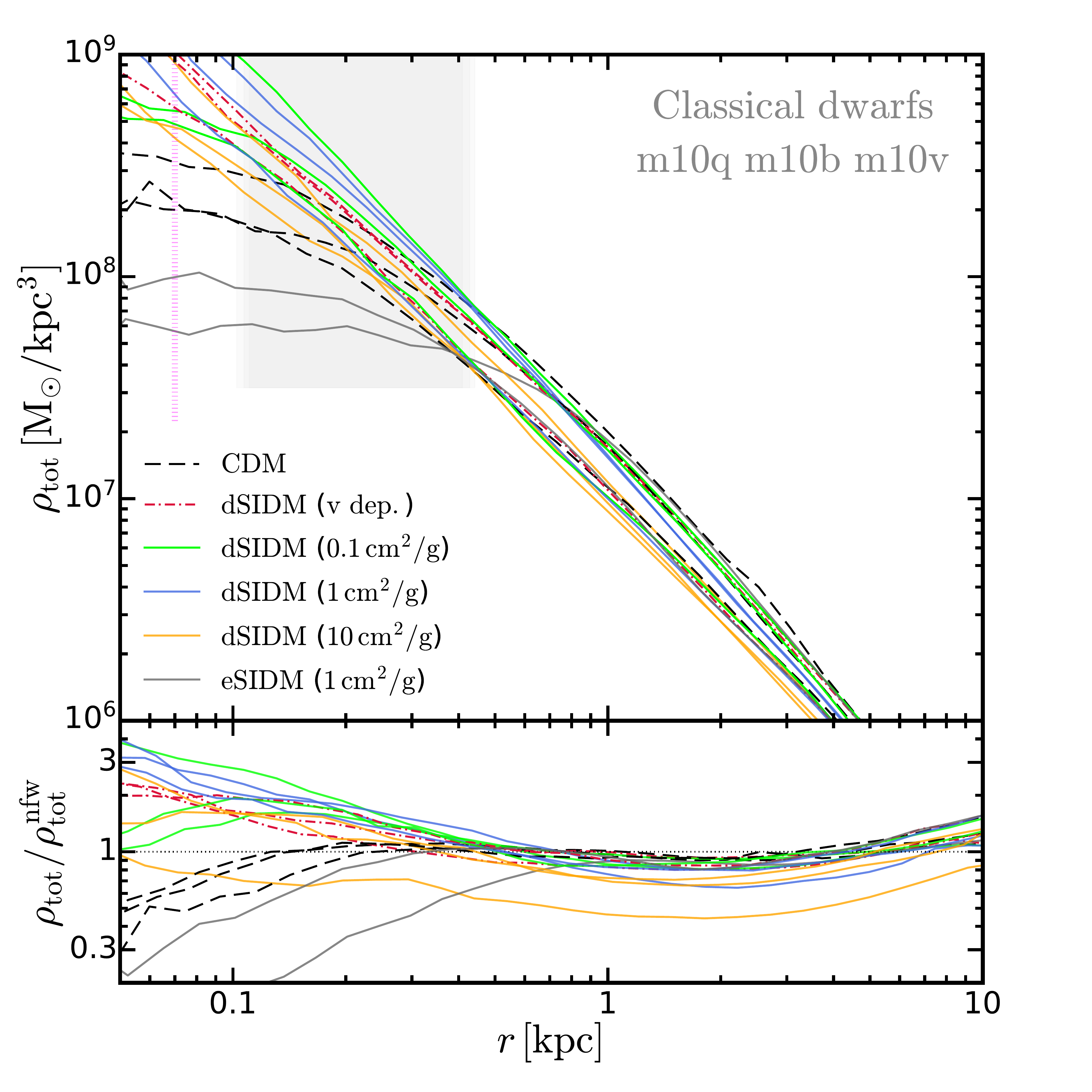}
    \includegraphics[trim=0.6cm 1cm 3.2cm 1.2cm, clip,width=0.49\textwidth]{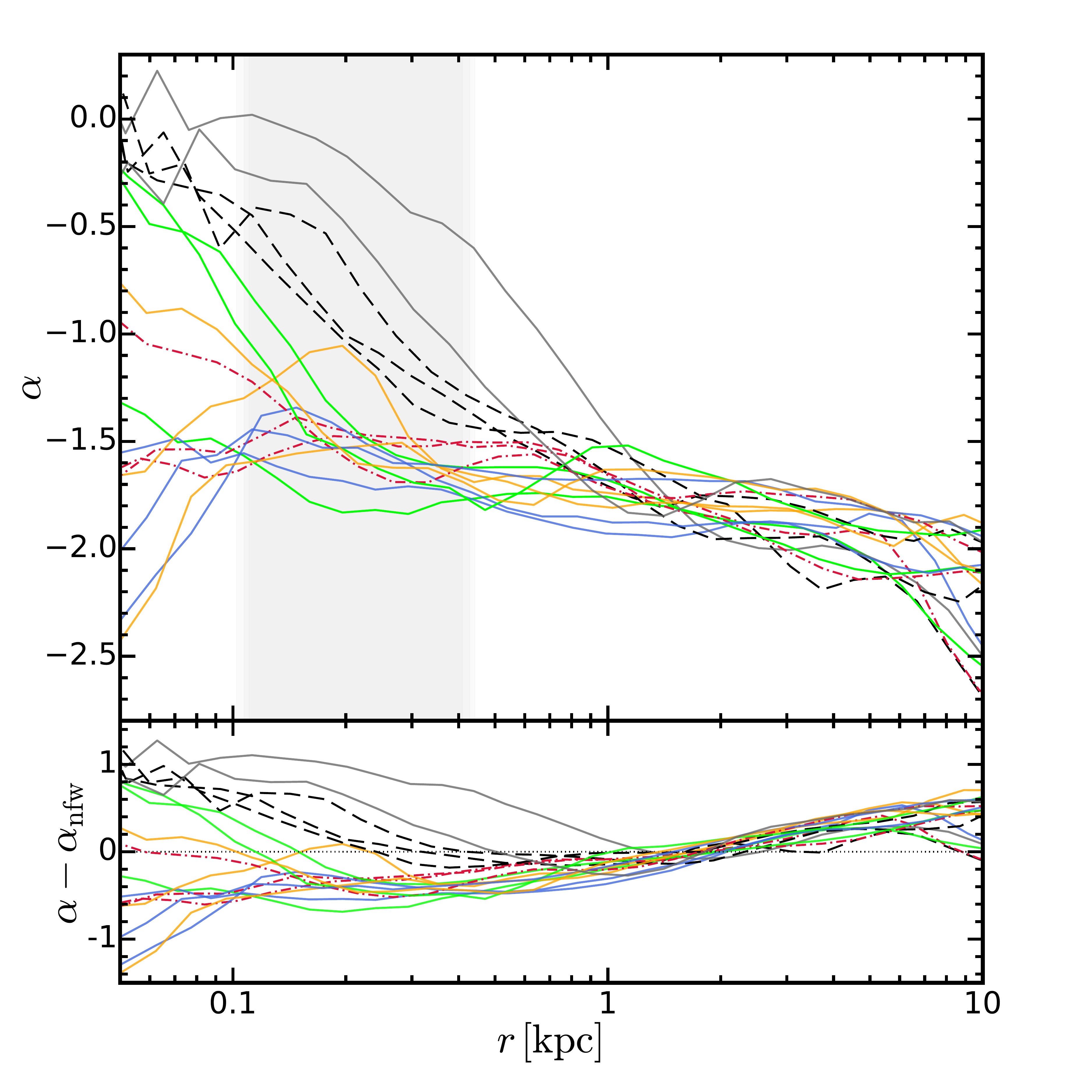}
    \caption{{\it Left}: \textbf{Total mass density profiles of the classical dwarf galaxies in simulations.} The three classical dwarfs presented here are \textbf{m10q}, \textbf{m10b} and \textbf{m10v}. The total mass density profiles in different dark matter models are shown (as labelled). They can be compared to the NFW profiles derived by fitting the density profiles at large radii of the haloes ($0.5\,r^{\rm cdm}_{1/2}<r<20\,r^{\rm cdm}_{1/2}$), and the ratios of the density profiles to the NFW fits are shown in the lower sub-panel. The gray shaded region denotes the range of radii where we measure the slopes of the density profiles below. The purple dotted vertical line indicates the average convergence radius ($\sim 70\pc$) of the classical dwarfs (see Table~\ref{tab:sims}). {\it Right}: \textbf{Local power-law slopes of density profiles of the classical dwarf galaxies.} The slopes are derived via fitting the nearby density profile with power-law. In these classical dwarfs, the CDM model predicts cored central density profiles due to baryonic feedback. The eSIDM model produces cores of slightly bigger sizes and shallower slopes. The dSIDM model with $(\sigma/m)=0.1\cpm$ still produces cored profiles but with higher central densities and steeper slopes than their CDM counterparts. The dSIDM models with effective cross-section $>0.1 \cpm$ all produce cuspy central density profiles with power-law slopes centering around $-1.5$. These profiles are even steeper than the NFW profiles. 
    }
    \label{fig:denpro_lowmass}
\end{figure*}

In addition to the density profile, the kinematic properties of haloes are also quite different in different dark matter models. Despite some variations, there are some important features shared by the simulations of different haloes. When the cross-section is high, the rotation curves of dwarf galaxies in dSIDM models are significantly higher at small radii compared to their CDM counterparts. The differences are consistent with the findings in density profiles. Again, an outlier is the dSIDM model with $(\sigma/m)=10\cpm$, with the normalization of rotation velocities lower than other models. For the velocity dispersion profile, the ones in eSIDM are flat at halo centers indicating an isothermal distribution of dark matter particles. The velocity dispersions in dSIDM models in general decreases towards halo centers. Particularly, the dSIDM model with $(\sigma/m)=10\cpm$ shows dramatic decrease in velocity dispersion at $r\lesssim 10\kpc$. This indicates more coherent motion of dark matter particles and a decreasing support from random velocity dispersion. For the velocity anisotropy profile, the dSIDM models with $(\sigma/m)\geq1\cpm$ have lower velocity anisotropies than their CDM counterparts at halo centers, indicating that the velocity dispersions are more dominated by the tangential component. At the same time, the coherent rotation is also stronger in these dSIDM models. An extreme case is the dSIDM model with $(\sigma/m)=10\cpm$ where the sub-kpc structure is clearly in transition from dispersion supported to coherent rotation supported. The ratio between coherent circular velocity and velocity dispersion is significantly higher than others. In Section~\ref{sec:kin}, the kinematic properties of simulated galaxies will be investigated in detail.

\subsection{Halo mass and galaxy stellar mass}
\label{sec:halo_prop}

We measure the bulk properties of the dark matter haloes and galaxies in simulations following what has been done for the standard FIRE-2 simulations as described in \citet{Hopkins2018}. We define the halo mass $M_{\rm halo}$ and the halo virial radius $R_{\rm vir}$ using the overdensity criterion introduced in \citet{Bryan1998}. We define the stellar mass $M_{\ast}$ as the total mass of all the stellar particles within an aperture of $0.1\,R_{\rm vir}$ and correspondingly define the stellar half-mass radius $r_{1/2}$ as the radius that encloses half of the total stellar mass. For the isolated dwarf galaxies in simulations, these definitions on the stellar mass and the stellar half-mass radius give similar results to what derived using the iterative approach described in \citet{Hopkins2018}. 

In Figure~\ref{fig:mstar-mhalo}, we compare the stellar mass versus halo mass of simulated dwarf galaxies with the scaling relations derived based on observations~\citep{Moster2014,Brooks2014,SGK2017}. The black dashed lines show $95\%$ inclusion contour assuming the scatter estimated in \citet{SGK2017}. The simulated dwarfs are consistent with observations in the stellar mass versus halo mass relation and the galaxies we sampled in the simulation suite well represent the "median" galaxies in the real Universe. With mild dark matter self-interaction ($(\sigma/m)\lesssim 1\cpm$), the halo and stellar masses of galaxies are not significantly affected compared to their CDM counterparts, in agreement with previous studies of eSIDM~\citep[e.g.,][]{Vogelsberger2014,Robles2017,Fitts2019}. However, in the dSIDM model with $(\sigma/m) = 10\cpm$, both the halo masses and the stellar masses decrease for about $\operatorname{0.1\,-\,0.2}\, {\rm dex}$ (compared to CDM) in dwarf galaxies with $M_{\rm halo}\lesssim 10^{11}\msun$. Although this level of differences is still minor compared to the scatter of the relation, it is worth to note that the model with $(\sigma/m) = 10\cpm$ behaves qualitatively different from other models explored. This aspect will be discussed in Section~\ref{sec:denpro} and Section~\ref{sec:kin} in the following.

\subsection{Total mass density profiles}
\label{sec:denpro}

In this section, we present the total mass density profiles (including the contribution from dark matter, stars and gas) of simulated dwarf galaxies in dSIDM models with different parameters and compare them with the CDM predictions. We note that, for the dwarf galaxies in simulations, the mass density profiles are dominated by dark matter. We divide the simulated dwarf galaxies into two categories: ({\it \rmnum{1}}) classical dwarfs, e.g. the \textbf{m10's}, with typical halo mass of $\lesssim 10^{10}\msun$ and sub-kpc stellar half-mass radius; ({\it \rmnum{2}}) bright dwarfs, e.g. the \textbf{m11's}, with typical halo mass of $\gtrsim 10^{10}\msun$ and stellar half-mass radius of several kpc. We will investigate the extent at which the dissipative dark matter self-interactions affect the structure of these dwarfs.

\begin{figure*}
    \centering
    \includegraphics[trim=0.6cm 1cm 3.2cm 1.2cm, clip,width=0.49\textwidth]{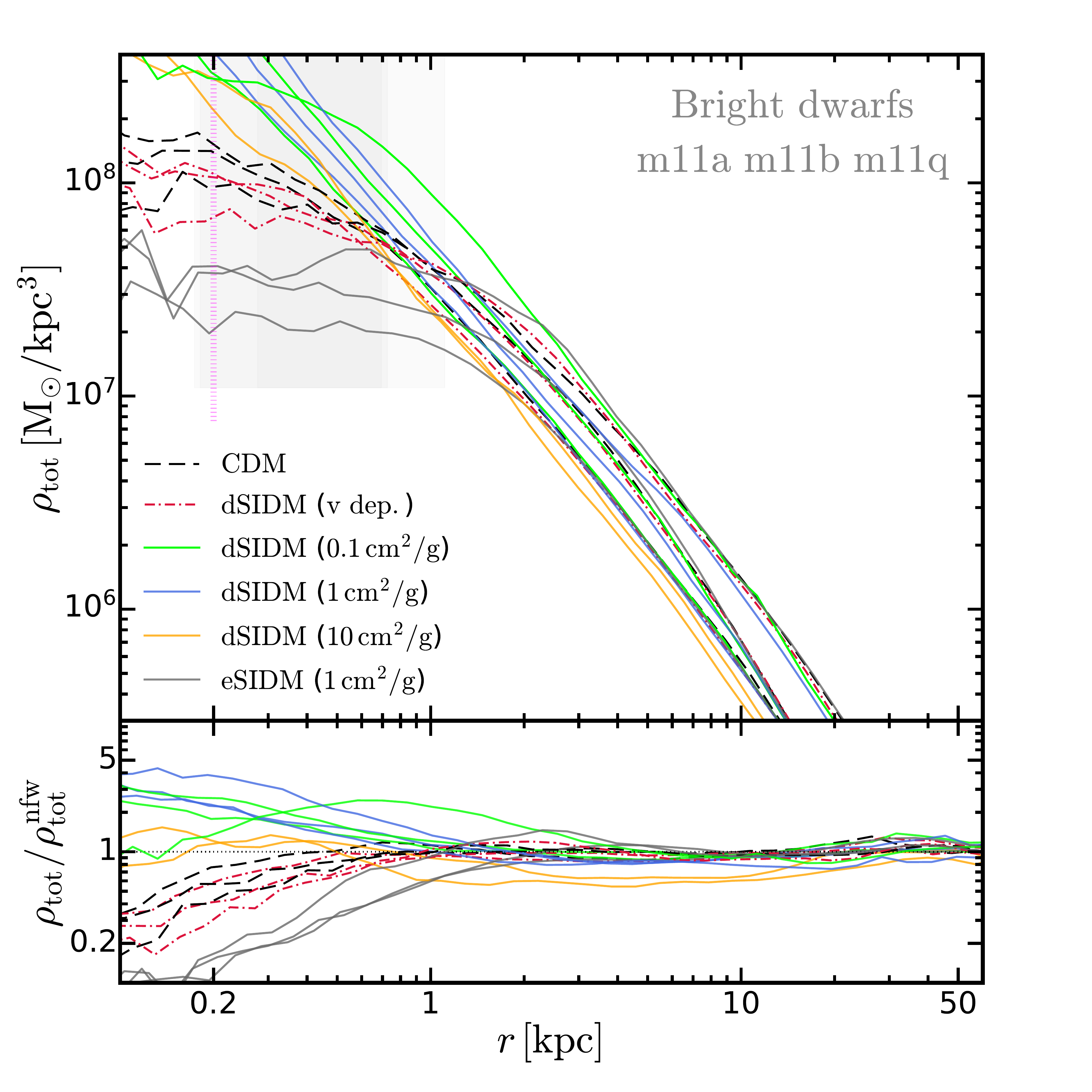}
    \includegraphics[trim=0.6cm 1cm 3.2cm 1.2cm, clip,width=0.49\textwidth]{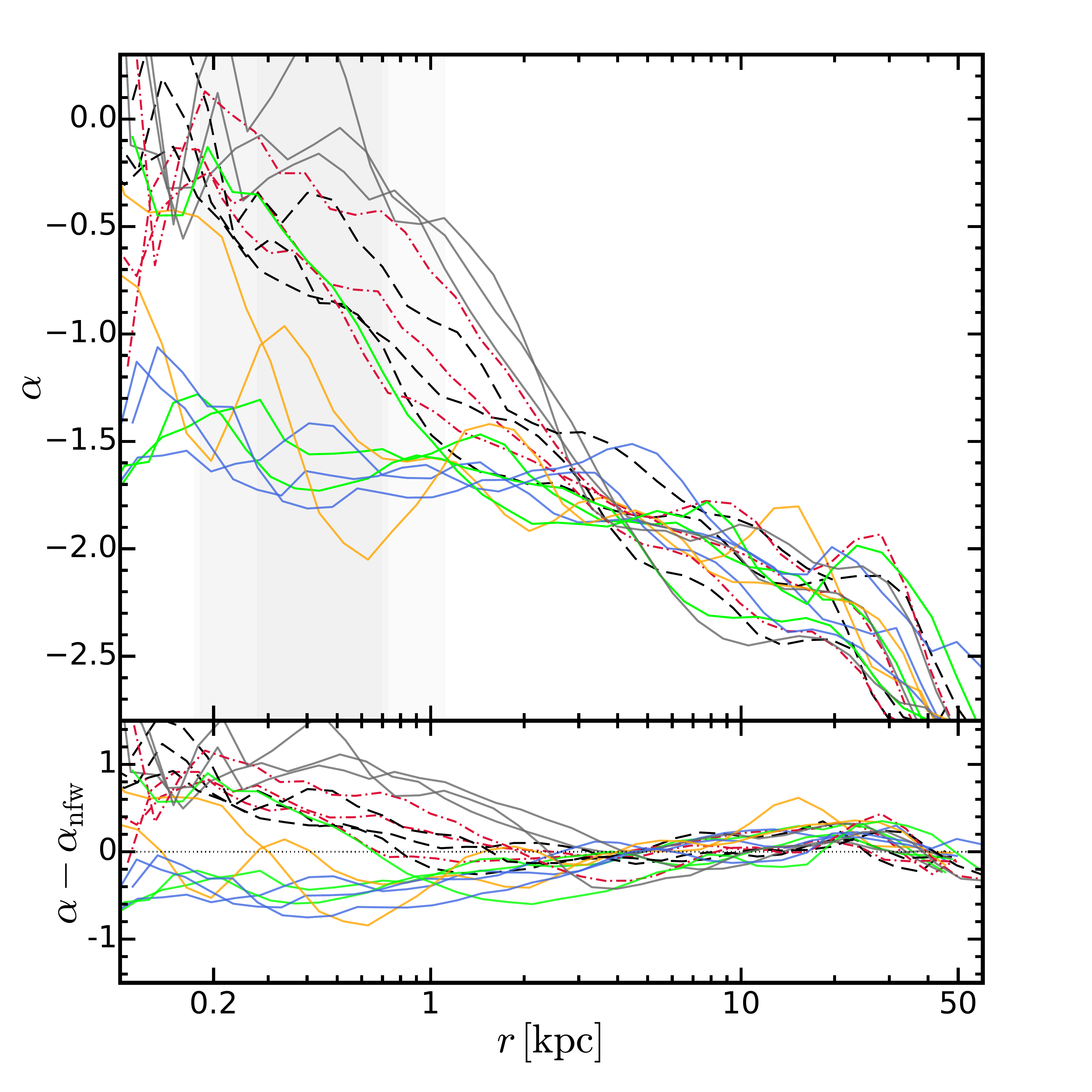}
    \caption{ {\it Left}: \textbf{Total mass density profiles of the bright dwarf galaxies in simulations.} The three bright dwarfs presented here are \textbf{m11a}, \textbf{m11b} and \textbf{m11q}. The notation is the same as Figure~\ref{fig:denpro_lowmass}. The purple dotted vertical line here indicates the average convergence radius ($\sim 200\pc$) of the bright dwarfs (see Table~\ref{tab:sims}). {\it Right}: \textbf{Local power-law slopes of the density profiles of the bright dwarf galaxies.} In these bright dwarfs, the CDM model again predicts cored central density profiles with even larger cores ($\sim \kpc$) than the classical dwarfs due to stronger baryonic feedback. The eSIDM model produces cores of similar sizes and slopes. The velocity-dependent dSIDM model has relatively low effective cross-sections ($\sim 0.01\cpm$) in these dwarfs. This model still produce cores but with slightly higher central densities than their CDM counterparts. The dSIDM models with relatively high effective cross-sections ($\gg 0.01\cpm$) still produce cuspy and power-law like central density profiles. The power-law slopes center around $-1.5$ with scatter from $-2$ to $-1$.}
    \label{fig:denpro_intemass}
\end{figure*}

In the left panel of Figure~\ref{fig:denpro_lowmass}, we show the total mass density profiles of the classical dwarf galaxies in simulations with CDM, eSIDM and dSIDM models at $z=0$ \footnote{The bursty star formation history in dwarf galaxies could create fluctuations in density profiles, which leads to uncertainties in the profile measured at the $z=0$ snapshot. But we have explicitly checked that the difference between the density profiles at $z=0$ and other four latest snapshots are minimal.}. The effective cross-section $(\sigma/m)_{\rm eff}$ of the velocity-dependent dSIDM model in these classical dwarfs is $\sim 0.3\cpm$ calculated using Equation~\ref{eq:sigma_eff}, plugging in the density and one-dimensional velocity dispersion of dark matter particles enclosed in a sphere of radius $1/3\, r^{\rm cdm}_{1/2}$, where $r^{\rm cdm}_{1/2}$ is the stellar half-mass radius in the CDM model. We fit the density profiles at large radii of the haloes ($0.5\,r^{\rm cdm}_{1/2}<r<20\,r^{\rm cdm}_{1/2}$) with the NFW profile. In the lower sub-panel, we show the ratios between the density profiles in different models and the NFW fits. In the right panel of Figure~\ref{fig:denpro_lowmass}, we show the local power-law slopes of the density profiles. In the lower sub-panel, we show the differences in the slopes versus the NFW fits. In the classical dwarfs, the central density profiles are cored in the CDM case due to baryonic feedback. The eSIDM model produces profiles with much larger cores and shallower slopes than CDM. However, the dSIDM models all predict cuspy and power-law like central density profiles at sub-kpc scale, except for the one with low self-interaction cross-section $0.1\cpm$. These profiles are even steeper than the NFW profiles, with power-law slopes $\sim -1.5$ compared to the $-1$ asymptotic power-law slope of the NFW profile at sub-kpc scale. The dSIDM model with low cross-section of $0.1\cpm$ still produces cored central profiles in two galaxies, but the central densities are higher, and the core sizes are smaller than their CDM counterparts. The profiles in the velocity-dependent dSIDM model lie between the profiles in the dSIDM models with $(\sigma/m)=0.1$ and $1\cpm$, which is consistent with the estimate of $(\sigma/m)_{\rm eff}$ in these systems. Surprisingly, increasing the self-interaction cross-section to $10\cpm$ does not lead to further contraction of the haloes. Instead, the density profiles in the model have lower normalization out to $\sim 10\kpc$, although the profiles still have cuspy shapes at galaxy centers. The classical dwarf galaxy that exhibits the strongest decrease in density profile normalization in this model is \textbf{m10q}. This decreased normalization of density profiles measured spherical shells is likely related to the deformation of haloes (e.g. with the same energy budget, a disk-like structure will have lower spherically averaged density than a spherical structure). Assuming that the radial contraction is adiabatic which preserves specific angular momentum, the radial contraction of dSIDM haloes will eventually be halted by the growing centrifugal force from coherent dark matter rotation. This will also make dSIDM haloes deform from spherical to oblate in shape and the density profiles will appear with lower normalization. In subsequent sections, we will see more evidence for this phenomenon from the analysis of kinematic properties (Section~\ref{sec:kin}) and shapes (Section~\ref{sec:shape}) of dark matter haloes.

In the left panel of Figure~\ref{fig:denpro_intemass}, we show the total mass density profiles of the bright dwarf galaxies in simulations with CDM, eSIDM and dSIDM models. The $(\sigma/m)_{\rm eff}$ of the velocity-dependent dSIDM model in these bright dwarfs is $\sim 0.01\cpm$. In the right panel of Figure~\ref{fig:denpro_intemass}, we show the local power-law slopes of the density profiles of the bright dwarfs. The phenomena in the bright dwarfs are qualitatively consistent with those in the classical dwarfs shown above. In the bright dwarfs, the central density profiles are cored in the CDM case. The decrease of the central density compared to the NFW profile is stronger than that in the classical dwarfs, due to stronger baryonic feedback in the bright dwarfs. The eSIDM model again produces larger cores and shallower slopes in these galaxies compared to the CDM case. In dSIDM models, the shapes of the density profiles vary with the self-interaction cross-section (or equivalently the efficiency of dissipation, assuming fixed $f_{\rm diss}$). The velocity-dependent dSIDM model has relatively low effective cross-section in the bright dwarfs and thus the central density profiles are still cored, similar to the CDM case. However, in the dSIDM model with $(\sigma/m)=0.1\cpm$, cuspy and power-law like central profiles show up in two out of the three bright dwarfs and the only cored one shows enhanced central densities at $r\lesssim\kpc$. In the dSIDM model with $(\sigma/m)=1\cpm$, the central profiles of all three bright dwarfs are cuspy with power-law slopes centering around $-1.5$ at sub-kpc scale. In the dSIDM model with $(\sigma/m)=10\cpm$, the density profiles have lower normalization although they are still cuspy, similar to the phenomenon we found in the classical dwarfs. Here, the bright dwarf galaxy that exhibits the strongest decrease in density profile normalization in this model is \textbf{m11b}.

\begin{figure*}
    \centering
    \includegraphics[width=0.98\textwidth]{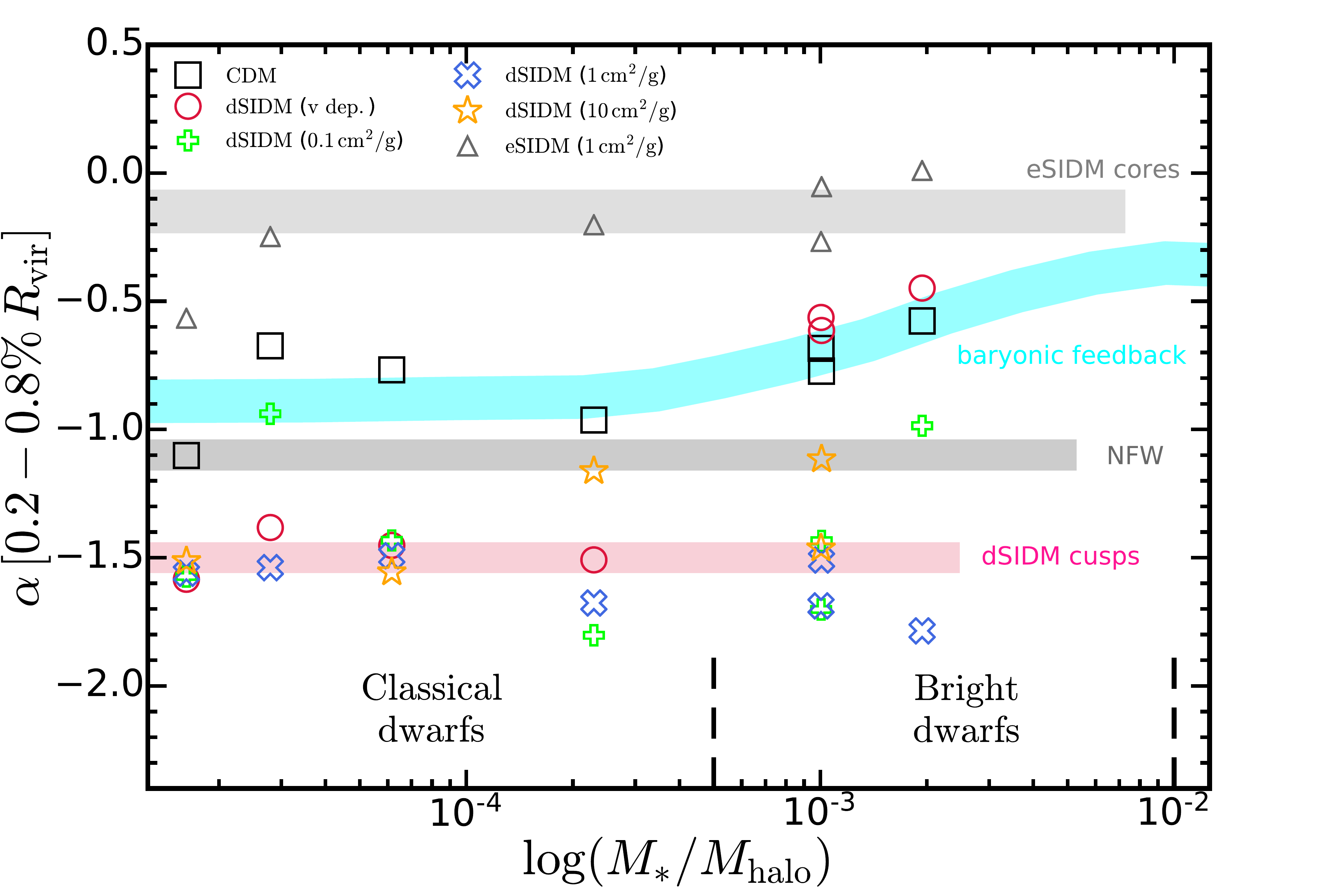}
    \caption{\textbf{Slopes of the central density profiles of dwarf galaxies in the simulation suite.} The slopes are measured at $0.2-0.8\%\, R^{\rm cdm}_{\rm vir}$. The slopes measured in simulations with different dark matter models are shown in open markers (as labelled). Galaxies are ordered from left to right based on their stellar-to-halo mass ratios ($M_{\ast}/M_{\rm halo}$), and are classified as classical dwarfs and bright dwarfs. (The ultra-faint dwarf \textbf{m09} in the suite also has its $M_{\ast}/M_{\rm halo}$ value lying in the classical dwarf regime.) The asymptotic behaviours of the slopes at the low mass end are clearly different between different dark matter models. In low-mass dwarf galaxies, the density profiles in dSIDM models with $(\sigma/m)\geq 1\cpm$ and the velocity-dependent model converge to a slope of $\sim -1.5$ (indicated by the thick red horizontal line). The slope is steeper than the asymptotic slope $-1$ of the NFW profile ($\sim -1.1$ at the radii we measure the slope, indicated by the thick black horizontal line). In contrast, the dSIDM model with $(\sigma/m)= 0.1\cpm$ can still produce small cores in some dwarf galaxies with relatively strong baryonic feedback, with $\alpha \sim -1$ at the radius of measurement and becoming even shallower at smaller radii as shown in the right panels of Figure~\ref{fig:denpro_lowmass} and \ref{fig:denpro_intemass}. In the bright dwarfs, the velocity-dependent dSIDM model produces cored profiles with $\alpha \sim -0.5$. The dSIDM models with constant cross-sections still produce cuspy density profiles with slopes centering around $-1.5$ but scattering from $-2$ to $-1$. Unlike dSIDM models, density profiles in CDM are shallower than the NFW profile and are shallower in more massive dwarf galaxies, due to stronger baryonic feedback there (indicated by the thick cyan line). The eSIDM model consistently produces cored density profiles with slope $\sim -0.2$ in most of the dwarf galaxies (indicated by the thick gray horizontal line). We note that all the thick reference lines are meant to label different ``tracks'' and are rigorous fits to the simulation results.
    }
    \label{fig:slope}
\end{figure*}

Comparing the density profiles of the classical dwarfs and bright dwarfs, we find that the dSIDM model with the same constant cross-section can behave qualitatively differently in galaxies of different masses. For example, the model with $(\sigma/m)=0.1\cpm$ produces cored central profiles in two of the classical dwarfs but produces cuspy central profiles in two of the bright dwarfs. As discussed in Section~\ref{sec:timescale}, the dissipation time scale of models with constant cross-section inversely depends on density and velocity dispersion of the system. The bright dwarfs typically have much higher velocity dispersion at their centers than the classical dwarfs while the central densities are comparable to the classical dwarfs. As expected, dissipation has stronger impact in the bright dwarfs. On the other hand, the velocity-dependent dSIDM model produces cuspy central profiles in the classical dwarfs but produces cored central profiles in the bright dwarfs. The dissipation time scale of the velocity-dependent model inversely depends on density but exhibits a $v^{3}$ asymptotic dependence on velocity dispersion. The opposite dependence on velocity dispersion makes the impact of dissipation stronger in the classical dwarfs.

\begin{figure}
    \centering
    \includegraphics[width=0.49\textwidth]{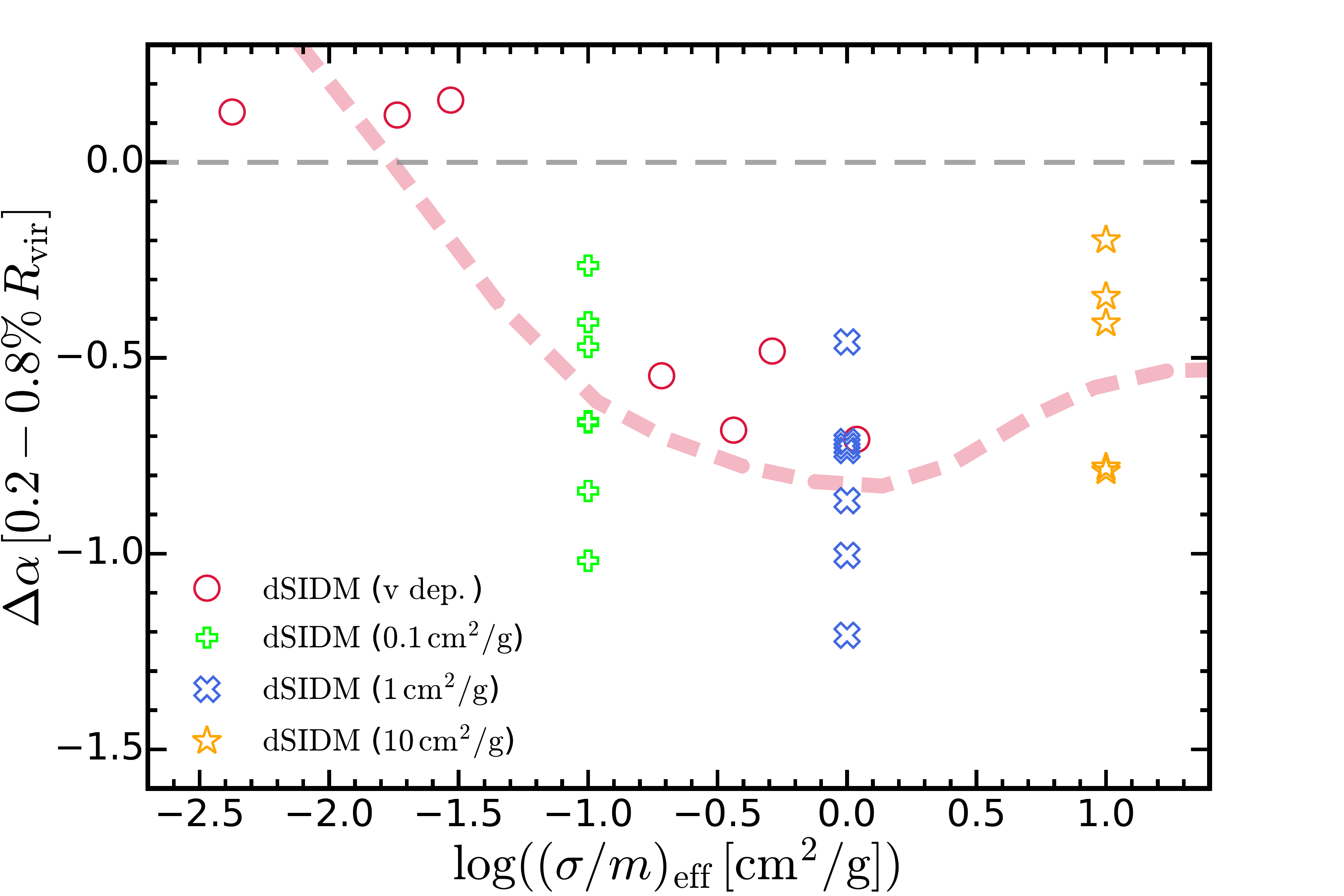}
    \includegraphics[width=0.49\textwidth]{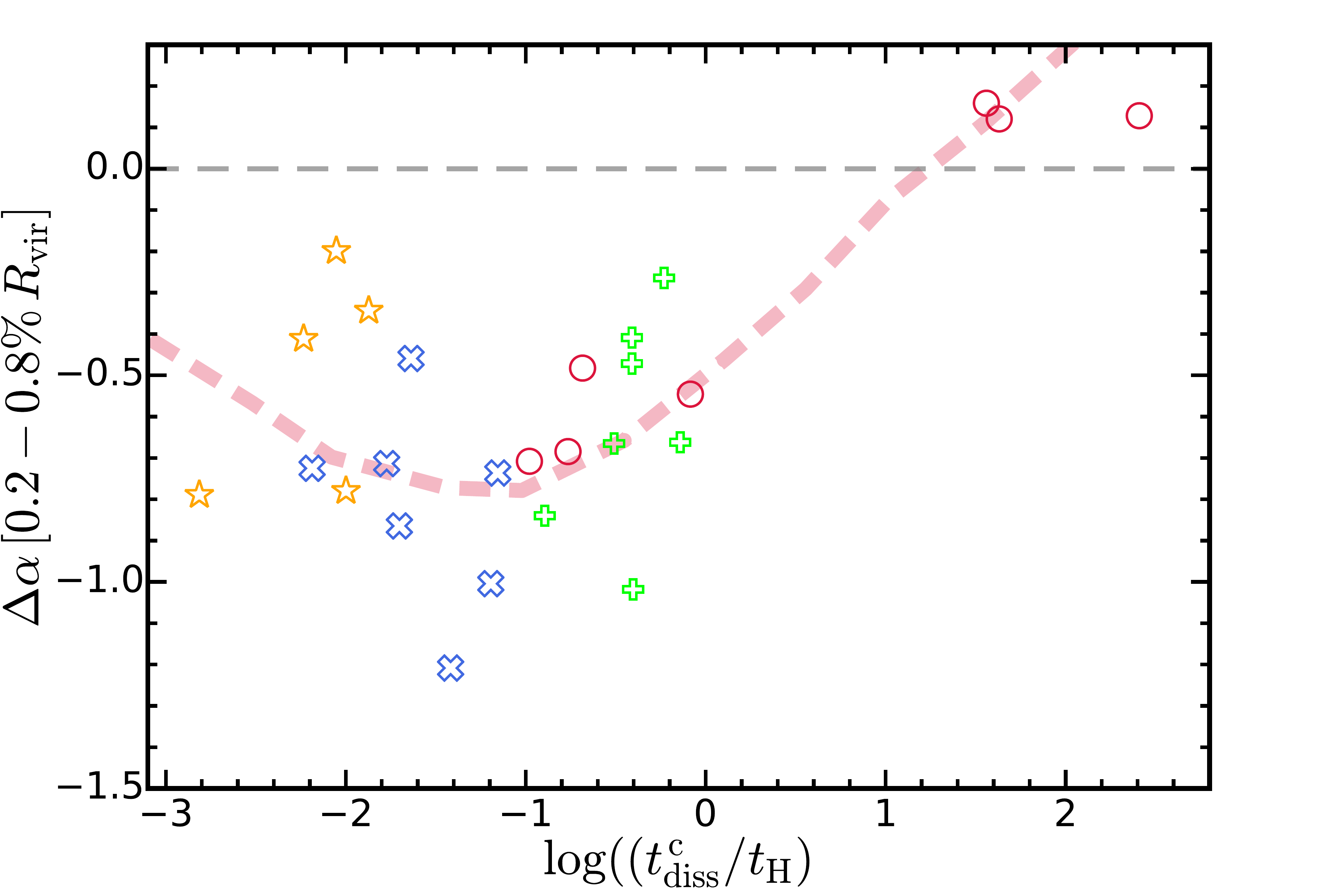}
    \caption{{\it Top}: \textbf{Slope change versus effective self-interaction cross-section of dwarf galaxies in simulations.} $\Delta \alpha$ is defined as the difference in slopes measured at $0.2-0.8\% R^{\rm cdm}_{\rm vir}$ between galaxies in dSIDM and CDM. The red dashed line labels the qualitative trend (not rigorous fitting). In the regime where $(\sigma/m)_{\rm eff}<1\cpm$, the steepening of central profiles induced by dissipative dark matter self-interactions becomes progressively stronger in systems with higher effective cross-sections. In the regime where $(\sigma/m)_{\rm eff}>1\cpm$, the steepening of central profiles saturates. {\it Bottom}: \textbf{Slope change versus dissipation time scale at halo center.} When $\log{(t^{\rm c}_{\rm diss}/t_{\rm H})}>-1$, the density profiles become steeper as $t^{\rm c}_{\rm diss}$ decreases while the steepening saturates when $\log{(t^{\rm c}_{\rm diss}/t_{\rm H})}<-1$.}
    \label{fig:slope_vs_sigma}
\end{figure}

To quantify the impact of dissipation on galaxy structures, we measure the slopes of the total mass density profiles at galaxy centers. The aperture we choose for this measurement is $0.2-0.8\%\,R^{\rm cdm}_{\rm vir}$ (as indicated by the gray bands in Figure~\ref{fig:denpro_lowmass} and \ref{fig:denpro_intemass}), where $R^{\rm cdm}_{\rm vir}$ is the virial radius of the halo in the CDM model.~\footnote{The virial radius does not vary much in simulations with different dark matter models. Using the virial radius in the CDM run is simply to ensure that the aperture is identical for different dark matter models.} This has been chosen since it is an appropriate aperture to illustrate the impact of dissipation at small radii while remaining larger than the convergence radii of dark matter profiles in these runs (rather conservative estimates, see Table~\ref{tab:sims}). In Figure~\ref{fig:slope}, we show the power-law slopes of the density profiles (measured at $0.2-0.8\% R^{\rm cdm}_{\rm vir}$) of simulated dwarf galaxies versus their stellar-to-halo mass ratios ($M_{\ast}/M_{\rm halo}$). The slopes of the density profiles in different models show four different ``tracks'': 
\begin{itemize}

\item The NFW profile has an asymptotic $-1$\,\footnote{The slope of the NFW profile varies with radius. At the radii we measure the slopes, the NFW profile has a slope of $\sim -1.1$.} power-law slope at galaxy centers. 

\item In CDM, baryonic feedback drives gas outflow and creates fluctuations in the central gravitational potential which significantly affects the distribution of dark matter. Dwarf galaxies have shallower density profiles than the NFW profile. The difference in slope peaks in most massive bright dwarfs where baryonic feedback is most efficient in perturbing galaxy structures, as has been found in previous studies~\citep[e.g.,][]{diCintio2014,Chan2015,Onorbe2015,Tollet2016,Lazar2020}. 

\item In eSIDM, elastic dark matter self-interaction drives the halo to thermal equilibrium and produces an isothermal density profile with a core at the center. The power-law slopes of the central profiles are close to zero in most of the simulated dwarf galaxies, regardless of their mass.

\item In dSIDM, dissipative dark matter self-interaction is a competing factor against baryonic feedback in shaping the central density profile. When $(\sigma/m)_{\rm eff} > 0.1\cpm$, dark matter dissipation becomes dominant and the central density profiles in dwarf galaxies are steeper than the ones in the CDM model.~\footnote{We verify that the impact of baryonic feedback becomes negligible in this regime through the comparison with DMO simulations in Section~\ref{sec:dmo}.} In the classical dwarfs, the power-law slopes are steeper than the $\rm -1$ of NFW profiles and asymptote to $\sim -1.5$. In the bright dwarfs, the power-law slopes have larger scatter, ranging from $-2$ to $-1$. When the $(\sigma/m)_{\rm eff}$ is relatively low (e.g. the model with $(\sigma/m)=0.1\cpm$ in the classical dwarfs and the velocity-dependent model in the bright dwarfs), the central density profiles are affected by a mixture of dark matter dissipation and baryonic feedback, which compete with each other. In some dwarfs with relatively strong feedback effects, the slopes become shallower than the $\sim -1.5$ value at the radius of measurement. They could even develop a core ($\alpha \gtrsim -0.5$) at smaller radii as shown in the right panels of Figure~\ref{fig:denpro_lowmass} and \ref{fig:denpro_intemass}.

\end{itemize}

To demonstrate the net impact of dissipation, in the top panel of Figure~\ref{fig:slope_vs_sigma}, we show the slope change $\Delta \alpha$ versus the effective self-interaction cross-section $(\sigma/m)_{\rm eff}$. $\Delta \alpha$ is defined as the difference in slopes measured at $0.2-0.8\% R^{\rm cdm}_{\rm vir}$ between galaxies in dSIDM and CDM, $\Delta \alpha = \alpha^{\rm dsidm} - \alpha^{\rm cdm}$. More negative $\Delta \alpha$ indicates stronger impact of dissipation on the steepness of the density profile. The effective self-interaction cross-section is calculated using Equation~\ref{eq:sigma_eff}, plugging in the density and one-dimensional velocity dispersion of dark matter particles enclosed in a sphere of radius $1/3\, r^{\rm cdm}_{1/2}$. The red dashed line shows the qualitative trend (not rigorous fitting) of $\Delta \alpha$ versus $(\sigma/m)_{\rm eff}$. When $(\sigma/m)_{\rm eff}\lesssim 1\cpm$, the steepening of the central density profiles induced by dissipation becomes progressively stronger in systems with higher effective cross-sections. The change of the power-law slope scales roughly linearly as the logarithm of the effective cross-section. When $(\sigma/m)_{\rm eff}$ is larger than $1\cpm$, the steepening of the central density profiles saturates. The $\Delta \alpha$ when $(\sigma/m)_{\rm eff}\simeq 10\cpm$ is comparable to the $(\sigma/m)_{\rm eff}\simeq 0.1\cpm$ case. In the bottom panel of Figure~\ref{fig:slope_vs_sigma}, we show the slope change $\Delta \alpha$ versus the dissipation time scale at halo center $t^{\rm c}_{\rm diss}$, calculated using Equation~\ref{eq:tdiss}. The steepening of the central density profiles occurs when $t^{\rm c}_{\rm diss}$ becomes comparable to $t_{\rm H}$. The slope difference becomes larger as $t^{\rm c}_{\rm diss}$ decreases when $t^{\rm c}_{\rm diss} \gtrsim 0.1\,t_{\rm H}$. When $t^{\rm c}_{\rm diss} \lesssim 0.1\, t_{\rm H}$, the steepening of the central profile saturates, similar to the trend in the top panel. This is likely related to the increasing rotation support of dark matter when $(\sigma/m)_{\rm eff}\gtrsim 1\cpm$, which will be shown in the following section.

\subsection{Kinematic properties}
\label{sec:kin}

In this section, we will explore the kinematic properties of dark matter particles in the simulated dwarf galaxies. These properties include velocity dispersion, coherent rotation velocity, velocity anisotropy and the velocity distribution function of dark matter. 

To evaluate these properties, we first divide a simulated halo into spherical shells with respect to the halo center. In each shell, we measure the total angular momentum of dark matter particles and align the z-axis of the coordinate system with the direction of the angular momentum. This helps us define the azimuthal and zenith directions (note that different shells could have different directions of angular momentum and thus different definitions of the z-axis). The velocities of dark matter particles are decomposed to the radial, zenith and azimuthal components ($v_{\rm r}$, $v_{\theta}$ and $v_{\phi}$) in spherical galactocentric coordinates. The coherent rotation velocity $V_{\rm rot}$ of particles in the shell is calculated as
\begin{align}
    & V_{\rm rot} = \dfrac{J_{\rm dm}}{I_{\rm shell}} R_{\rm shell}, \nonumber \\
    & I_{\rm shell} = \dfrac{2}{5} M_{\rm dm} \dfrac{r^5_{\rm o}-r^5_{\rm i}}{r^3_{\rm o}-r^3_{\rm i}}, \hspace{0.3cm} R_{\rm shell} = \dfrac{r_{\rm o}+r_{\rm i}}{2}
\end{align}
where $J_{\rm dm}$ is the total angular momentum of dark matter particles in the shell, $M_{\rm dm}$ is the total mass of dark matter in the shell, $I_{\rm shell}$ is the moment of inertia of the shell, $r_{\rm o}$ and $r_{\rm i}$ are the outer and inner radii of the shell, $R_{\rm shell}$ is the median radius of the shell. Here, we have assumed that the mass is uniformly distributed in the shell in the calculation of moment of inertia. We also measure the mean inflow/outflow velocity ($\overline{v_{\rm r}}$) of dark matter particles in the shell. We subtract both the coherent rotation velocity and the mean inflow/outflow velocity before measuring the velocity dispersion $\sigma_{\rm r}$, $\sigma_{\theta}$ and $\sigma_{\phi}$ corresponding to the radial direction, and the azimuthal and zenith angles, respectively. Finally, the three-dimensional velocity dispersion is calculated as: $\sigma_{\rm 3d} = \sqrt{\sigma^2_{\rm r} + \sigma^2_{\theta} + \sigma^2_{\phi}}$. The one-dimensional velocity dispersion is estimated as: $\sigma_{\rm 1d} = \sqrt{(\sigma^2_{\rm r} + \sigma^2_{\theta} + \sigma^2_{\phi})/3}$. The degree of velocity anisotropy is calculated as
\begin{equation}
    \beta = 1- \dfrac{\sigma_{\phi}^2 + \sigma_{\theta}^2}{2\sigma_{\rm r}^2}, 
    \label{eq:vanis}
\end{equation}
Under this definition, $\beta=0$ corresponds to an isotropic velocity dispersion, $\beta=1$ to a velocity dispersion purely dominated by the radial component, and negative $\beta$ to a velocity dispersion dominated by the tangential component. 

\begin{figure}
    \centering
    \includegraphics[width=0.49\textwidth]{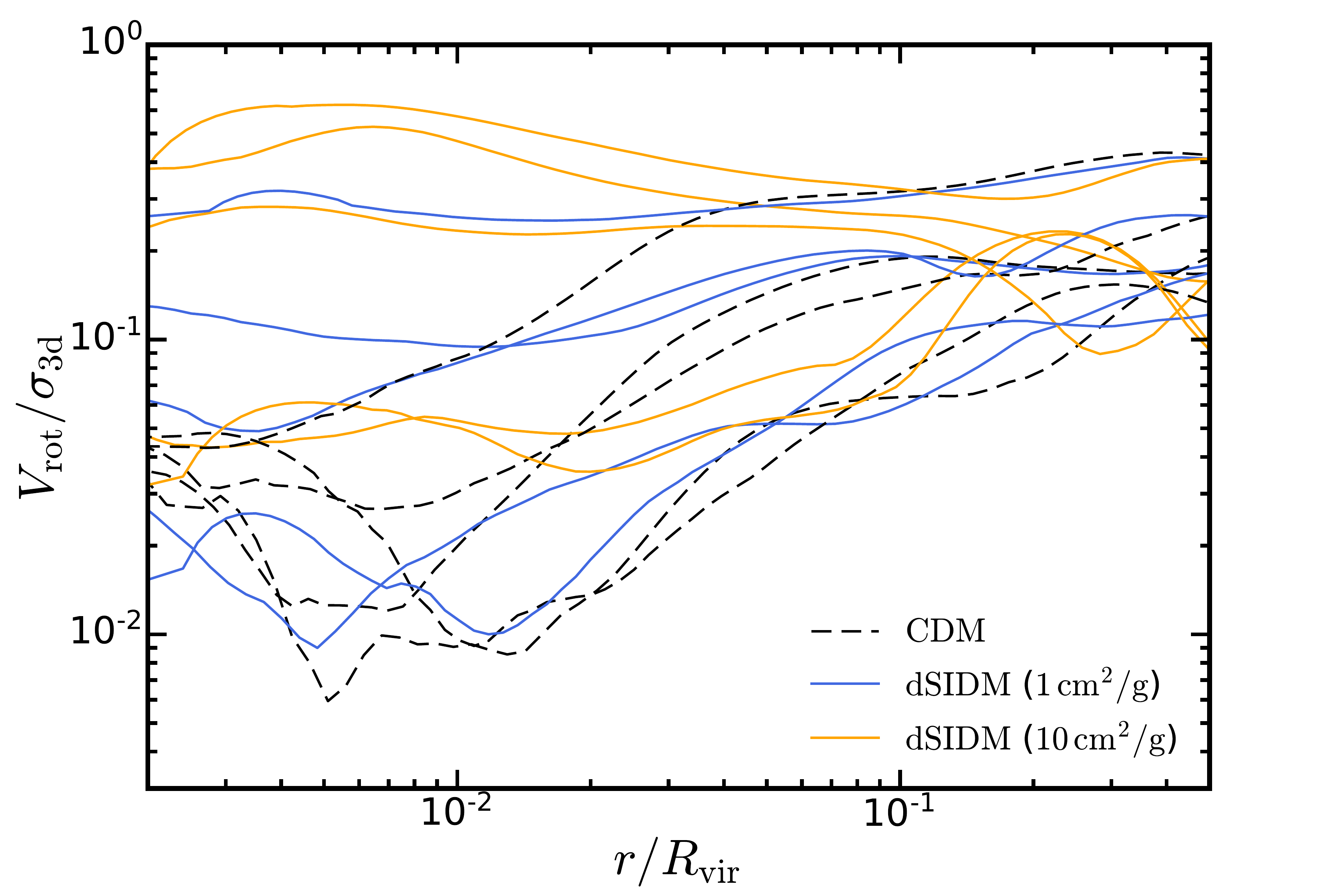}
    \caption{\textbf{Coherent rotation velocity relative to velocity dispersion of dark matter in simulations.} The coherent rotation velocities and the velocity dispersions are measured in spherical shells as discussed in the main text. We present the results in CDM and dSIDM with $(\sigma/m)=1$ and $10 \cpm$. For each model, we show the results of five dwarf galaxies: \textbf{m10q}, \textbf{m10b}, \textbf{m10v}, \textbf{m11a} and \textbf{m11b}. The coherent rotation becomes more prominent inside $\sim 1\%\,R_{\rm vir}$ as the self-interaction cross section increases, but not in every galaxy. The two galaxies that have rotation velocities comparable to velocity dispersions are \textbf{m10q} and \textbf{m11b}.}
    \label{fig:circ_motion}
\end{figure}

\begin{figure}
    \centering
    \includegraphics[width=0.49\textwidth]{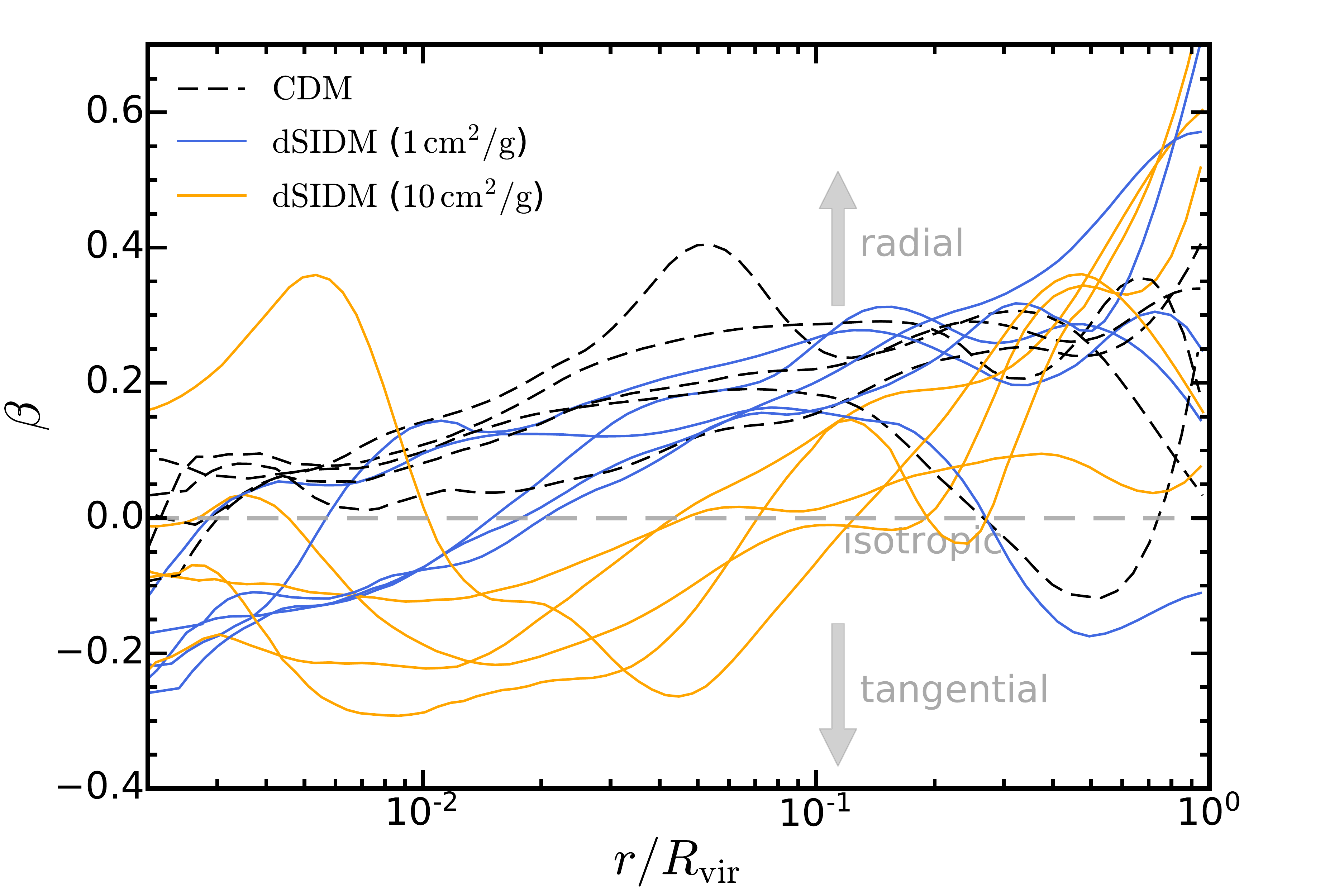}
    \caption{\textbf{Velocity anisotropy profiles of dark matter in simulated dwarf galaxies.} The velocity anisotropies are calculated using Equation~\ref{eq:vanis}. We present the results in CDM and dSIDM with $(\sigma/m)=1$ and $10 \cpm$. For each model, we show the results of the same five galaxies as in Figure~\ref{fig:circ_motion}. The velocity anisotropy decreases as the self-interaction cross-section increases and eventually becomes negative, suggesting that the velocity dispersion is more dominated by the tangential component. This is consistent with more coherent rotation found in Figure~\ref{fig:circ_motion}.}
    \label{fig:anisotropy}
\end{figure}

\begin{figure*}
    \centering
    \textbf{central ($\sim 100-200{\rm pc}$)} \hspace{3.0cm}  \textbf{intermediate ($\sim\kpc$)} \hspace{3.0cm} \textbf{outskirt ($\gtrsim 10\kpc$)} \\
    \includegraphics[trim=0.4cm 0.3cm 3cm 1cm, clip, width=0.32\textwidth]{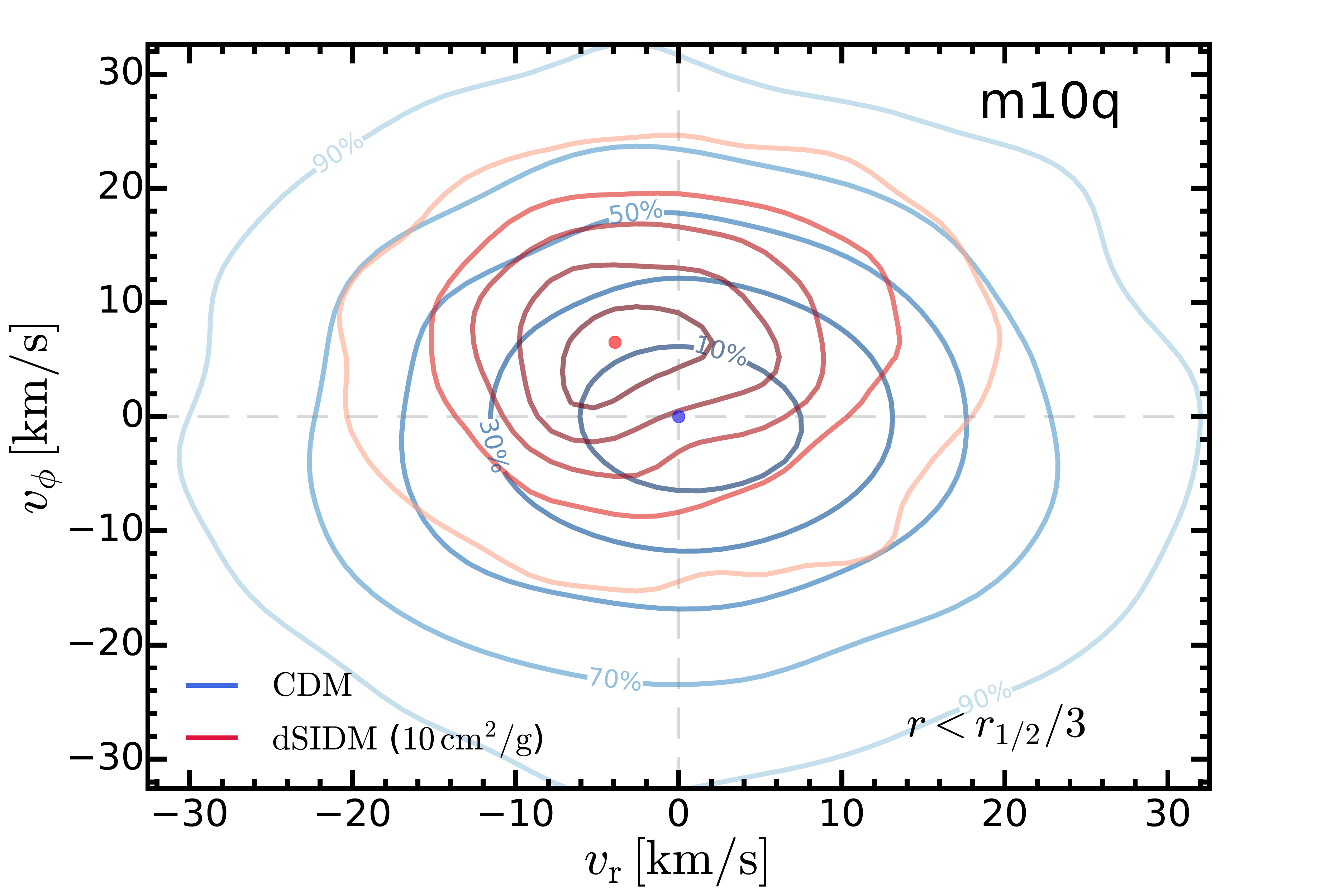}
    \includegraphics[trim=0.4cm 0.3cm 3cm 1cm, clip, width=0.32\textwidth]{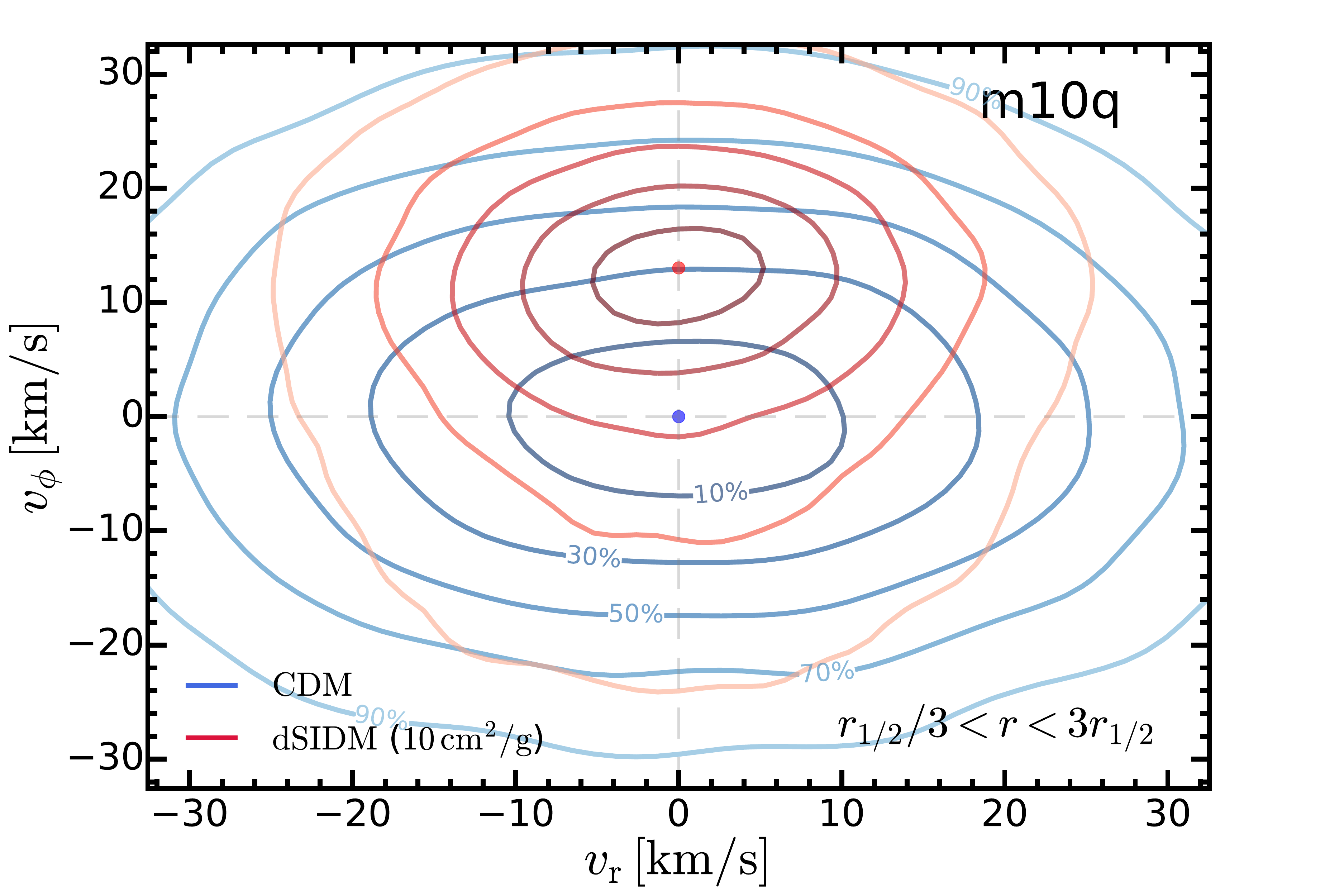}
    \includegraphics[trim=0.4cm 0.3cm 3cm 1cm, clip, width=0.32\textwidth]{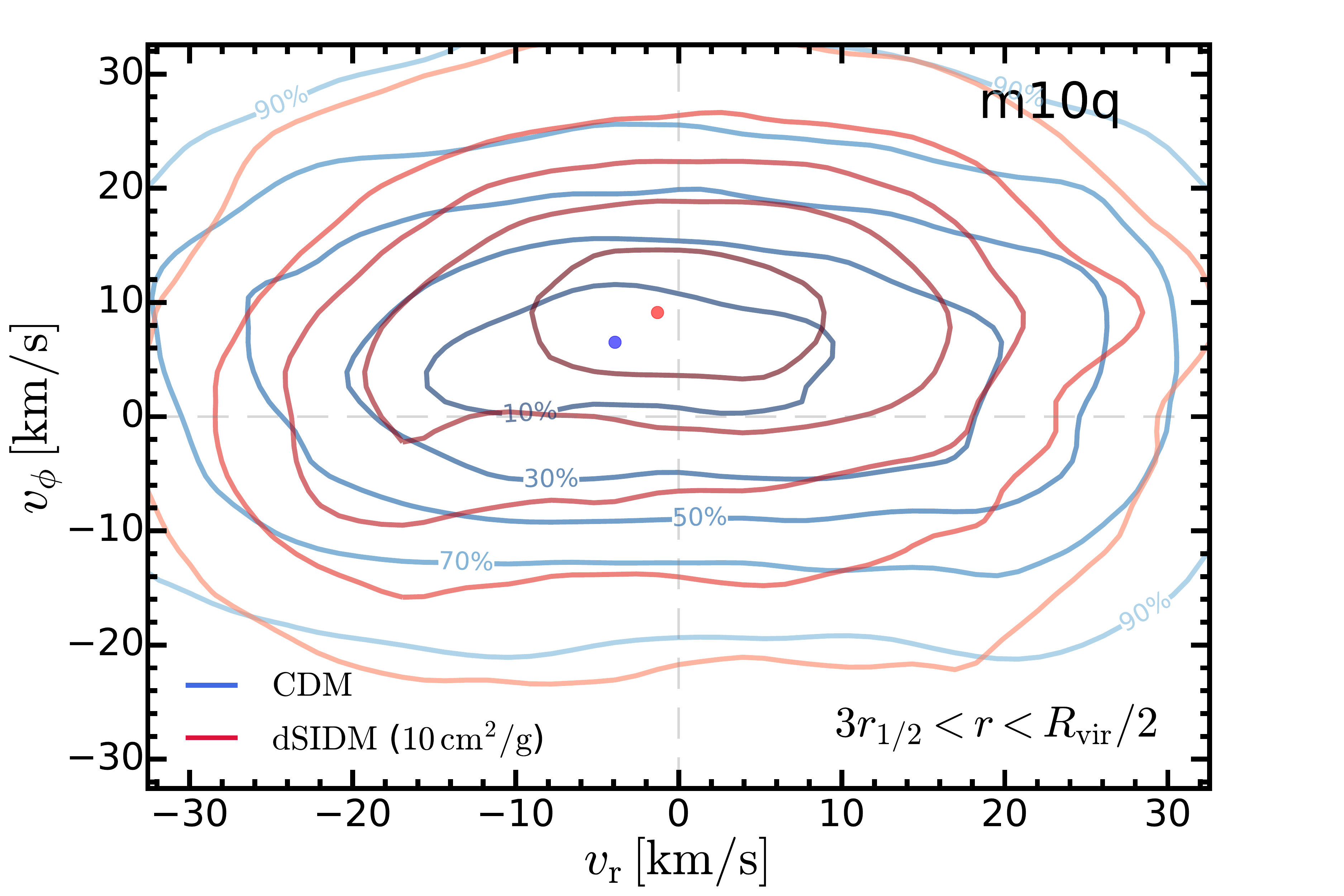}
    
    \includegraphics[trim=0.4cm 0.3cm 3cm 1cm, clip, width=0.32\textwidth]{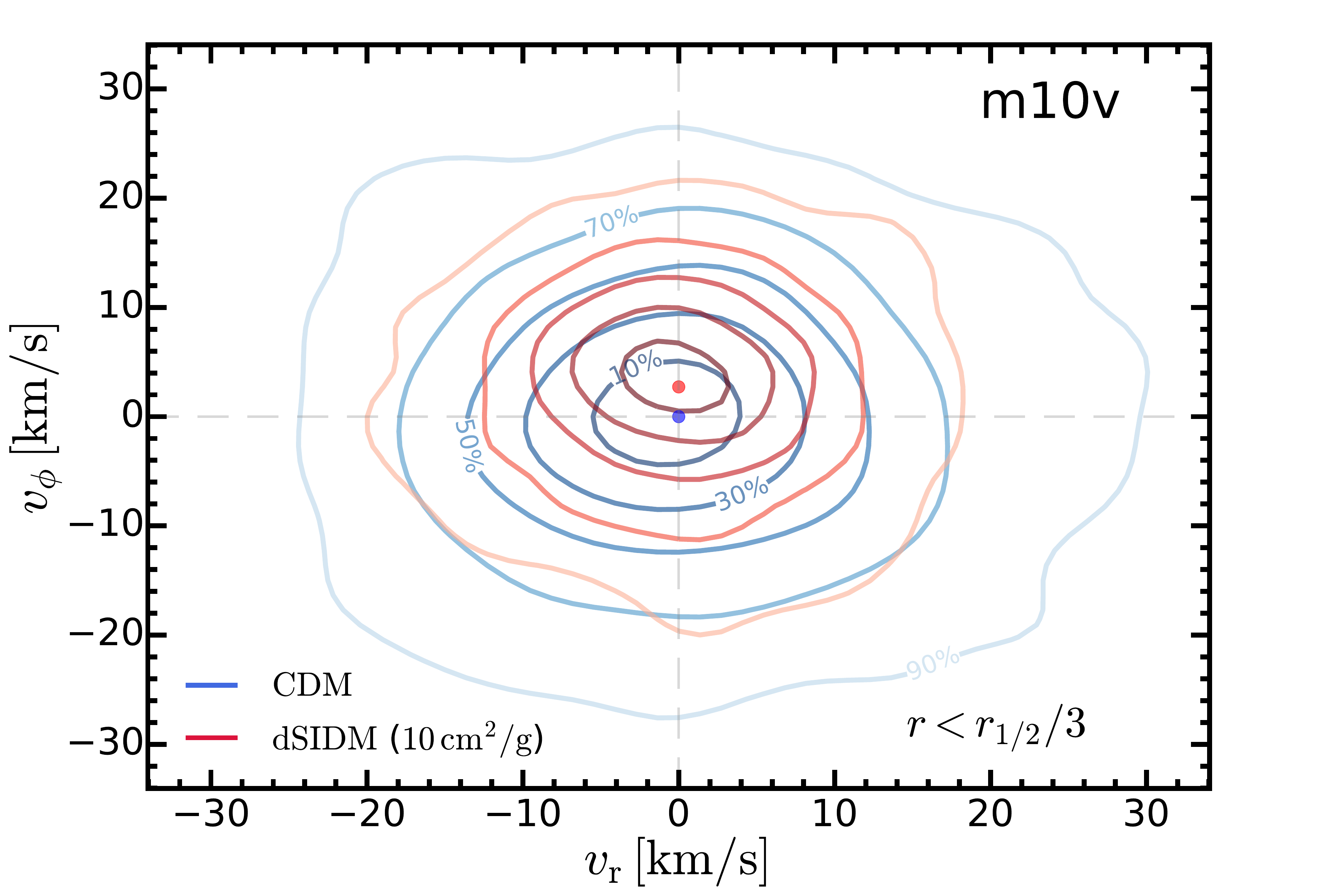}
    \includegraphics[trim=0.4cm 0.3cm 3cm 1cm, clip, width=0.32\textwidth]{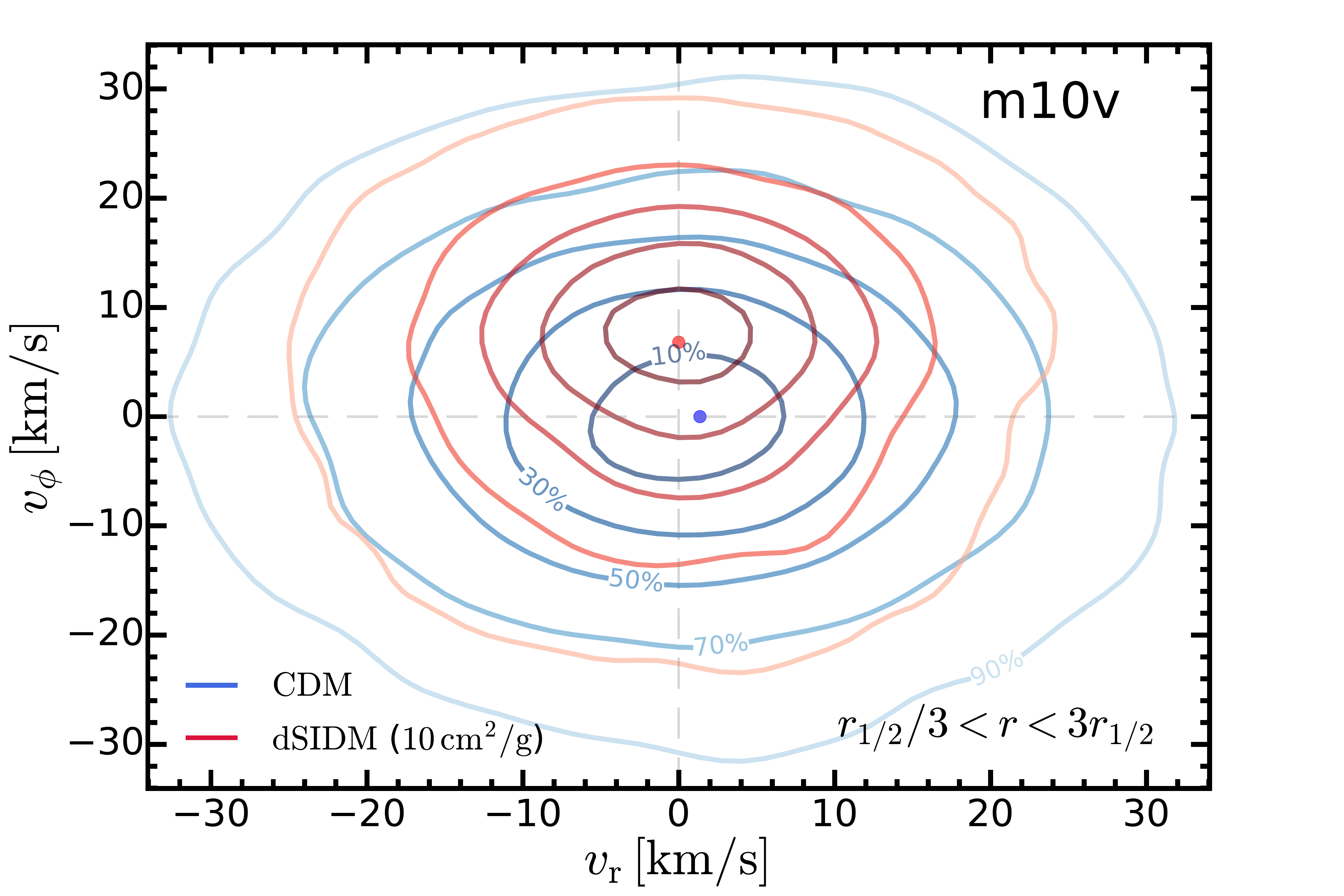}
    \includegraphics[trim=0.4cm 0.3cm 3cm 1cm, clip, width=0.32\textwidth]{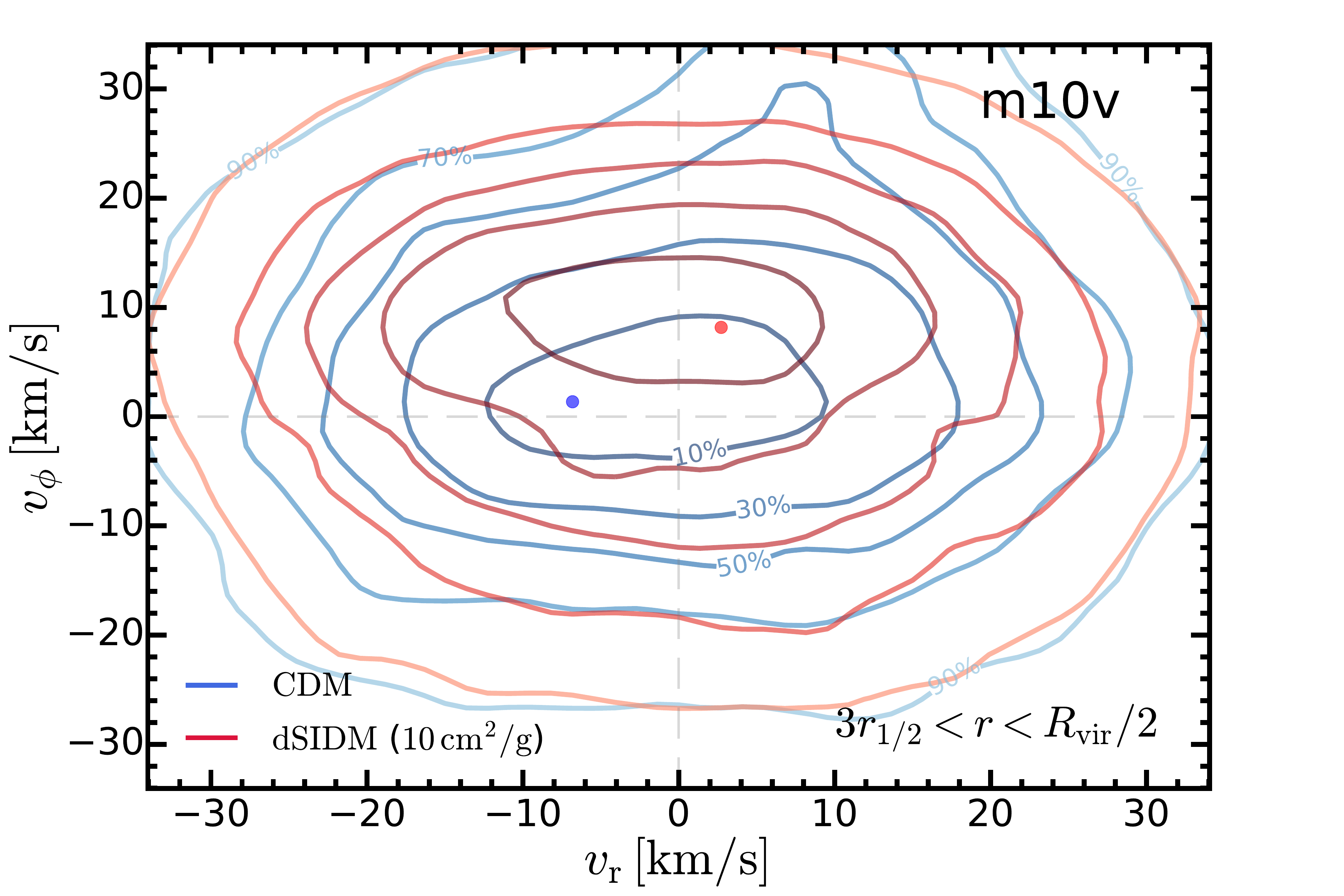}
    
    \caption{\textbf{Phase space distribution function of dark matter in simulated classical dwarfs.} We present the two-dimensional density distribution of dark matter in the $v_{\phi}-v_{\rm r}$ phase space, ${\rm d}\rho_{\rm dm}/{\rm d}v_{\rm r} {\rm d}v_{\phi}$. In the three columns, we show the distribution in three radial bins: central, $r<r^{\rm cdm}_{1/2}/3$, intermediate, $r^{\rm cdm}_{1/2}/3<r<3r^{\rm cdm}_{1/2}$, and ``outskirt'', $3r^{\rm cdm}_{1/2}<r<0.5R^{\rm cdm}_{\rm vir}$, respectively. From inside out, each contour is determined such that it encloses a certain percentile of dark matter particles in the bin. The percentiles range from $10\%$ to $90\%$ with $20\%$ as interval, as labeled on the contours. The dots represent the locations where the velocity distribution function peaks. Dark matter in dSIDM models exhibit positive median $v_{\phi}$ while the phase space distribution is almost isotropic in CDM. The differences consistently show up in the three radial bins and suggest a coherent rotation built up in dSIDM haloes. The phase space distribution in the dSIDM model is also more peaky than the CDM case, at least for the central and intermediate radial bins.}
    \label{fig:phasespace}
\end{figure*}

\begin{figure*}
    \centering
    \includegraphics[trim=1.2cm 0.3cm 1cm 0.7cm, clip,width=0.49\textwidth]{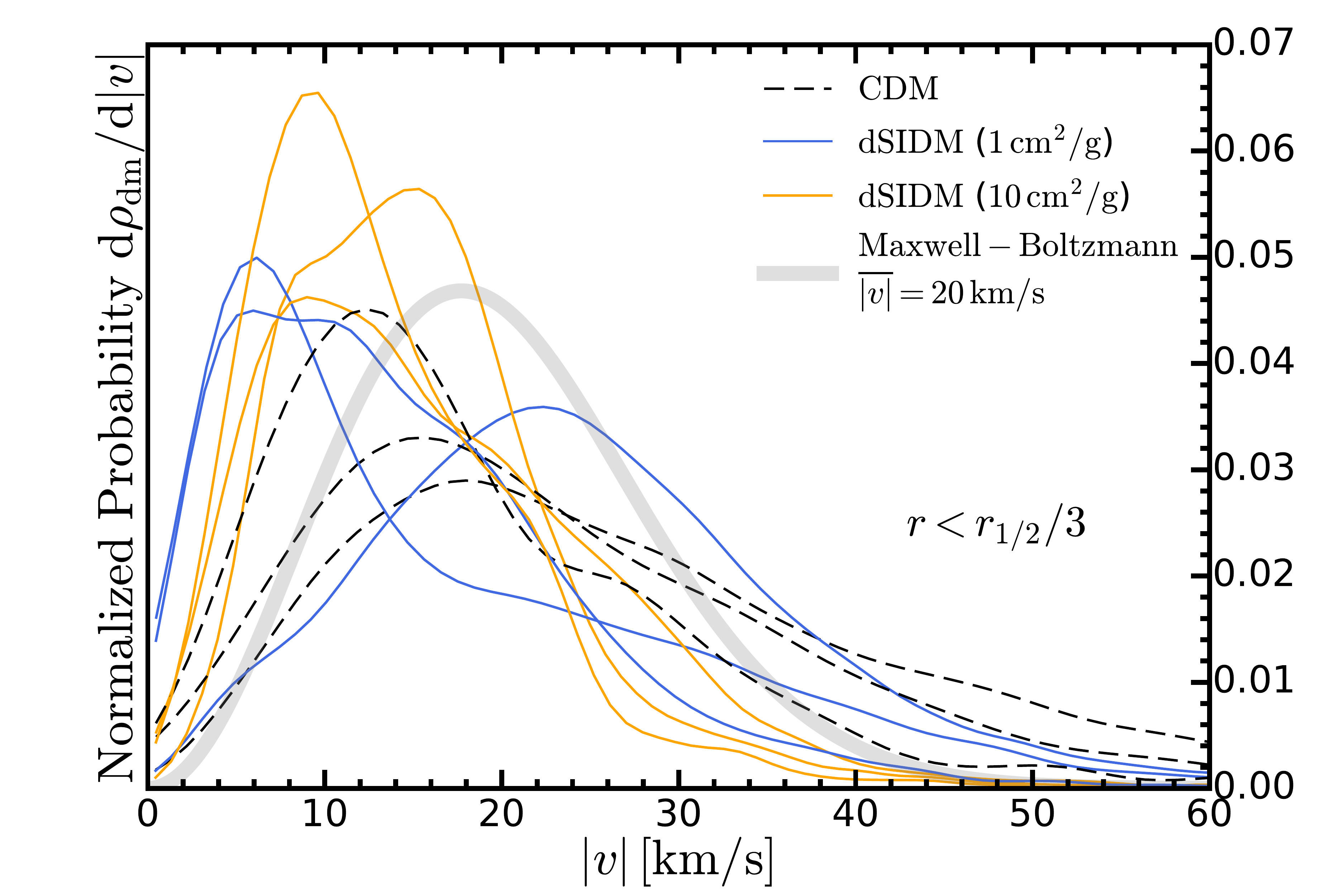}
    \includegraphics[trim=1.2cm 0.3cm 1cm 0.7cm, clip,width=0.49\textwidth]{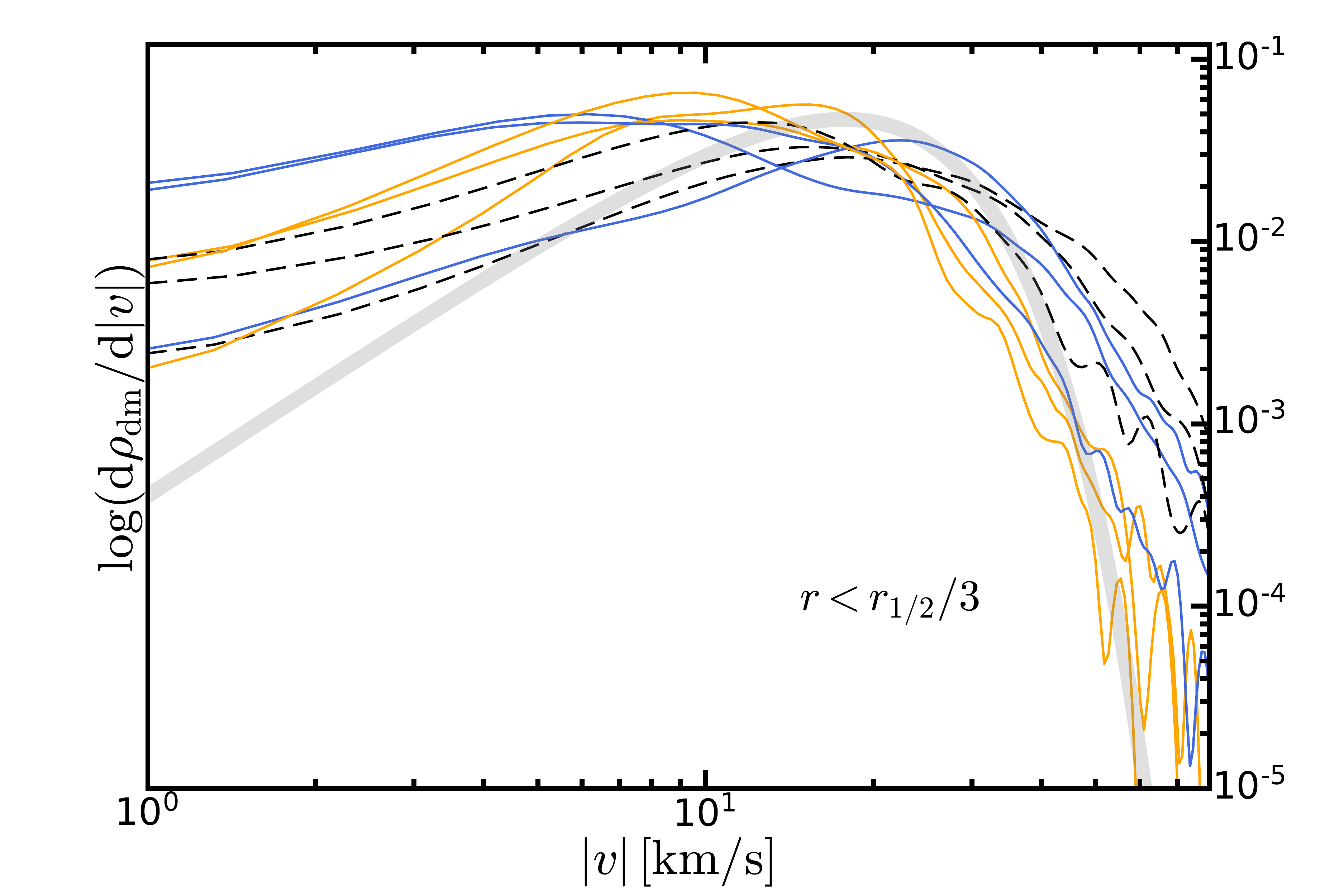}
    \includegraphics[trim=1.2cm 0.3cm 1cm 0.7cm, clip,width=0.49\textwidth]{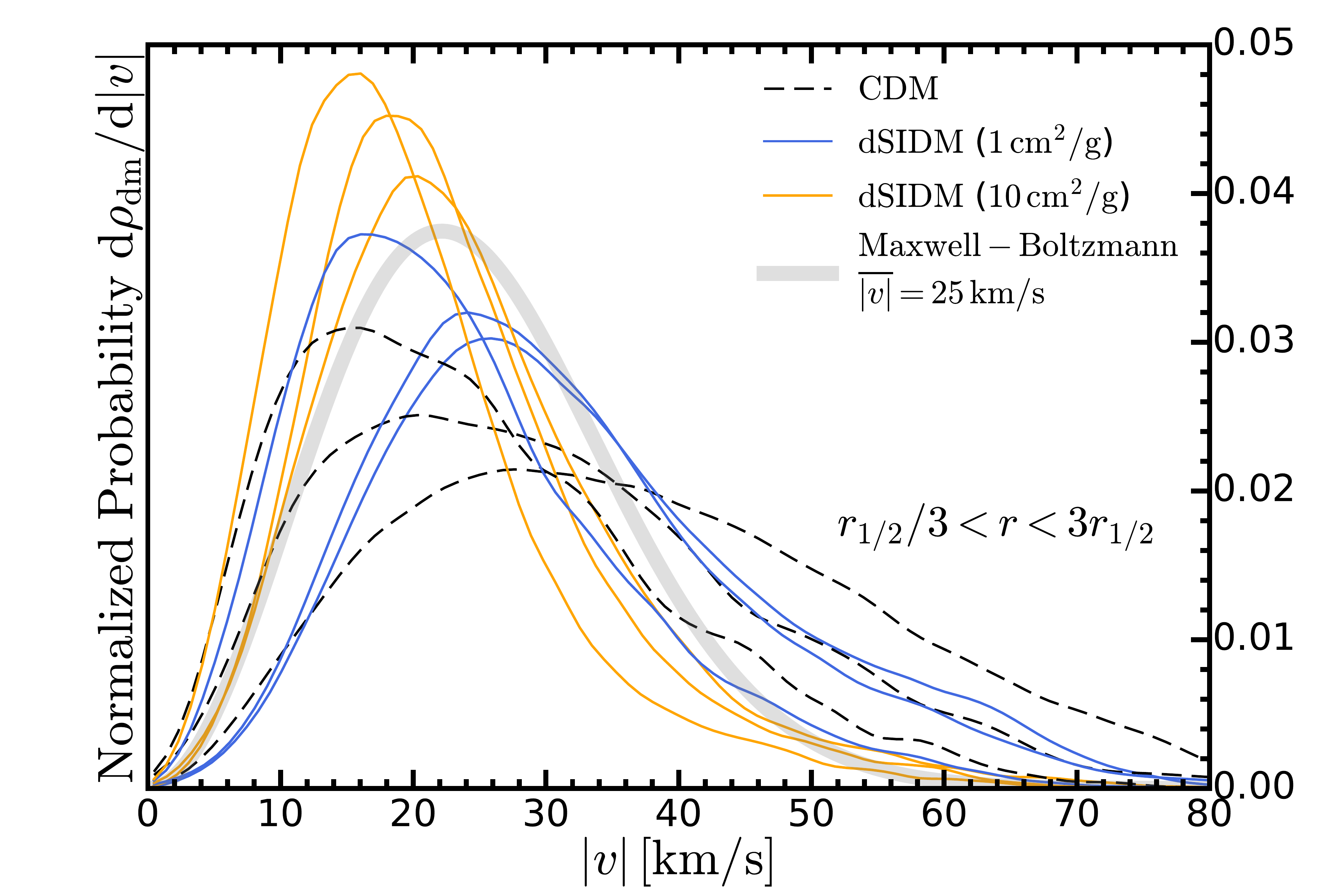} 
    \includegraphics[trim=1.2cm 0.3cm 1cm 0.7cm, clip,width=0.49\textwidth]{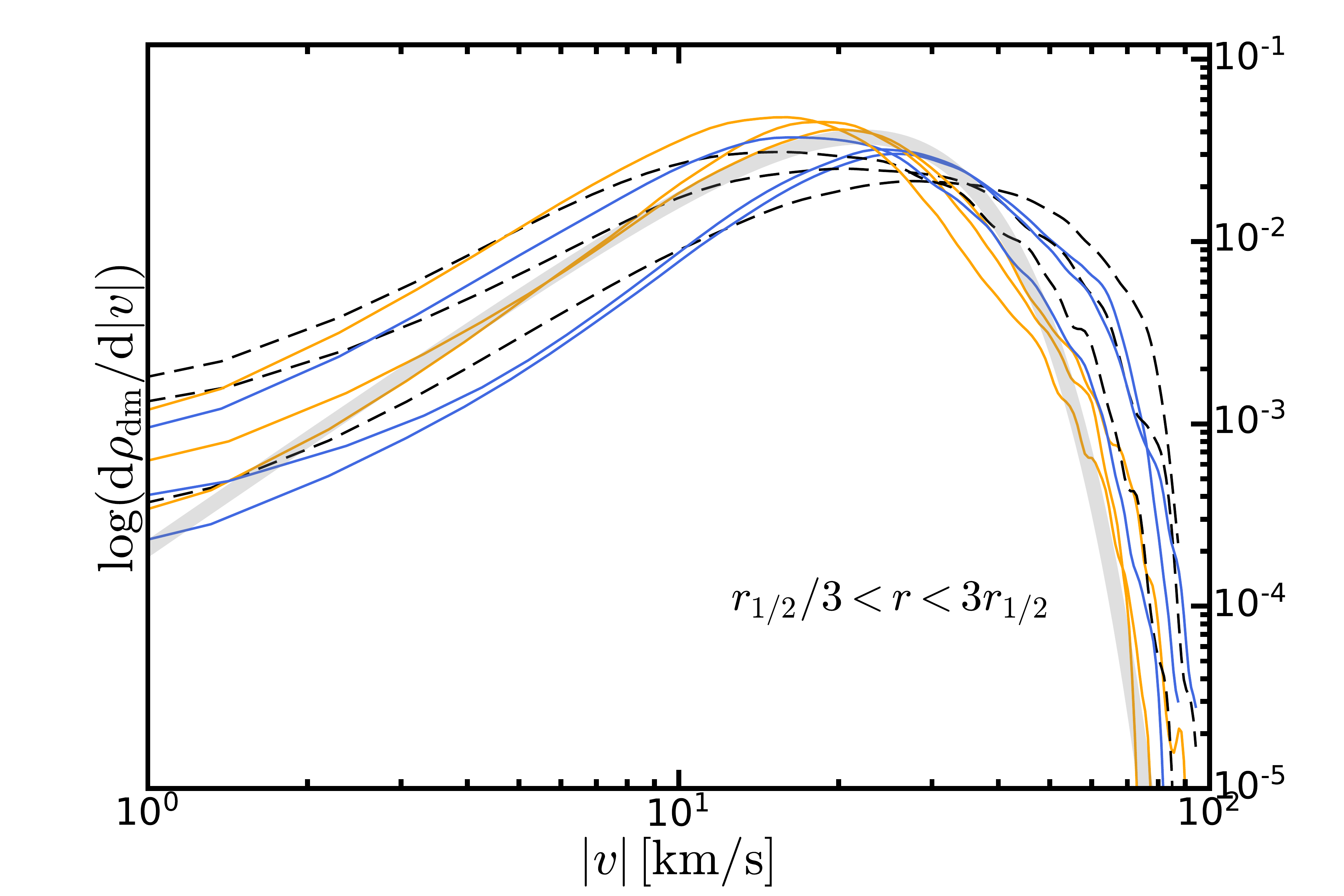}
    \caption{\textbf{Velocity distribution functions of dark matter in the classical dwarfs.} {\it Top left}: Velocity distribution function at small galactocentric radii ($r<r^{\rm cdm}_{\rm 1/2}/3$). We show the velocity distributions in CDM and dSIDM with $(\sigma/m)=1$ and $10\cpm$ (as labelled). As a reference, a Maxwell-Boltzmann distribution is shown with the thick gray line. Compared to CDM, the velocity distribution functions in dSIDM models are more suppressed at the high velocity tail as the cross-section increases and the peaks of the distributions also decrease systematically. {\it Top right}: Same velocity distribution functions as the top left panel but in log-log scale to highlight the asymptotic behaviour at the low velocity tail. Both CDM and dSIDM models have velocity distribution functions that decreases slower than the Maxwell-Boltzmann distribution at the low velocity tail. Dissipation has limited impact at low velocities due to small interaction rates there. {\it Bottom left}: Velocity distribution function at intermediate galactocentric radii ($r^{\rm cdm}_{1/2}/3<r<3r^{\rm cdm}_{1/2}$). Similar differences in the velocity distribution of CDM and dSIDM are found compared to the one at small radii. {\it Bottom right}: The same velocity distribution function as the bottom left panel but in log-log scale. Both CDM and dSIDM models have velocity distributions that overall resemble the Maxwell-Boltzmann distribution at the low velocity tail.
    }
    \label{fig:vdist}
\end{figure*}

\medskip
\noindent {\textbf{Coherent rotation:}} A natural consequence of dissipative interactions is that particles tend to move in a more coherent fashion, rather than in random dispersion. If the energy dissipation is faster than the relaxation processes (either through dark matter self-interactions or gravitational interactions), the coherent rotation would gradually become prominent in the system if angular momentum is conserved. In Figure~\ref{fig:circ_motion}, we show the ratio between coherent rotation velocity and three-dimensional velocity dispersion of dark matter measured in spherical shells in CDM and dSIDM with $(\sigma/m)=1$ and $10 \cpm$. For each model, each line corresponds to one of the simulated dwarf galaxies: \textbf{m10q}, \textbf{m10b}, \textbf{m10v}, \textbf{m11a} and \textbf{m11b}. Qualitatively, the coherent rotation velocity at small galactocenric radii becomes progressively more prominent as self-interaction cross-section becomes higher (and dissipation becomes more efficient). At large radii, the systematic difference becomes negligible. Quantitatively, there are apparent galaxy-to-galaxy variations. The ratio can reach $\sim 0.5$ inside $\sim 1\%\,R_{\rm vir}$ (roughly sub-kpc scale in dwarfs) in \textbf{m10q} and \textbf{m11b} in dSIDM with $(\sigma/m)=10 \cpm$, while in \textbf{m11a} and \textbf{m10b}, the ratio remains $\lesssim 0.1$ inside $\sim 1\%\,R_{\rm vir}$ in any models. These evidences suggest that, at the centers of galaxies, some dSIDM realizations are in a transition from a pure dispersion supported system to a system supported by a mixture of random velocity dispersion and coherent rotation. The radial scale for this transition to take place is a few percent of the virial radius. Such scale is quite consistent with the centrifugal barrier $\sim s R_{\rm vir}$ ($s$ is the halo spin parameter with typical value $\sim \operatorname{0.01\,-\,0.1}$) found for dissipative gas in CDM haloes~\citep[e.g.,][]{Mo1998}. 

\medskip
\noindent {\textbf{Velocity anisotropy:}} In Figure~\ref{fig:anisotropy}, we show the velocity anisotropy of dark matter measured in spherical shells in CDM and dSIDM with $(\sigma/m)=1$ and $10 \cpm$. The velocity anisotropies are calculated using Equation~\ref{eq:vanis}. The measured anisotropy is not sensitive to the bulk motion of dark matter in the shell since we have subtracted the mean rotation/inflow/outflow velocities. For each model, we show the results of the same five galaxies as in Figure~\ref{fig:circ_motion}. CDM haloes are almost isotropic at the centers with mild radial velocity dispersion anisotropy at the outskirt, which is consistent with previous studies~\citep[e.g,][]{Lemze2012,Sparre2012,Wojtak2013}. In dSIDM models, it is similar to the CDM case that the velocity anisotropy increases towards larger galactocentric radii. However, as dissipation becomes more efficient, the normalization of the velocity anisotropy decreases and eventually becomes negative at small radii. In the dSIDM model with $(\sigma/m)=10\cpm$, the velocity anisotropy drops to $\sim -0.2$ at $r \sim 1\%\,R_{\rm vir}$, suggesting that the tangential component of the velocity dispersion is relatively stronger there. This phenomenon is inline with the more prominent coherent rotation developed in dSIDM haloes. 
 
\medskip
\noindent {\textbf{Phase space distribution:}} In Figure~\ref{fig:phasespace}, we present the density distribution function of dark matter in the $v_{\phi}-v_{\rm r}$ phase space, ${\rm d}\rho_{\rm dm}/{\rm d}v_{\rm r} {\rm d}v_{\phi}$, of \textbf{m10q} and \textbf{m10v}. We compare the results in CDM and dSIDM with $(\sigma/m)=10 \cpm$ to better illustrate the contrast. The phase space distributions are measured in three radial bins: central, $r<r^{\rm cdm}_{1/2}/3$ ($\sim \operatorname{100\,-\,200}{\rm pc}$); intermediate, $r^{\rm cdm}_{1/2}/3<r<3r^{\rm cdm}_{1/2}$ ($\sim \kpc$) and ``outskirt'', $3r^{\rm cdm}_{1/2}<r<0.5R^{\rm cdm}_{\rm vir}$ ($\gtrsim 10\kpc$). The azimuthal and zenith directions are defined based on the direction of the total angular momentum of dark matter in each radial bin respectively. From inside out, each contour is determined such that it encloses a certain percentile (as labelled on the contour line) of dark matter particles in the bin. We note that, different from the measurement of velocity dispersions, the coherent rotation or inflow/outflow velocity has not been subtracted when determining $v_{\rm r}$ and $v_{\phi}$. Dark matter at small and intermediate radii in the dSIDM model with $(\sigma/m)=10\cpm$ exhibits a median $v_{\phi} \simeq 5-10\kms$ contrary to the almost zero median $v_{\phi}$ in the CDM case. The distribution in the dSIDM model is also more peaky than the CDM case. The differences here is consistent with the coherent rotation of dark matter in dSIDM found above. At the outskirt of the galaxy, the increase in the median of $v_{\phi}$ is still visible but the scatter in the phase space also becomes larger. 

In Figure~\ref{fig:vdist}, we show the velocity ($|v|$) distribution functions of dark matter in the classical dwarfs in CDM and dSIDM with $(\sigma/m)=1$ and $10 \cpm$. We present the results at small ($r<r^{\rm cdm}_{1/2}/3$) and intermediate galactocentric radii ($r^{\rm cdm}_{1/2}/3<r<3r^{\rm cdm}_{1/2}$), respectively. We also show the distribution function in log-log scale to emphasize the low velocity tail. Compared to the CDM case, the velocity distributions in dSIDM models show apparent suppression at the high velocity tail and bumps at lower velocities, due to relatively high interaction rates of particles with high absolute velocities. The low velocity tail is less affected by dissipation due to relatively low interaction rates there. The peak velocity decreases as self-interaction cross-section becomes larger. The phenomenon is actually opposite to the prediction of the ``gravothermal collapse'' in SIDM haloes~\citep[e.g.,][]{Balberg2002,Essig2019}. The difference reflects the deviation of dSIDM haloes from both dynamical and thermal equilibrium in the phase of radial contraction, as well as the fact that one cannot assume velocity distributions as purely isotropic in relaxed dSIDM haloes. Compared with the Maxwell-Boltzmann distribution, the velocity distributions in CDM have extended tails at both the low and high velocity tail, since CDM particles are collisionless and are not locally thermalized. The distributions in the dSIDM models are suppressed the high velocity tail. At small galactocentric radii, the asymptotic behaviour of the velocity distribution function in CDM and dSIDM are quite different from the Maxwell-Boltzmann distribution, decreasing slower towards lower velocities. However, at intermediate radii, both CDM and dSIDM have distributions that resemble the Maxwell-Boltzmann distribution at the low velocity tail.

\begin{figure}
    \centering
    \includegraphics[width=0.49\textwidth]{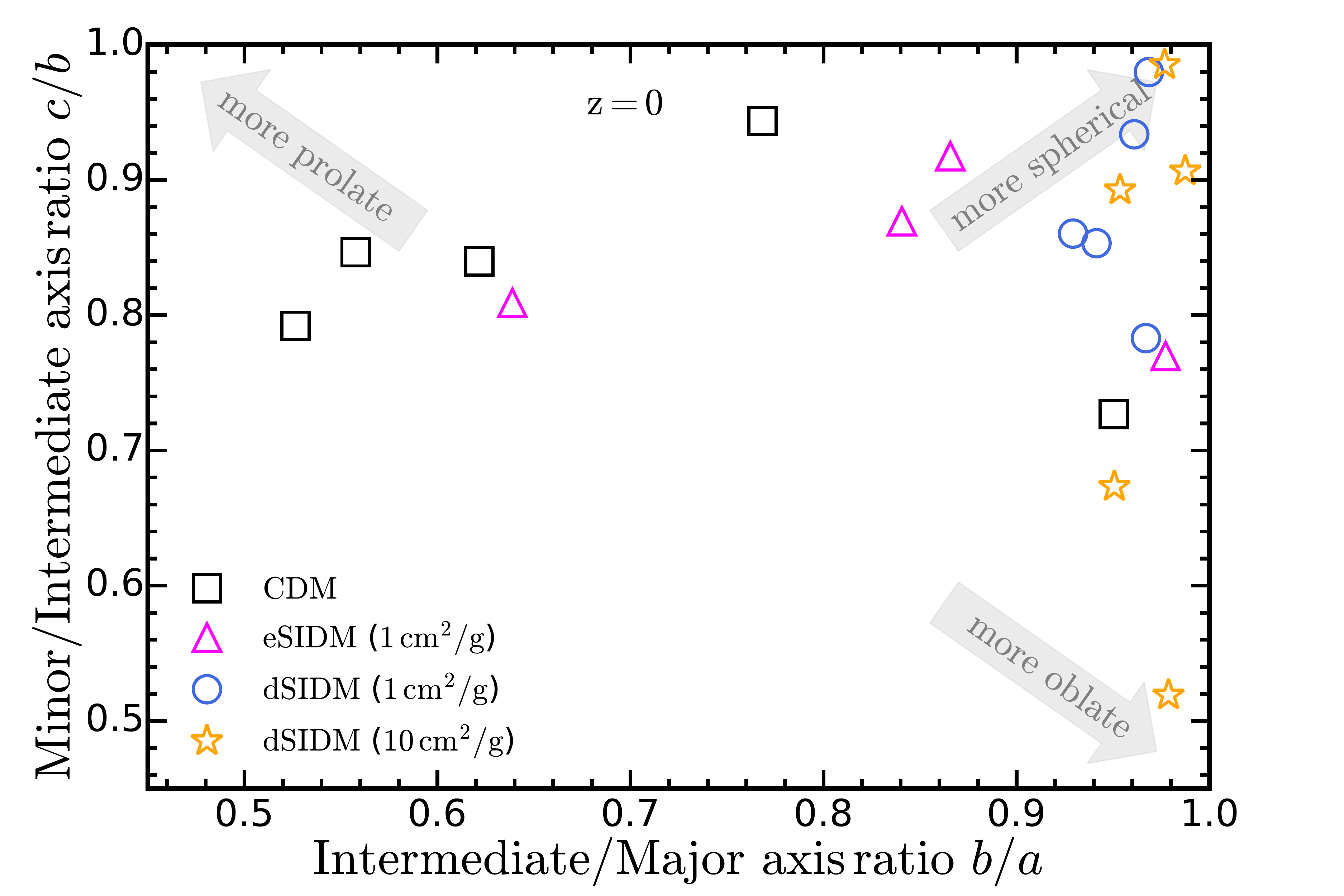}
    \includegraphics[width=0.49\textwidth]{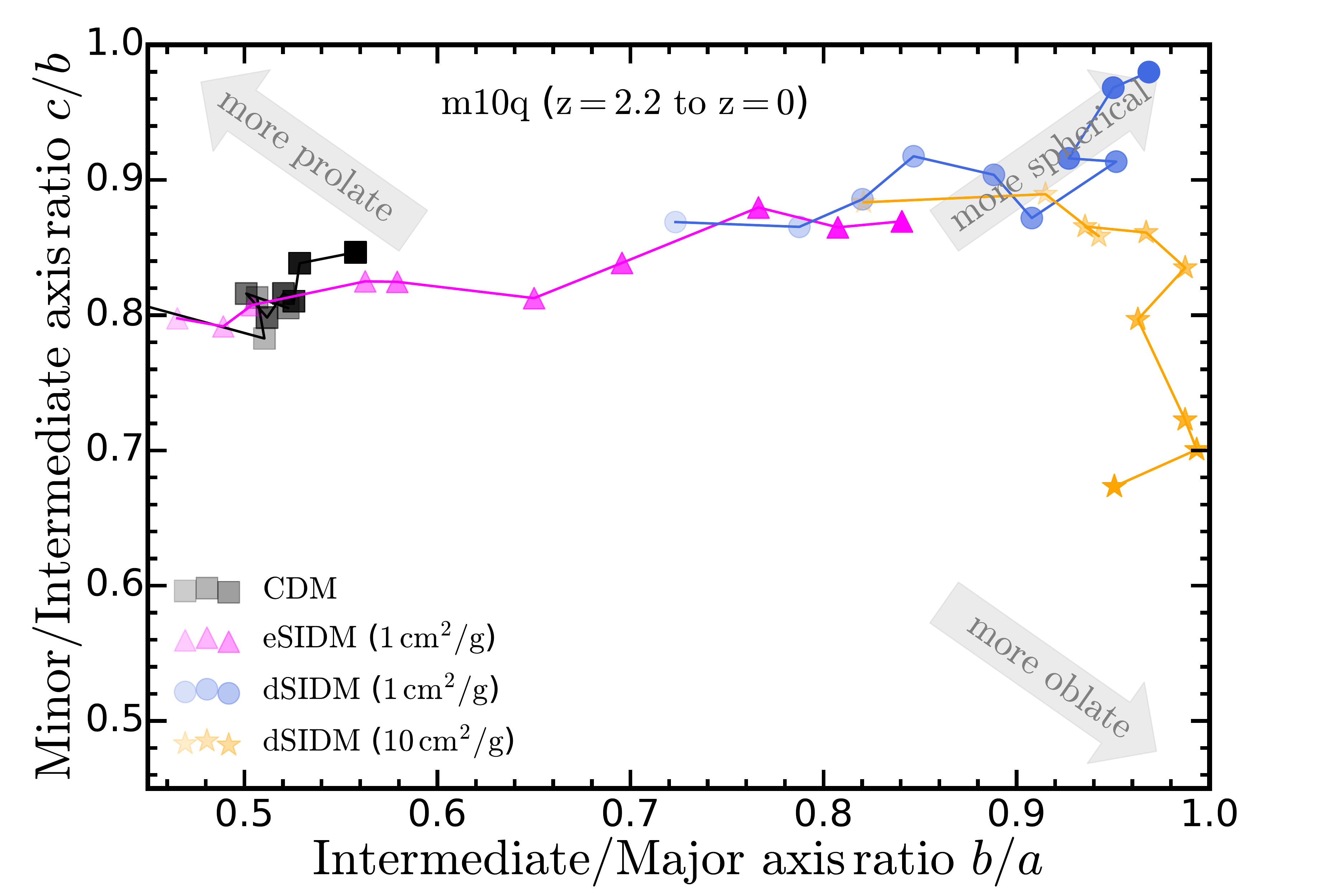}
    \caption{{\it Top}: \textbf{Axis ratios of dark matter haloes at central $\kpc$ in simulations at $z=0$.} We show the minor/intermediate axis ratio ($c/b$) versus the intermediate/major axis ratio ($c/a$) of dark matter mass distribution in different simulations. The axes are measured iteratively while fixing the volume of an ellipsoid as $4\pi/3\, r_{\rm lim}^3$, where $r_{\rm lim}$ is chosen to be $1\kpc$. When $c/b$ ($b/a$) is close to unity, the system is a prolate (oblate) spheroid. When both $c/b$ and $b/a$ are close to unity, the system is spherically symmetric. In CDM, dark matter haloes are triaxial ellipsoids with a clear hierarchy of minor, intermediate and major axes. The CDM haloes lean towards prolate shapes, driven by mild radial dispersion anisotropy. In the dSIDM model with $(\sigma/m)=1$ and $10\cpm$, dark matter haloes behave as oblate spheroids, driven by the coherent rotation of dark matter. In the extreme cases (e.g., \textbf{m10q} in dSIDM with $(\sigma/m)=10\cpm$), $c/b$ drops to as low as $\sim 0.5$ while $b/a$ stays around unity. At larger radii ($r\gg \kpc$), the qualitative trends are similar but the differences between dark matter models become rapidly smaller. {\it Bottom}: \textbf{Evolution of the axis ratios of m10q at central $\kpc$ from $z\simeq 2.2$ to $z=0$.} The markers with darker colors represent measurements at lower redshifts. The CDM halo stays triaxial since $z\simeq 2.2$ while the eSIDM halo becomes more spherical at late times. The halo in dSIDM with $(\sigma/m)=1\cpm$ is already more spherical than CDM and eSIDM counterparts at $z\simeq 2.2$ and it becomes extremely spherical at $z=0$. However, the halo in dSIDM with $(\sigma/m)=10\cpm$ initially follows the track of becoming more spherical but then turns oblate in shape.}
    \label{fig:axis_ratio}
\end{figure}

\section{Halo shape}
\label{sec:shape}

The change in halo shape is another important signature for alternative dark matter physics. This aspect has been explored in detail for the eSIDM case~\citep[e.g.,][]{Zemp2011,Peter2013,Robles2017,Brinckmann2018,Sameie2018}. In dSIDM haloes, morphological changes in response to the energy dissipation are also expected, inline with the steepening of the density profile and the increased rotation support found in previous sections. 

To measure the shape of dark matter haloes, we determine the orientation and magnitude of the principal axes of dark matter distribution by computing the eigenvectors and eigenvalues of the shape tensor of dark matter mass distribution, defined as
\begin{equation}
    \textbf{S}= \dfrac{\int_{V} \rho(\textbf{r})\, \textbf{r}\, \textbf{r}^{\rm T}\, {\rm d}V }{\int_{V} \rho(\textbf{r})\, {\rm d}V},
    \label{eq:shape_tensor}
\end{equation}
where $\rho(\textbf{r})$ is the dark matter mass density at position $\textbf{r}$ with respect to halo center. In terms of discrete dark matter particles, each element of the tensor is calculated as
\begin{equation}
    S_{\rm ij} = \dfrac{\sum_{\rm k} m_{\rm k}\, (r_{\rm k})_{\rm i}\, (r_{\rm k})_{\rm j}  }{\sum_{\rm k} m_{\rm k}}.
\end{equation}
where $m_{\rm k}$ is the mass of the k-th dark matter particle and $(r_{\rm k})_{\rm i}$ is the spatial coordinate of the k-th particle. The three eigenvectors of the shape tensor give the three axes of the mass distribution. Specifically, the major, intermediate and minor axes will be denoted as $\textbf{a}$, $\textbf{b}$ and $\textbf{c}$, respectively. The ratios between the eigenvalues of the shape tensor give the axis ratios of the mass distribution. 

For the simulated dark matter haloes, we perform this measurement in a fixed volume of $V=4\pi r_{\rm lim}^3/3$, where $r_{\rm lim}$ is chosen to be $1\kpc$. The volume is an ellipsoid with its major, intermediate and minor axes ($\textbf{a}$, $\textbf{b}$ and $\textbf{c}$ are set to $r_{\rm lim}$ initially) updated iteratively until convergence is reached. This gives an estimation of the shape of the dark matter halo at $\kpc$ scale. In the top panel of Figure~\ref{fig:axis_ratio}, we show the minor/intermediate axis ratio ($c/b$) versus the intermediate/major axis ratio ($b/a$) of dark matter mass distribution at $z=0$ in simulations. Most of the CDM haloes are triaxial, with a clear hierarchy of minor, intermediate and major axes, and lean towards prolate shapes likely driven by mild radial velocity dispersion anisotropy~\citep[e.g.,][]{Warren1992,Bett2007,Hayashi2007}. The eSIDM haloes overall become more spherical than CDM haloes. Despite some galaxy-to-galaxy variations, it is clear that haloes in the dSIDM models behave as oblate or spherical spheroids, with the intermediate axes always comparable to the major axes. In the model with $(\sigma/m)=1\cpm$, haloes are quite spherical with $b/a \gtrsim 0.9$ and $c/b \gtrsim 0.8$. The radial contraction washes the initial triaxiality of the haloes and the increased central force makes haloes more spherical. However, in the model with $(\sigma/m)=10\cpm$, two of the haloes become oblate in shape, with $c/b$ drops to around $0.5$ and $0.7$, while the other three are still quite spherical in the end. 

In the bottom panel of Figure~\ref{fig:axis_ratio}, we show the evolution of the axis ratios of \textbf{m10q} from $z\simeq 2.2$ to $z=0$ as an example. The halo shape is again measured at central $\kpc$ scale, invariant of redshift. We choose \textbf{m10q} as an example, since it has dramatic changes in its shape in dSIDM models. The markers with darker colors represent measurements at lower redshifts. The CDM halo stays triaxial since $z\simeq 2.2$ with little change in its shape subsequently. The eSIDM halo are initially triaxial but becomes progressively more spherical at late times due to elastic scattering of dark matter. The halo in dSIDM with $(\sigma/m)=1\cpm$ is already more spherical than CDM and eSIDM counterparts at $z\simeq 2.2$ and it becomes extremely spherical ($c/b,\, b/a > 0.95$) at $z=0$. However, the halo in dSIDM with $(\sigma/m)=10\cpm$ initially follows the track of becoming more spherical but then turns oblate in shape. We note that, though not shown explicitly here, the other halo (\textbf{m11b}) which ends up oblate ($c/b \sim 0.5$ at $z=0$) in the model with $(\sigma/m)=10\cpm$ has similar evolutionary track in the axis ratio plane. However, the three haloes (\textbf{m10b}, \textbf{m10v}, \textbf{m11a}) that end up spherical ($c/b,\, b/a \gtrsim 0.9$ at at $z=0$) are still in the phase of turning spherical.

The morphological differences found here are consistent with our findings in the previous sections that coherent rotation develops in dSIDM haloes with $(\sigma/m)=10\cpm$ and could also result in the lower normalization of the density profiles (measured in spherical shells) found in Section~\ref{sec:denpro}. In the model with $(\sigma/m)=10\cpm$, the two haloes that become oblate in shape at $z=0$ (\textbf{m10q} and \textbf{m11b}) are the haloes with the most significant coherent rotation (as presented in Section~\ref{sec:kin}) and also with the most significant decrease in density profile normalization (as presented in Section~\ref{sec:kin}). When the coherent rotation velocity becomes comparable to the velocity dispersion, a self-gravitating spheroidal system consisting of collisionless particles flattens. This is a well-known behaviour in the stellar distribution of elliptical galaxies~\citep[e.g.,][]{Davies1983,Cappellari2007} and models of isotropic oblate rotating spheroids~\citep{Binney1978,Binney1987,Binney2008}. Similar to these previous studies, the response of the ellipticity of the spheroid to $V_{\rm rot}/\sigma_{\rm 3d}$ is weak. In the simulated dwarfs \textbf{m10q} and \textbf{m11b}, significant coherent rotation of $V_{\rm rot}/\sigma_{\rm 3d} \sim 0.5$ results in only modest ellipticity of the halo ($c/b,\, c/a \sim 0.5-0.7$ at $r\lesssim \kpc$). However, the coherent rotation and halo deformation are weaker in other simulated dwarfs and this is likely related to the differences in the mass assembly history of the dwarfs.

We note that, for the oblate spheroids we found here, the minor and major axes are still comparable to each other. The shape is qualitatively different from the thin "dark disk" discussed in the literature (albeit for Milky Way-sized galaxies) regarding dissipative dark matter~\citep{Fan2013,Fan2013b,Fan2014,Randall2015,Foot2013,Foot2015b,Foot2016}. The dissipation time scale in the model studied here is still orders of magnitude longer than the dynamical time scale of the system, which prevents fragmentation of the dark matter into e.g., ``dark stars'' and other compact structures~\citep[e.g.,][]{Hoyle1953,Rees1976,Gammie2001}. This is qualitatively different from baryon-like dissipative dark matter models. In addition, unlike those models that assume dissipative dark matter is a sub-component of all the dark matter, the model studied here assumes that all the dark matter are dissipative. In our case, there would be no external gravitational force that can suppress the growth of secular gravitational instabilities~\citep[e.g.,][]{Ostriker1973,Chris1995}, which prevents the formation of a cold and thin "dark disk" completely supported by rotation.

\section{Discussion}
\label{sec:discussion}

In previous sections, we have presented several signatures of dSIDM models in dwarf galaxies that differ from their CDM counterparts. In this section, we discuss these phenomena in more detail and provide some physical explanations to the behaviours using simple analytical arguments.

\subsection{Slope of the density profile}
\label{sec:discussion-slope}

When $\sigma/m$ becomes large enough such that the dissipation time scale is comparable or lower than the Hubble time scale ($1/H_{0}$), all the dSIDM haloes in simulations first undergo radial contraction, accompanied by the steepening of the central density profiles. It is surprising that, during this phase, the asymptotic power-law slopes of the central density profiles of dwarf galaxies converge to $\sim -1.5$ (though with significant scatter $\sim 0.5$ in the bright dwarfs), insensitive to the detailed value of effective cross-section. 

The cooling and contraction of dSIDM haloes here share some similarities with the cooling and collapse of gas clouds in the baryonic sector, which have been well studied in the context of star formation. However, compared to dSIDM haloes studied here, there are notable differences in the hierarchy of relevant time scales, which result in different evolution patterns. Gas clouds exhibit much higher particle scattering rates and less energy dissipation per scattering, so the collisional relaxation time scale is orders of magnitude shorter than the cooling time scale, which means that global thermal equilibrium is easier to be established in gas clouds. During the early contraction of gas clouds, it is often assumed that the compressional heating will offset the radiative loss of thermal energy and keep the cloud nearly isothermal~\citep[e.g.,][]{Gaustad1963,Shu1977}. However, in dSIDM haloes, since the dissipation time scale is comparable to the collision time scale (see Section~\ref{sec:timescale}), the dSIDM fluid cannot adjust itself to global thermal equilibrium during the contraction of the system, which is qualitatively different from the isothermal contraction of gas clouds. This is supported by the fact that the velocity dispersion profiles (shown in Figure~\ref{fig:gallery}) at the centers of simulated dwarfs in are never flat in dSIDM models, contrary to the isothermal profiles in eSIDM cases.

For gas clouds, the isothermal contraction will gradually increase the imbalance of gravitational forces over thermal pressure forces, which eventually results in the free-fall collapse of the central part of the cloud~\citep[e.g.,][]{Bodenheimer1968,Penston1969b,Penston1969a,Larson1969,Shu1977,Hunter1977,Foster1993}. In terms of time scales, the free-fall collapse will happen when the cooling time scale becomes shorter than the dynamical time scale of the cloud. However, in dSIDM haloes, this is also prohibited, since the dissipation time scale (in the surveyed parameter space) is orders of magnitude larger than the dynamical time scale of the system. As the dissipation of thermal/kinetic energy drives the contraction of the halo on the dynamical time scale, dark matter particles could be gravitationally accelerated again, which would effectively increase the thermal pressure and slow down the collapse. Moreover, on the dynamical time scale, dark matter particles from different radii can ``mix'' because they are only weakly collisional as oppose to gas. As a consequence, even though the global thermal equilibrium of the system is broken, the contraction would still be much slower than the free-fall collapse of gas clouds (as found in Figure~\ref{fig:phasespace}).

We find the behaviour of our systems can be reasonably described by the solution for a ``slow'' quasi-equilibrium cooling flow (with negligible thermal conduction) rather than isothermal or rapid free-fall ``collapse''. Following \citet{Stern2019}, the continuity equation of a steady slow-cooling halo, that is spherically symmetric, isotropic and pressure supported, can be written as
\begin{equation}
     \dfrac{{\rm d}\ln{\rho}}{{\rm d}\ln{r}} + \dfrac{{\rm d}\ln{v_{\rm r}}}{{\rm d}\ln{r}} = -2,
     \label{eq:mass_cons}
\end{equation}
where $\rho$ is the density of the fluid and $v_{\rm r}$ is the radial inflow velocity. The momentum equation and the entropy equation of the system can be reduced to~\citep{Stern2019}
\begin{equation}
    \dfrac{{\rm d}\ln{v_{\rm r}}}{{\rm d}\ln{r}} \Big(\dfrac{v_{\rm r}^2}{c_{\rm s}^2} -  1 \Big) = 2 - \dfrac{v_{\rm c}^2}{c_{\rm s}^2} - \dfrac{r/v_{\rm r}}{\gamma t_{\rm cool}},
\end{equation}
where $v_{\rm c}$ is the circular velocity, $c_{\rm s}$ is the adiabatic sound speed, $\gamma$ is the adiabatic index and $t_{\rm cool}$ is the cooling time scale of the fluid. Applying the solution to the cooling flow of dark matter, we replace the sound speed $c_{\rm s}$ with the one-dimensional velocity dispersion of dark matter $\sigma_{\rm 1d}$ and the cooling time scale $t_{\rm cool}$ with the dissipation time scale $t_{\rm diss}$ of dark matter self-interactions. In the "subsonic" limit ($v_{\rm r} \ll \sigma_{\rm 1d} $), the second equation becomes
\begin{equation}
    - \dfrac{{\rm d}\ln{v_{\rm r}}}{{\rm d}\ln{r}} = 2 - \dfrac{v_{\rm c}^2}{\sigma_{\rm 1d}^2} - \dfrac{r/v_{\rm r}}{\gamma t_{\rm diss}}.
\end{equation}
A simple self-similar solution exists by requiring that all the logarithmic derivatives of dark matter properties are constants. Then $v_{\rm c}^2/\sigma_{\rm 1d}^2$ and $(r/v_{\rm r})/t_{\rm diss}$ also need to be constants. If we assume $\rho \sim r^{\alpha}$, we obtain the scaling of the one-dimensional velocity dispersion as
\begin{equation}
    \sigma_{\rm 1d} \sim v_{\rm c} \sim \sqrt{GM_{\rm enc}(r)/r} \sim r^{1+\alpha/2}.
\end{equation}
In the meantime, Equation~\ref{eq:mass_cons} implies that $v_{\rm r} \sim r^{-\alpha-2}$. According to Equation~\ref{eq:tdiss}, the dissipation time scale $t_{\rm diss}$ scales with density and velocity dispersion as $\rho^{-1} \sigma_{\rm 1d}^{-1} \sim r^{-(1+3\alpha/2)}$. If we plug in the scaling of $v_{\rm r}$ and $t_{\rm diss}$ to the term $(r/v_{\rm r})/t_{\rm diss}$, we obtain
\begin{equation}
        \dfrac{r/v_{\rm r}}{t_{\rm diss}} \sim \dfrac{ r\, r^{\alpha+2}}{r^{-(1+3\alpha/2)}} \sim r^{4+5\alpha/2}.
\end{equation}
So the power-law solution (which requires the term to be a constant at all radii) has $\alpha = -8/5$. Quantitatively, the slope of the density profile given by this ``dark cooling flow'' solution is consistent with the finding in dSIDM simulations that the asymptotic slopes of the density profiles converge to around $-1.5$. It also predicts $\sigma_{\rm 1d} \sim r^{0.2}$, which is consistent with the central velocity dispersions of simulated dwarfs that mildly increase with radii. 

A similar solution for self-gravitating gaseous spheres with a polytropic equation of state has been presented in \citet{Suto1988}, as a generic study of the solution proposed in \citet{Shu1977}. They considered spherical gaseous systems with the same equations for mass and momentum conservation. Purely aiming at finding self-similar solutions and without involving a detailed description of cooling/heating, they derived an asymptotic self-similar density profile $\rho \sim r^{-1.5}$ that is independent of the assumed polytropic index of gas. The solution is not restricted to a steady-state, subsonic inflow of gas and still holds even when there is no cooling term.

\subsection{Dark matter energy transfer in dSIDM}
\label{sec:discussion-heat}

In general, ``thermal conduction'' and dissipation are the two main mechanisms in SIDM haloes to transfer kinetic energy of dark matter. ``Thermal conduction'' is dark matter collisional energy transfer. The detailed form of the heat conductivity depends on the nature of the heat conduction. In the theory of thermal conductivity of an ideal fluid, the heat flux is the averaged one-way flux of particles across an imaginary surface multiplied by the difference in energy per particle between the starting and ending points. Up to order unity corrections, this gives
\begin{equation}
    \kappa \simeq \dfrac{3}{2} \dfrac{k_{\rm B}}{m} \rho \dfrac{l^2}{\tau}
\end{equation}
where $k_{\rm B}$ is the Boltzmann constant, $l$ is the characteristic distance between the starting and ending points and $\tau$ is the time between collisions. In SIDM haloes, the collision (or close encounters) between particles is governed by dark matter self-interactions since the collision time scale of dark matter self-interaction is significantly lower than the two-body gravitational relaxation time scale. Thus, we have $\tau = t_{\rm coll}$. If the mean free path between collisions is significantly shorter than the physical size of the system~(referred to as the Short Mean Free Path (SMFP) regime), dark matter will behave like a fluid and the heat conductivity is fully regulated by the mean free path of dark matter particles ($l = \lambda = 1/(\rho \sigma/m)$). Therefore, in this regime, the thermal conductivity is:
\begin{equation}
     \kappa = \dfrac{3}{2} \dfrac{k_{\rm B}}{m} C_{1} \rho \dfrac{\lambda^2}{t_{\rm coll}},
\end{equation}
where $C_{1}$ is an order unity constant and has been found to be $(25\sqrt{\pi}/32)/(4/\sqrt{\pi})$ in the Chapman-Enskog theory~\citep[e.g.,][]{Chapman1970, Lifshitz1981} and $0.25/(4/\sqrt{\pi})$ in numerical simulations~\citep{Koda2011}. 

\begin{figure}
    \centering
    \includegraphics[width=0.49\textwidth]{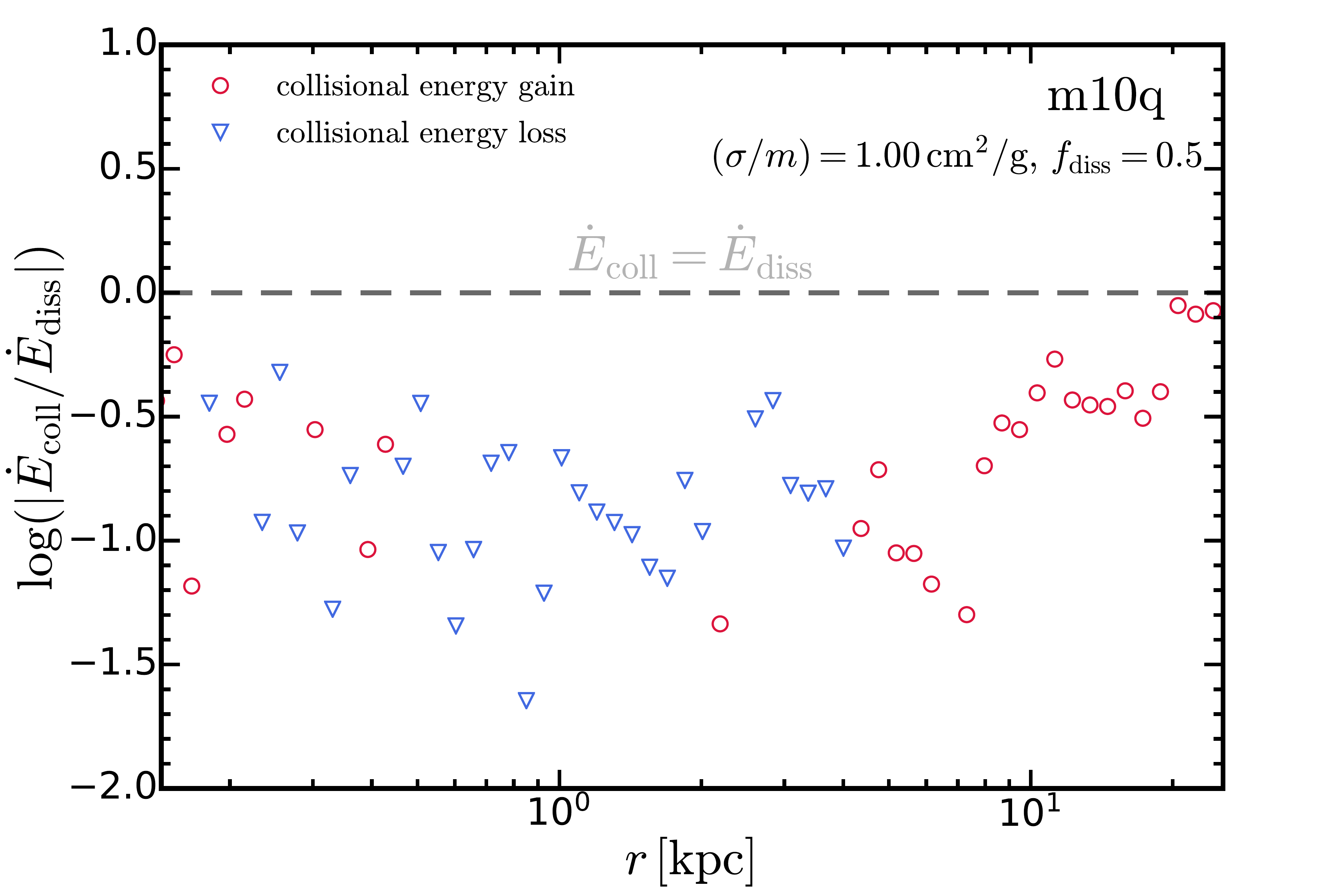}
    \includegraphics[width=0.49\textwidth]{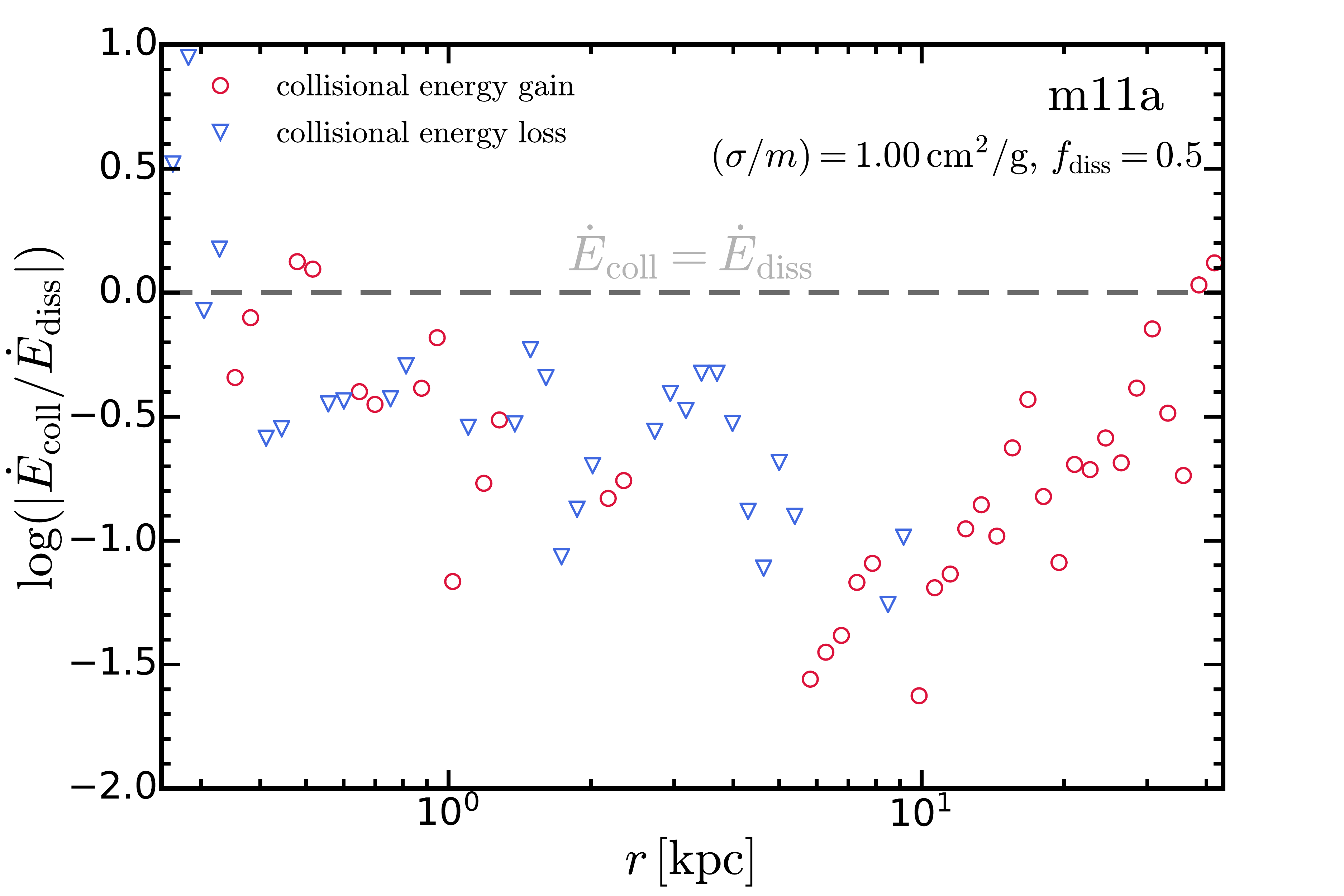}
    \caption{\textbf{Dark matter energy transfer rates via ``thermal conduction'' (dark matter collisional energy transfer) versus dissipation energy loss rates, measured in spherical shells, as a function of galactocentric radii.} We show the heat gain or loss of dark matter via collisions ($\dot{E}_{\rm coll}$, Equation~\ref{eq:conduction}) versus the energy dissipation rate ($\dot{E}_{\rm diss}$, Equation~\ref{eq:dissipation}) in circles (red for $\dot{E}_{\rm coll}>0$, blue for $\dot{E}_{\rm coll}<0$). We present the results in one of the classical dwarfs \textbf{m10q} and in one of the bright dwarfs \textbf{m11a}. In both galaxies, with $f_{\rm diss}=0.5$, the collisional energy transfer rate is always roughly an order of magnitude lower than the energy dissipation rate. }
    \label{fig:heat_flow}
\end{figure}

On the other hand, this picture is not valid when the mean free path between collisions is much larger than the gravitational scale height $H$ of the system~(referred to as the Long Mean Free Path (LMFP) regime), defined as:
\begin{equation}
    H = \sqrt{\dfrac{\sigma_{\rm 1d}^2}{4 \pi G \rho}}.
\end{equation}
In this regime, particles can travel several orbits before experiencing a collision. \citet{Bell1980} found that the characteristic distance between encounters in this limit (for weakly collisional fluid) can be roughly described by the gravitational scale height ($l=H$). In this case, the thermal conductivity is:
\begin{equation}
     \kappa = \dfrac{3}{2} \dfrac{k_{\rm B}}{m} C_{2} \rho \dfrac{H^2}{t_{\rm coll}},
     \label{eq:kappa_lmfp}
\end{equation}
where $C_{2}$ is an order unity constant and has been found to be $0.75$ in numerical simulations~\citep{Koda2011}. For the fiducial model studied in the paper, the mean free path of dark matter self-interaction is always orders of magnitudes larger than the gravitational scale height of the systems (or translated to time scale, the collision time scale of dark matter self-interaction is orders of magnitudes larger than the dynamical time scale of the system). So, these haloes all stay in the LMFP regime.

The flux of thermal energy transferred outward through a sphere of radius $r$ can be calculated as:
\begin{equation}
    j_{\rm coll}(r) = - \kappa \dfrac{\partial T(r)}{\partial r} = - \kappa \dfrac{m}{k_{\rm B}} \dfrac{\partial \sigma^{2}_{\rm 1d}(r)}{\partial r},
\end{equation}
where $\kappa$ takes the conductivity in the LMFP regime defined in Equation~\ref{eq:kappa_lmfp}. The net collisional energy gain per unit volume in a spherical shell can be calculated as:
\begin{equation}
    \dot{E}_{\rm coll}(r) = - \dfrac{1}{4\pi r^2} \dfrac{\partial (4\pi r^{2} j_{\rm coll}(r) )}{\partial r}.
    \label{eq:conduction}
\end{equation}

The second mechanism of energy transfer is energy dissipation due to dark matter self-interactions. Different from ``thermal conduction'', the dissipation we modelled here is not regulated by any characteristic length scale, since the dissipated energy will not be reabsorbed and effectively has an infinite mean free path. The dissipation energy loss per unit volume in a spherical shell is the volumetric cooling rate:
\begin{equation}
    \dot{E}_{\rm diss}(r) = C(r) =\dfrac{3}{2}\rho(r) \sigma_{\rm 1d}^2(r) / t_{\rm diss}(r).
    \label{eq:dissipation}
\end{equation}
The relative importance of collisional energy transfer and dissipation is determined by the comparison between $t_{\rm coll}$ and $t_{\rm diss}$. For the dSIDM model studied in this paper, $t_{\rm coll}$ and $t_{\rm diss}$ always have similar dependence on density and velocity dispersion. Thus, their ratio is almost a constant over the evolution of the halo and only depends on $f_{\rm diss}$. For the fiducial model with $f_{\rm diss}=0.5$, $t_{\rm diss}$ is of the same order of magnitude as $t_{\rm coll}$ (e.g., $t_{\rm diss} = 0.75\, t_{\rm coll}/f_{\rm diss}$ for the models with constant cross-sections). In this regime, dissipation is always the dominant mechanism for energy transfer and is responsible for triggering the contraction of the halo. Collisional energy transfer is negligible. Therefore, the evolution pattern of dSIDM haloes in this regime will be qualitatively different from the canonical gravothermal collapse of eSIDM haloes.

In Figure~\ref{fig:heat_flow}, we demonstrate the dominance of dissipation over collisional energy transfer in simulations. We show the collisional energy transfer rate, $\dot{E}_{\rm coll}$, relative to the energy loss rate due to dissipation, $\dot{E}_{\rm diss}$, of spherical shells as a function of galactocentric radii. In the classical and bright dwarfs, assuming the fiducial choice of $f_{\rm diss}$, the rate of energy transfer via collisions is always roughly an order of magnitude lower than the energy dissipation rate. 

\begin{figure*}
    \centering
    \includegraphics[width=0.49\textwidth]{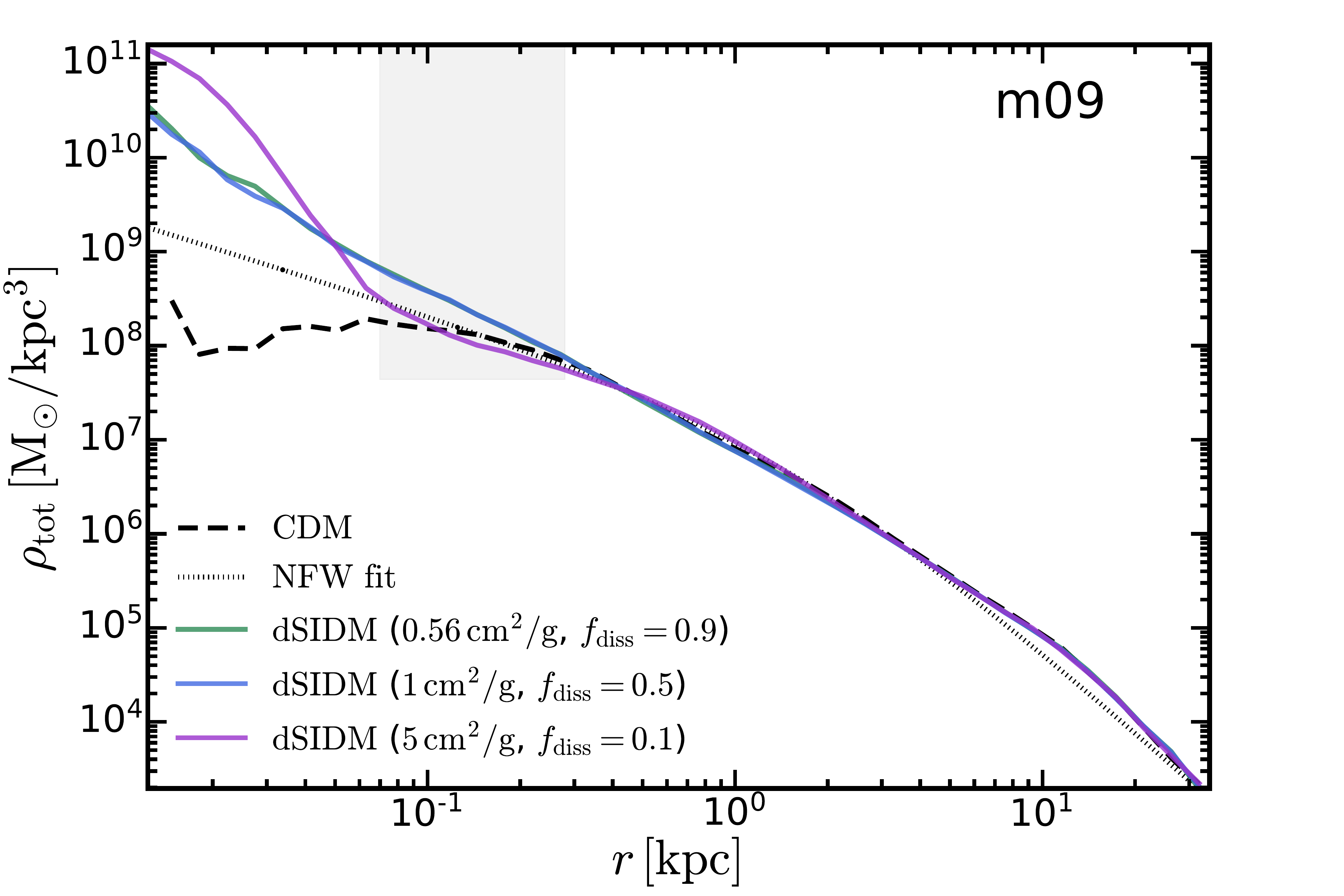}
    \includegraphics[width=0.49\textwidth]{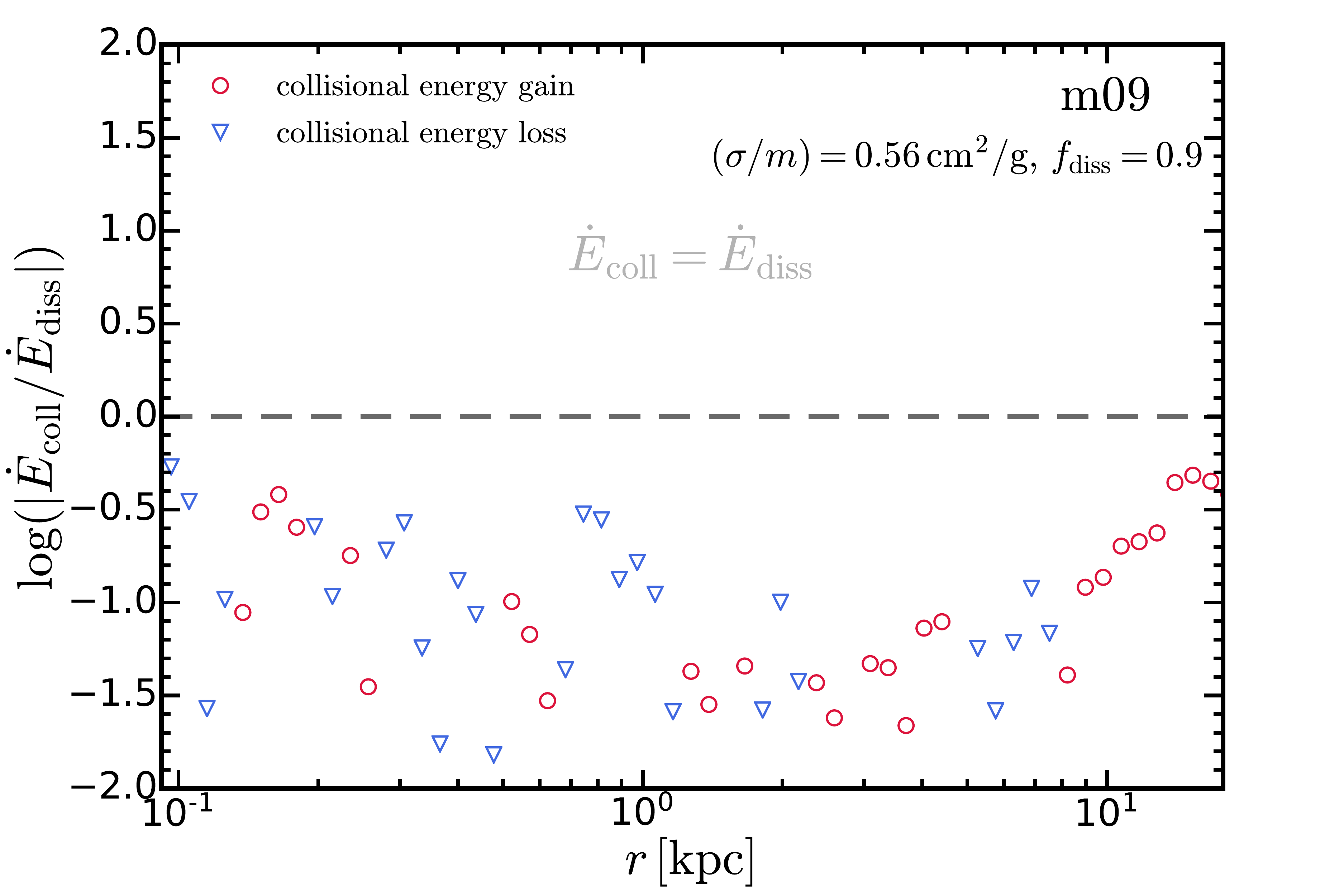}
    \includegraphics[width=0.49\textwidth]{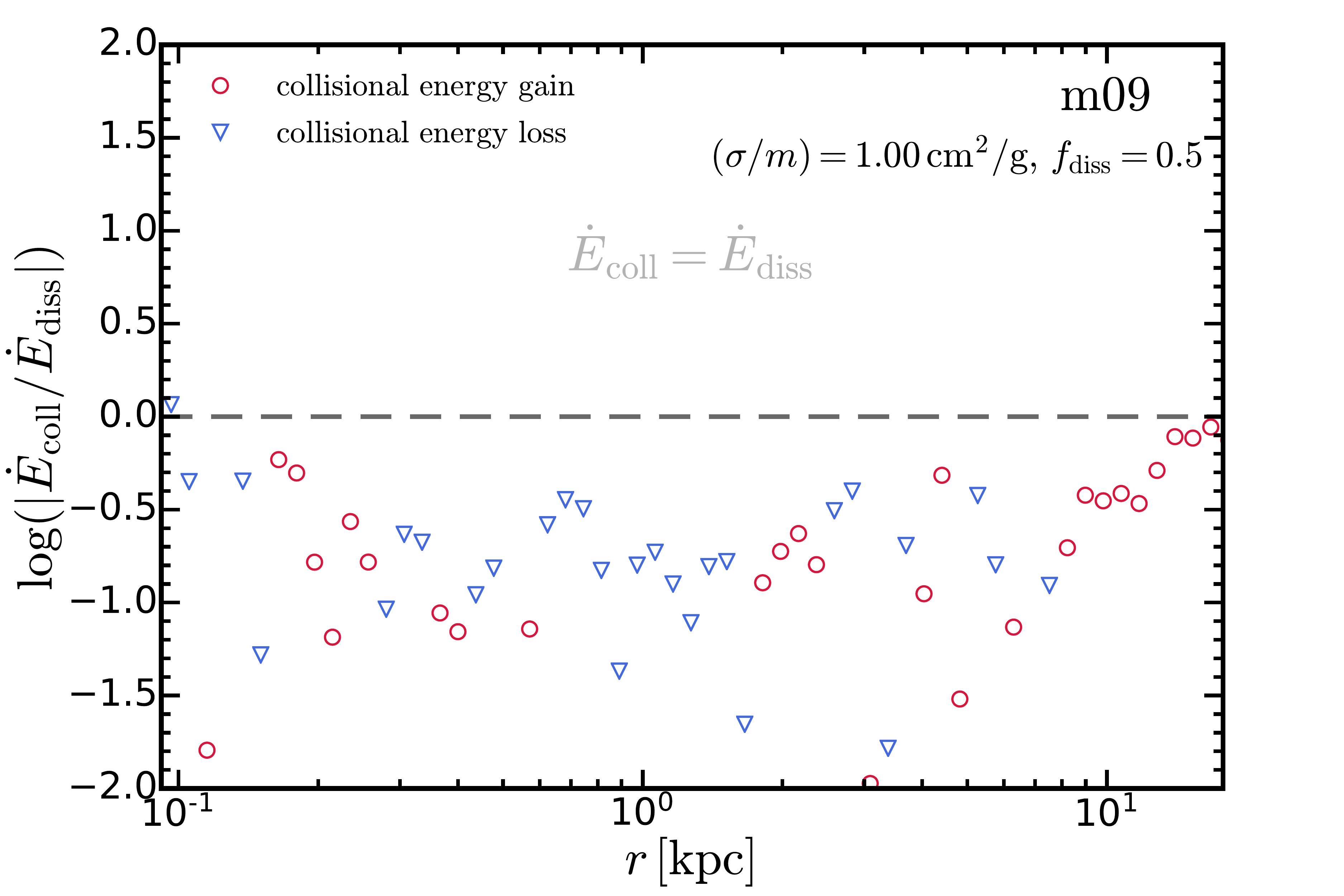}
    \includegraphics[width=0.49\textwidth]{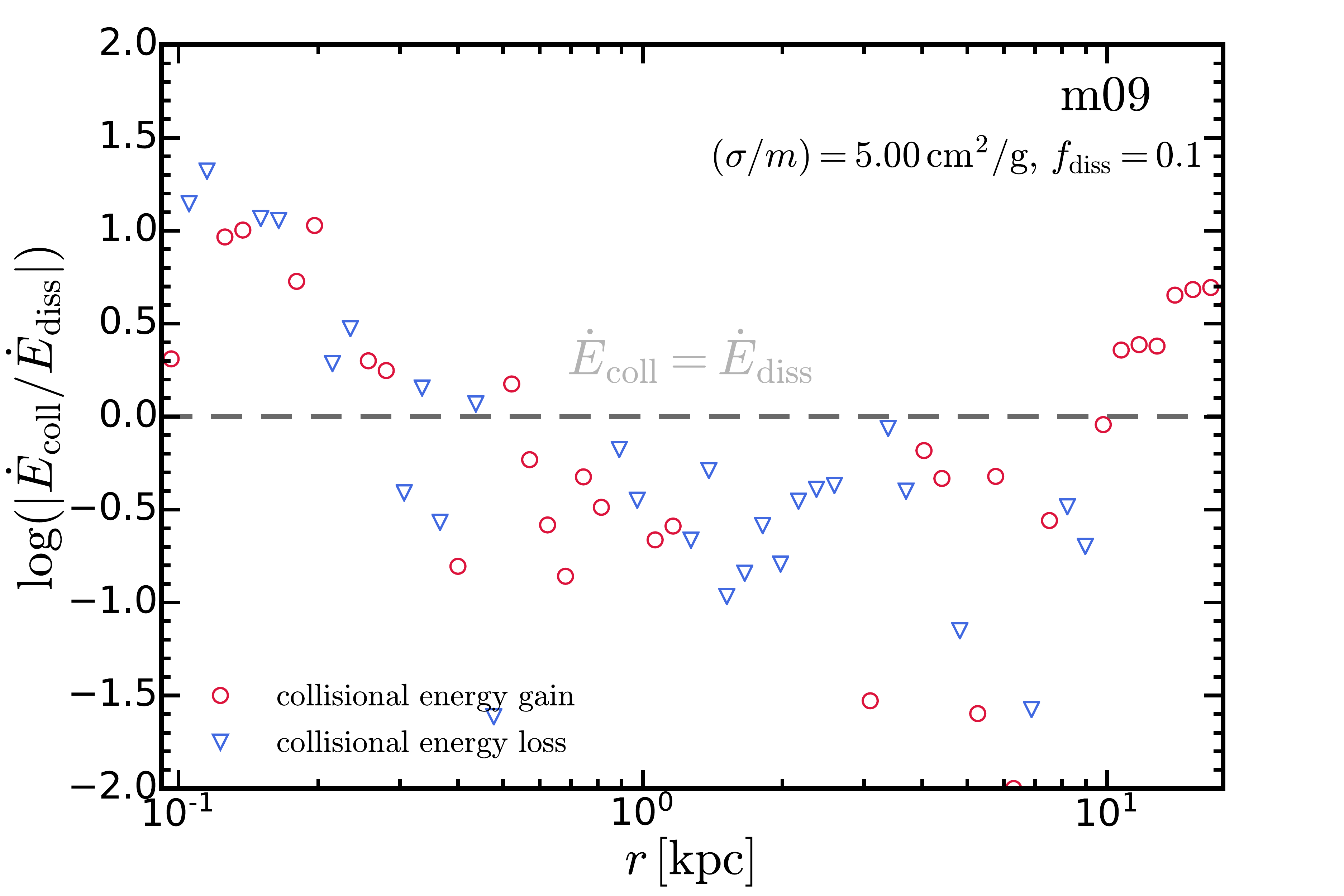}
    \caption{{\it Top left:} \textbf{Total mass density profiles of \textbf{m09} in dSIDM models with other combinations of $f_{\rm diss}$ and $\sigma/m$.} We choose three combinations of $f_{\rm diss}$ and $\sigma/m$ that give the same dissipation time scale: $f_{\rm diss}=0.5, \, \sigma/m = 1 \cpm$; $f_{\rm diss}=0.1, \, \sigma/m = 5 \cpm$; $f_{\rm diss}=0.9, \, \sigma/m = 0.56 \cpm$. {\it Other panels:} \textbf{Collisional energy transfer rates versus energy dissipation rate of dark matter (as Figure~\ref{fig:heat_flow})}. The energy transfer rate via collisions is subdominant compare to dissipation in the model with $f_{\rm diss}=0.5$ or $0.9$. In the model with $f_{\rm diss}=0.1$, collisional heating overtakes dissipation at the center of the galaxy. This model actually produces denser and cuspier central density profile, as the halo experiences the gravothermal collapse and a dense core in the SMFP regime emerges at the center. In all models, at large radii ($\sim 10\kpc$), collisional energy transfer rates become comparable to the dissipation rate, but the absolute value of both terms at these radii are too small to make a difference. 
    }
    \label{fig:m09}
\end{figure*}

\subsection{Evolution of a dSIDM halo}

When dissipation dominates over collisional energy transfer of dark matter, the evolution track of an isolated dSIDM halo can be divided into four regimes, depending on the dissipation time scale $t_{\rm diss}$
\begin{itemize}
    \item Regime A ($t_{\rm diss}\gg t_{\rm H}$): The halo evolves in the same way as analogous CDM halo since both $t_{\rm diss}$ and $t_{\rm coll}$ are significantly longer than the lifetime of the system.

    \item Regime B ($ t_{\rm H} \gtrsim t_{\rm diss} \gtrsim 0.1\, t_{\rm H}$): The halo undergoes radial contraction. The density profile within the radius where $t_{\rm H} \gtrsim t_{\rm diss}$ steepens and becomes cuspy with power-law slopes asymptoting to $\sim -1.5$. The shape of the halo becomes more spherical in this phase.
    
    \item Regime C ($ 0.1\, t_{\rm H} \gtrsim t_{\rm diss} \gg t_{\rm dyn}$ at the halo center): At a certain stage of the radial contraction, prominent coherent rotation of dark matter will develop in the system. The system is in a transition from purely dispersion supported to being supported by a mixture of random velocity dispersion and coherent rotation. During this transition, the radial contraction of the halo and the steepening of the density profile are stopped by centrifugal forces. The halo becomes oblate in shape during this phase and the normalization of the density profile measured in spherical shells decreases. 

    \item Regime D ($t_{\rm dyn} \gtrsim t_{\rm diss}$): Local instability starts to build up and results in fragmentation of the halo. Numbers of dark "clumps" would start to form within the local free-fall time scale. None of our simulations has reached this regime and it would require order-of-magnitude larger self-interaction cross-sections to test.
    
\end{itemize}

\section{Comparison with other simulation physics}
\label{sec:vary}

\subsection{Varying the energy dissipation fraction }
\label{sec:m09}

We note that the specific simulations studied in this paper have assumed that the dimensionless degree of dissipation is $f_{\rm diss}=0.5$. However, the results can be extrapolated to other slices of the dSIDM parameter space based a simple time scale argument. In Section~\ref{sec:timescale}, we show that the energy dissipation time scale only depends on the product of $f_{\rm diss}$ and $\sigma/m$. Therefore, when dissipation is the dominant mechanism for energy transfer, different combinations of $f_{\rm diss}$ and $\sigma/m$ should give rise to similar predictions as long as the dissipation time scale is the same. In this section, we vary the dissipation fraction $f_{\rm diss}$ and test how the results are affected in explicit simulations.

We use the ultra faint dwarf \textbf{m09} as the test halo. The halo is ideal for the test since the density profile is dark matter dominated and baryonic feedback is weak considering its $M_{\ast}/M_{\rm halo}\lesssim 3\times 10^{-5}$. We choose three combinations of $f_{\rm diss}$ and $\sigma/m$ that give the same dissipation time scale: $f_{\rm diss}=0.5, \, \sigma/m = 1 \cpm$; $f_{\rm diss}=0.1, \, \sigma/m = 5 \cpm$; $f_{\rm diss}=0.9, \, \sigma/m = 0.56 \cpm$. In Figure~\ref{fig:m09}, we show the total mass density profile of \textbf{m09} in these three models compared with the CDM counterpart and the NFW profile. The models with $f_{\rm diss}=0.5$ and $f_{\rm diss}=0.9$ produce exactly the same density profile, which justifies that, when dissipation dominates energy transfer, the evolution of the halo is determined by the dissipation time scale and is independent of the detailed combination of parameters. However, we find the model with $f_{\rm diss}=0.1$ (and a large cross-section of $\sigma/m = 5 \cpm$) produces a qualitatively different profile from the other two models. The density follows the NFW profile at $\gtrsim 100\pc$ while gets enhanced by about two orders of magnitude at the scale $\lesssim 100\pc$ compared to the extrapolation of the NFW profile, and is even denser than the cuspy profile in the other two models. It is counterintuitive that the model with a lower degree of dissipation gives rise to higher central densities. The phenomenon can be explained by the increased importance of collisional energy transfer in this model. When $f_{\rm diss}=0.1$, the collision time scale becomes an order of magnitude lower than the dissipation time scale and the halo is no longer purely dominated by dissipation. Under the influence of collisional energy transfer, the evolution track of the halo resembles the ``gravothermal catastrophe'' of eSIDM haloes, where ``thermal conduction'' is responsible for energy transfer. The analytical model of the ``gravothermal catastrophe'' of SIDM haloes~\citep[e.g.,][]{Balberg2002} predicts that a halo initially in the LMFP regime will contract while maintaining a cored, self-similar density profile until the central part of the halo reaches the SMFP regime. Subsequently, a dense, optical thick core (in the SMFP regime) will form while the outskirt of the halo stays in the LMFP regime. In the simulation with $f_{\rm diss}=0.1$, at the center of \textbf{m09}, the density reaches $10^{11}\,\msun/ \kpc^{3}$ and the collision time scale there is comparable to the dynamical time scale (assuming a typical one-dimensional velocity dispersion $\sim 10\kms$) which indicates that the center of the halo is indeed in the SMFP regime. It is striking that the enhanced central density due to the gravothermal evolution is even higher than that produced by models with higher degree of dissipation.

We verify that the phenomenon discussed above is indeed caused by increased importance of ``thermal conduction'' by showing the collisional energy transfer rates versus dissipation rates in simulations in Figure~\ref{fig:m09}. In the model with $f_{\rm diss}=0.9, \, \sigma/m = 0.56 \cpm$ or $f_{\rm diss}=0.5, \, \sigma/m = 1 \cpm$, the collisional energy transfer rate is always subdominant compared to dissipation. However, in the model with $f_{\rm diss}=0.1, \, \sigma/m = 5 \cpm$, the collisional energy transfer rate overtakes dissipation at small radii ($\lesssim 0.2\kpc$). This is in very good agreement with the radii where we find the differences in density profiles between the two models. In summary, when $|\dot{E}_{\rm coll}| \gg |\dot{E}_{\rm diss}|$ at halo centers, which occurs for $f_{\rm diss}\lesssim 0.1$, the halo behaves more like an eSIDM halo and the higher central density is primarily due to the gravothermal evolution driven by collisional energy transfer (but potentially accelerated by dissipation).

\begin{figure}
    \centering
    \includegraphics[trim=1.2cm 0.1cm 2cm 1cm, clip, width=0.49\textwidth]{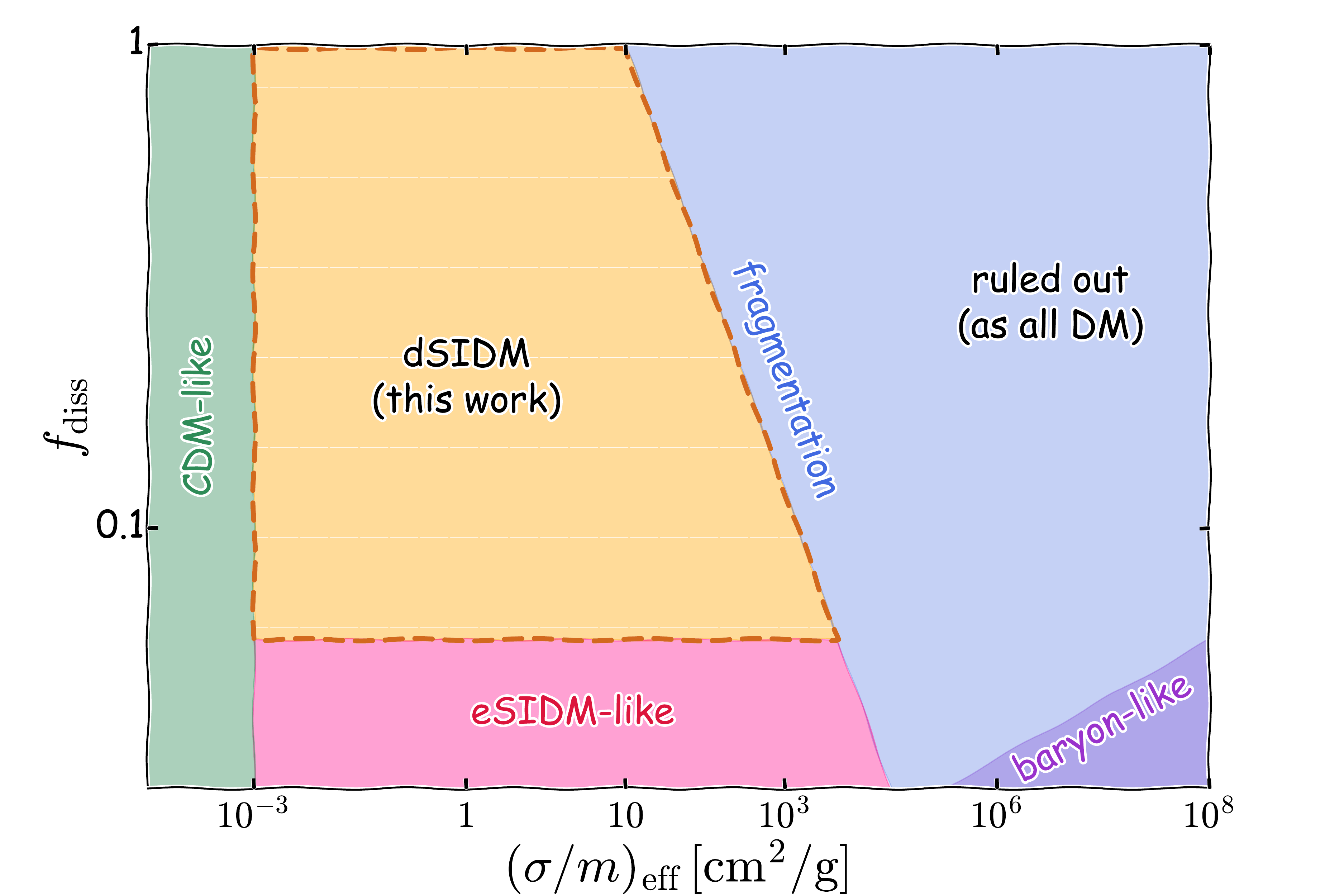}
    \caption{\textbf{A cartoon of the dSIDM parameter space.} The dSIDM model is parameterized with $\sigma/m$ and $f_{\rm diss}$. When $\sigma/m$ is small enough, both elastic and dissipative SIDM models become analogous to CDM in the lifetime of the Universe. When $f_{\rm diss}$ becomes small enough, dSIDM becomes essentially eSIDM-like since collisional energy transfer dominates over dissipation in this regime. When the product of $\sigma/m$ and $f_{\rm diss}$ becomes large enough, the dissipation time scale could drop below the local dynamical time scale of the system and results in fragmentation of dSIDM into compact dark objects. Effectively, baryon-like models are located at the low $f_{\rm diss}$, high $\sigma/m$ corner of the plot. The dSIDM models studied in this paper live in the parameter space, which is not immediately ruled out but can still give rise to unique phenomena different from CDM or eSIDM models. 
    }
    \label{fig:cartoon}
\end{figure}

 To better illustrate the parameter space of dSIDM (including the space that haven't been explored in this paper), we create a cartoon image (Figure~\ref{fig:cartoon}) which qualitatively divides the dSIDM parameter space into several regions. The dSIDM models are parameterized with $\sigma/m$ and $f_{\rm diss}$. Both eSIDM and dSIDM models become CDM-like when $\sigma/m$ is small enough such that the collision time scale becomes much longer than the lifetime of the Universe. dSIDM becomes essentially eSIDM-like when $f_{\rm diss}$ becomes small enough, since collisional energy transfer dominates over dissipation in this regime. When the product of $\sigma/m$ and $f_{\rm diss}$ becomes large enough, the dissipation time scale could drop below the local dynamical time scale of the system and results in fragmentation of dSIDM into compact dark objects. For higher value of $f_{\rm diss}$ and $\sigma/m$, the scenario that all dark is dissipative would be ruled out by observations (e.g., constraints from merger clusters \citep{Markevitch2004,Randall2008}; lensing constraints on compact dark matter substructures). If we put baryons (and baryon-copy dSIDM models) in this space effectively, they will be located at the low $f_{\rm diss}$, high $\sigma/m$ corner of the plot. Thus the interesting dSIDM parameter space that gives unique phenomena but is not immediately ruled out is roughly around $f_{\rm diss} \simeq 0.1 - 1$, $(\sigma/m)\simeq 0.01-100 \cpm$.

\begin{figure}
    \centering
    \includegraphics[width=0.49\textwidth]{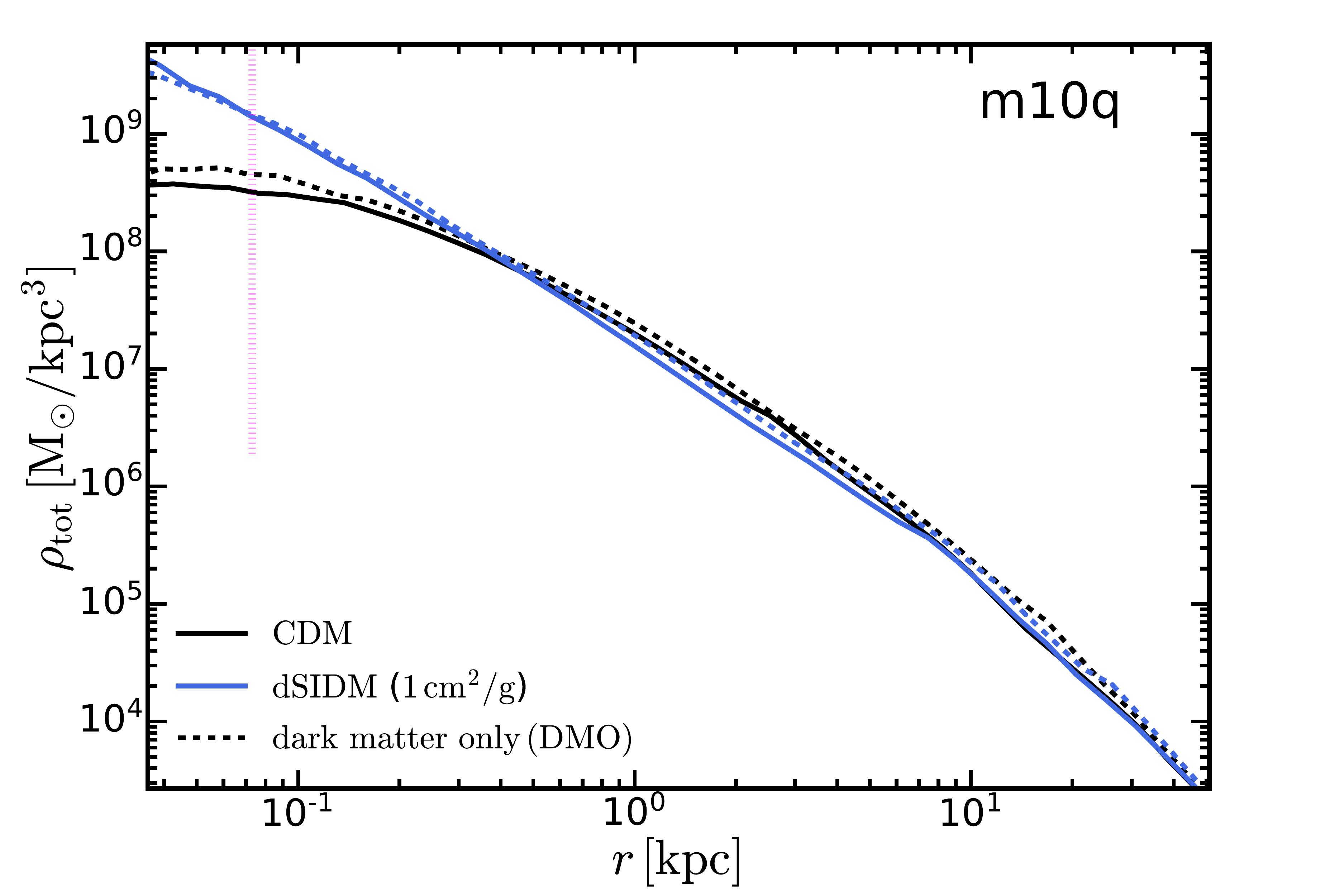}
    \includegraphics[width=0.49\textwidth]{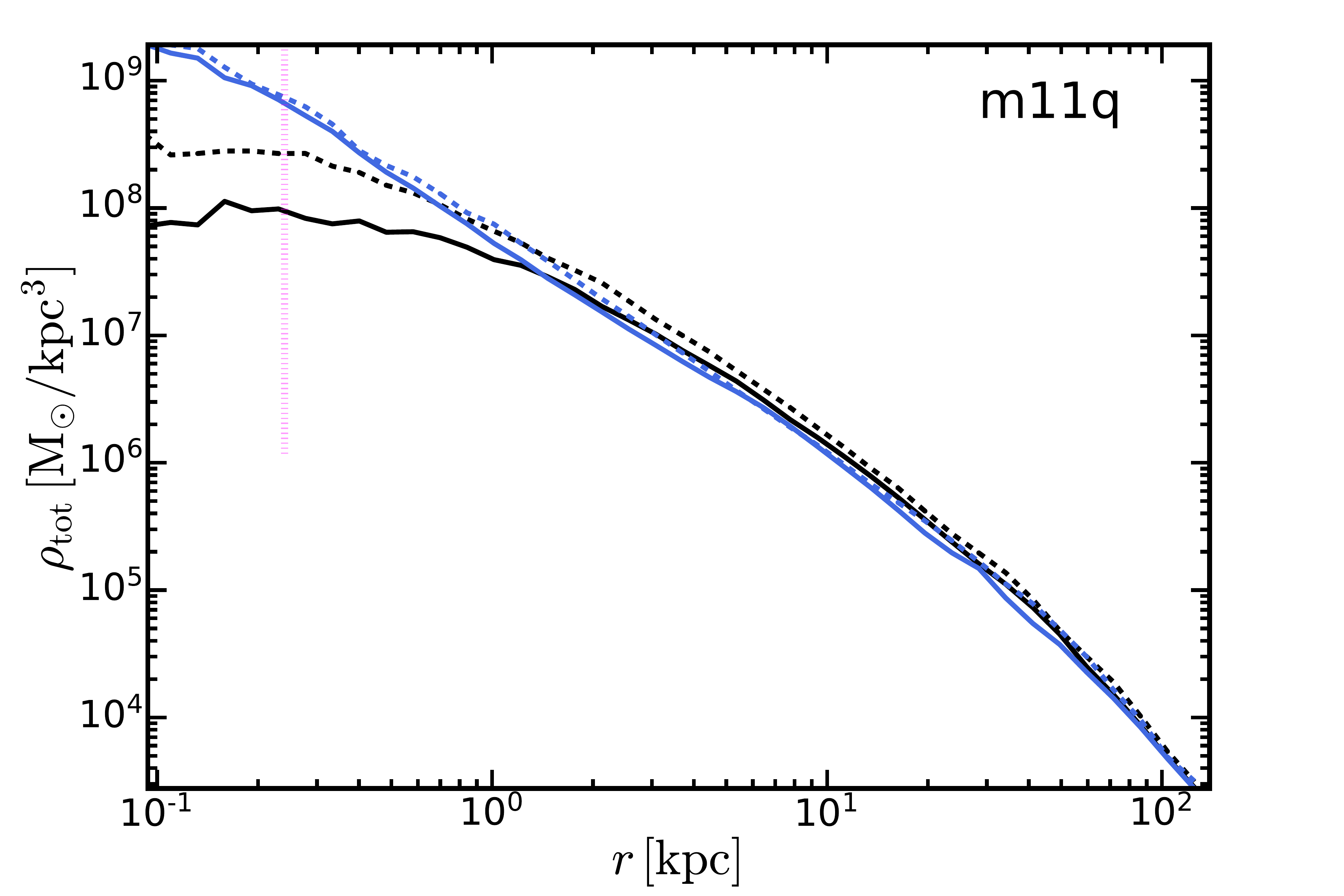}
    \caption{\textbf{Total mass density profiles of galaxies in DMO simulations and full physics simulations.} We present the density profiles of \textbf{m10q} and \textbf{m11q} in CDM and dSIDM with $(\sigma/m)=1\cpm$. The results of full physics simulations are shown in solid lines while the results of DMO simulations are shown in dashed lines. The purple dotted vertical line indicates the convergence radius in DMO runs (see Table~\ref{tab:sims}). In CDM, the central density profiles in DMO simulations are similar to the NFW profile before reaching the convergence radii. The full physics simulation of \textbf{m11q} produces a $\kpc$ size core at the center due to strong baryonic feedback there. However, in the dSIDM model with $(\sigma/m)=1\cpm$, the DMO and full physics simulations produce almost identical results, indicating that dissipative interactions of dark matter completely determine the evolution of the dark matter halo and the impact of baryonic feedback becomes negligible. This is generally true when the dissipation time scale becomes significantly shorter than the Hubble time scale.
    }
    \label{fig:dmo}
\end{figure}

\subsection{Dark matter only versus full physics simulations}
\label{sec:dmo}

The analysis and discussion in the main paper revolve around the impact of dissipative dark matter interactions on galaxy structures. However, baryonic physics could also impact galaxy structures in various ways. For instance, the gas outflow driven by stellar/supernovae feedback could irreversibly transfer energy to dark matter and induce cores at galaxy centers~\citep[e.g.,][]{Governato2010,Pontzen2012,Madau2014}; the gravitational influence of baryons condensed at galaxy centers could induce adiabatic contraction of dark matter haloes~\citep[e.g.,][]{Blumenthal1986,Gnedin2004}. The contamination of baryonic physics processes is an important factor when studying the influence of alternative dark matter physics.

We explore this aspect by performing dark matter only (DMO) simulations of the same haloes in the simulation suite and comparing the results. In Figure~\ref{fig:dmo}, we compare the total mass density profiles of dwarf galaxies \textbf{m10q} and \textbf{m11q} in DMO simulations and full physics simulations. It is not surprising that, in the CDM case, the density profiles produced by DMO simulations are cuspy and NFW-like before reaching the convergence radii. In full physics simulations, \textbf{m11q} exhibits a $\kpc$ size core while \textbf{m10q} still exhibits a cuspy profile like its DMO counterpart. The difference results from the different level of baryonic feedback in the two galaxies. However, in the dSIDM model with $(\sigma/m)=1\cpm$, the DMO and full physics simulations produce almost the same density profiles, indicating that {\it \textbf{baryonic physics no longer affect the density profiles of dwarf galaxies once dissipation is strong enough}}. This check also validates the results presented in this paper against uncertainties in modelling the baryonic physics processes in simulations.

\section{Summary and conclusion}
\label{sec:conclusion}

In this paper, we present the first suite of cosmological baryonic (hydrodynamical) zoom-in simulations of galaxies in dSIDM. We adopt a dSIDM model where a constant fraction $f_{\rm diss}$ of the kinetic energy is lost during dark matter self-interaction. We sample models with different constant self-interaction cross-sections as well as a model with velocity-dependent cross-section. The dSIDM models explored here are weakly collisional ($\sigma/m \lesssim 10 \cpm$) but strongly dissipative ($f_{\rm diss} \gtrsim 0.1$) and are qualitatively different from some previously proposed baryon-like dSIDM models~\citep[e.g.,][]{Fan2013,Foot2013,Randall2015}, which are limited to explain a subset of all dark matter in the Universe. The simulations utilize the FIRE-2 model for hydrodynamics and galaxy formation physics, which allows for realistic predictions on the structural and kinematic properties of galaxies. This simulation suite consists of various galaxies, from ultra faint dwarfs to Milky Way-mass galaxies. In this paper, we primarily focus on the analysis of dwarf galaxies in dSIDM and explore galaxy/halo's response to dissipative self-interactions of dark matter. The following signatures of dSIDM models in dwarf galaxies are identified and explored:
\begin{itemize}

    \item The dark matter halo masses and galaxy stellar masses are not significantly affected in dSIDM models with $(\sigma/m)\lesssim 1\cpm$ compared to the CDM case (see Figure~\ref{fig:mstar-mhalo}). The dwarf galaxies in the dSIDM model with $(\sigma/m)= 10\cpm$ have slightly lower ($\operatorname{0.1\,-\,0.2}{\rm dex}$) halo/galaxy stellar masses. But the results of this model are still within the scatter of the relation constrained in observations as well as the stochastic run-to-run scatter of simulations of different dwarf galaxies.
    
    \item Energy dissipation due to dark matter self-interactions induces radial contraction of dark matter halo. This mechanism competes with baryonic feedback in shaping the central profiles of dwarf galaxies (see Figure~\ref{fig:denpro_lowmass} and \ref{fig:denpro_intemass}). When the effective self-interaction cross-section is low, the central profiles are still cored despite higher densities and smaller core sizes. When the effective self-interaction cross-section is larger than $\sim 0.1\cpm$, assuming $f_{\rm diss}=0.5$, the central density profiles of dwarf galaxies become cuspy and power-law like. The resulting asymptotic power-law profile is steeper than the NFW profile. The power-law slopes asymptote to $\sim -1.5$ in the classical dwarfs and range from $-2$ to $-1$ in the bright dwarfs (see Figure~\ref{fig:slope}). The slope of the profile can be well explained by the stead-state solution of a ``dark cooling flow'' (see Section~\ref{sec:discussion-slope}), which predicts a density profile with power-law slope $-1.6$. 
    
    \item Interestingly, further increasing the effective cross-section to $10\cpm$ does not lead to further contraction of the halo or steepening of the density profile. Instead, the normalization of the density profiles drops. A likely explanation is that the centrifugal force increases faster than the gravitational attraction as the halo contracts with specific angular momentum conserved. This eventually halts the contraction, increases the rotation support of the halo and drives the halo deformation (to oblate), which makes the density measured in spherical shell decreased.
    
    \item Through time scale analysis (Section~\ref{sec:timescale}), we show that the dSIDM models with constant cross-sections will have stronger impact in more massive galaxies while the velocity-dependent model has the opposite dependence. This is demonstrated by the simulations of classical dwarfs and bright dwarfs with the same dark matter model (see Figure~\ref{fig:denpro_lowmass} and \ref{fig:denpro_intemass}). The dSIDM model with $(\sigma/m)=0.1\cpm$ produces small cores in two of the classical dwarfs but produces cuspy profiles in two of the bright dwarfs. The velocity-dependent dSIDM model produces cuspy profiles in all the classical dwarfs while producing cored profiles in the bright dwarfs that are almost identical to the CDM case.
    
    \item The kinematic properties of the dark matter change in parallel to the contraction of dark matter halo (see Section~\ref{sec:kin}). As the self-interaction cross-section of dSIDM increases, the coherent rotation becomes more prominent compared to random velocity dispersion. In the meantime, the velocity dispersions are more dominated by the tangential component than the radial component, reflected by the negative velocity anisotropies in dSIDM haloes. The central parts of the galaxies are in transition from dispersion supported to rotation supported. Meanwhile, the velocity distribution function is suppressed at high velocities while it increases at low velocities in dSIDM models. As the cross-section increases, the median velocity is also shifted lower. 
    
    \item  The shape of the halo is affected by dissipation (see Figure~\ref{fig:axis_ratio}). In the dSIDM model with $(\sigma/m) = 1\cpm$, the halo becomes more spherical towards lower redshifts, contrary to the triaxial shape of CDM haloes. The spherical ``dark cooling flow'' washes out the initial triaxiality of the halo and makes the halo compact and spherical in the end. However, in the dSIDM model with $(\sigma/m) = 10\cpm$, the halo shape shows a response to the more prominent coherent rotation of dark matter. Haloes are initially on the track of becoming more spherical, but later turn oblate in shape due to the halt of spherical contraction and increased rotation support. 
    
    \item As shown in Section~\ref{sec:discussion-heat}, the energy transfer in dSIDM haloes (with the degree of dissipation $f_{\rm diss}=0.5$) is dominated by dissipation rather than ``thermal conduction'' (collisional energy transfer). When we vary $f_{\rm diss}$ to $\lesssim 0.1$, collisional energy transfer becomes important and the density at small radii ($r\lesssim 100\pc$) is significantly enhanced (see Figure~\ref{fig:m09}), which resembles the gravothermal collapse of eSIDM haloes. This gives the counterintuitive prediction that a model with a lower degree of dissipation (but higher cross-section to make the dissipation time scale invariant) can produce even denser haloes than models with higher degrees of dissipation.  
    
    \item The density profiles in full physics simulations of CDM are more cored than the ones in DMO simulations, caused by the inclusion of baryonic physics. However, the DMO simulations of dSIDM models show little difference from the full physics simulations (see Figure~\ref{fig:dmo}), likely due to the dominance of dark matter energy dissipation over perturbations from baryonic feedback. This shows that the structural properties of dSIDM haloes is insensitive to baryonic physics in this regime and demonstrates the robustness of our results against various uncertainties in the baryonic sector in simulations.
    
\end{itemize}

In this paper, we present the first study of dwarf galaxies in dSIDM models using cosmological hydrodynamical simulations. We find several observable signatures of dSIDM models in dwarf galaxies and systematically study the evolution patterns of dSIDM haloes, which differs from canonical astrophysical systems. Analytical explanations are provided to explain the phenomena found in simulations. The findings in this paper could serve as effective channels to constrain dSIDM models when compared to observations. This aspect will be considered in follow-up work in this series.


\section*{Acknowledgements}

Support for PFH was provided by NSF Research Grants 1911233 \& 20009234, NSF CAREER grant 1455342, NASA grants 80NSSC18K0562, HST-AR-15800.001-A. Numerical calculations were run on the Caltech compute cluster ``Wheeler'', allocations FTA-Hopkins/AST20016 supported by the NSF and TACC, and NASA HEC SMD-16-7592.
LN is supported by the DOE under Award Number DESC0011632, the Sherman Fairchild fellowship, the University of California Presidential fellowship, and the fellowship of theoretical astrophysics at Carnegie Observatories.
FJ is supported by the Troesh scholarship.
MBK acknowledges support from NSF CAREER award AST-1752913, NSF grant AST-1910346, NASA grant NNX17AG29G, and HST-AR-15006, HST-AR-15809, HST-GO-15658, HST-GO-15901, HST-GO-15902, HST-AR-16159, and HST-GO-16226 from the Space Telescope Science Institute, which is operated by AURA, Inc., under NASA contract NAS5-26555. 
AW received support from NASA through ATP grants 80NSSC18K1097 and 80NSSC20K0513; HST grants GO-14734, AR-15057, AR-15809, and GO-15902 from STScI; a Scialog Award from the Heising-Simons Foundation; and a Hellman Fellowship. 

\section*{Data Availability}
The simulation data of this work was generated and stored on the supercomputing system Frontera \footnote{https://frontera-portal.tacc.utexas.edu/about/} at the Texas Advanced Computing Center (TACC), under the allocations FTA-Hopkins/AST20016 supported by the NSF and TACC, and NASA HEC SMD-16-7592. The data underlying this article cannot be shared publicly immediately, since the series of paper is still in development. The data will be shared on reasonable request to the corresponding author.






\appendix 
\section{Convergence tests}
\label{appsec:conv}

In the main text, we have used the convergence radii found for CDM DMO runs, quoted for the standard FIRE-2 simulations \citet{Hopkins2018}. However, it is possible that weakly-collisional dSIDM has different convergence properties from collisionless particles. In additional, in full physics simulations, the convergence properties could be set by the resolution of baryonic processes rather than collisionless particles. For CDM, the simulations with baryons usually give better convergence than their DMO counterparts \citep{Hopkins2018}, but this aspect has never been studied in the dSIDM case.

\begin{figure}
    \centering
    \includegraphics[width=0.49\textwidth]{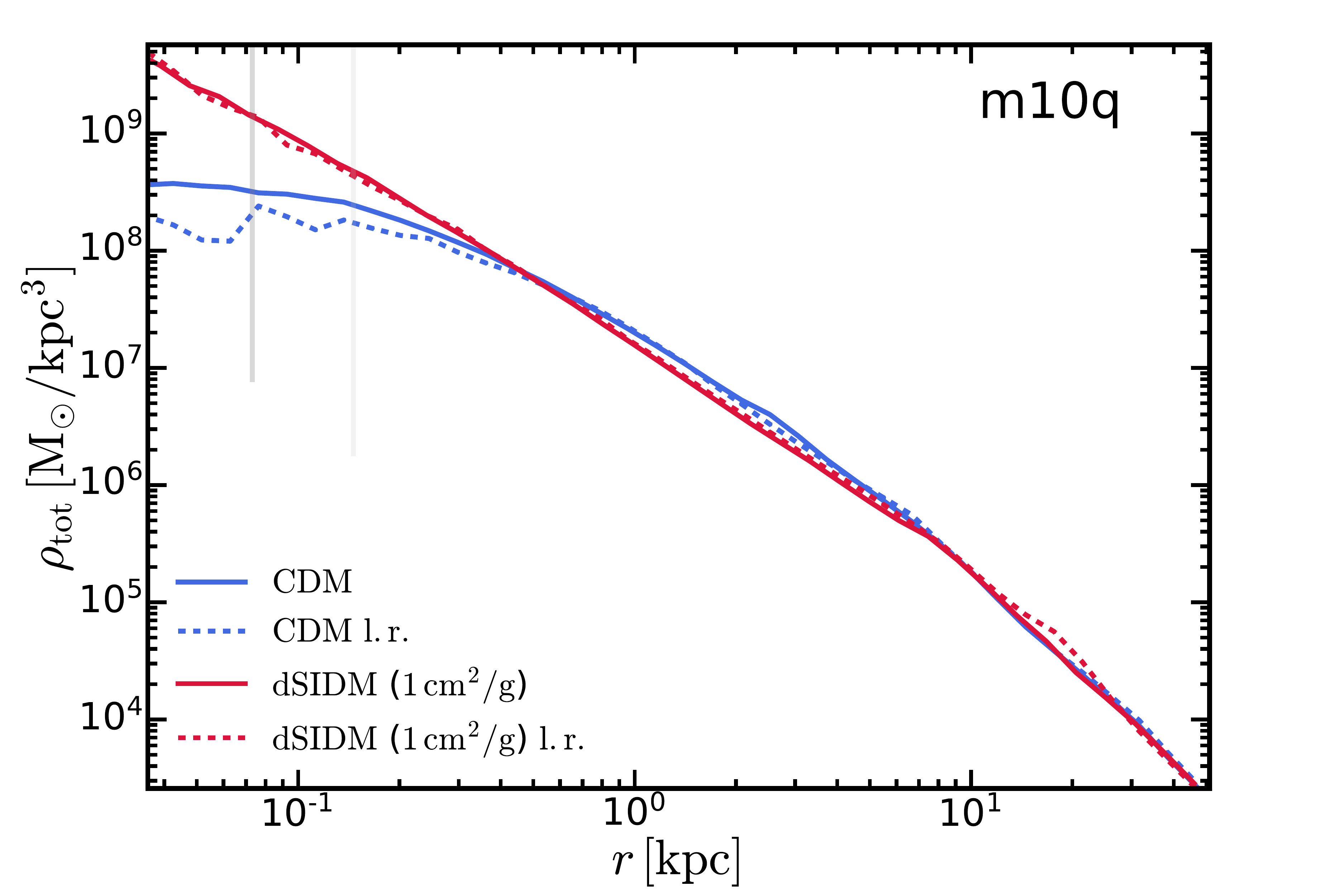}
    \includegraphics[width=0.49\textwidth]{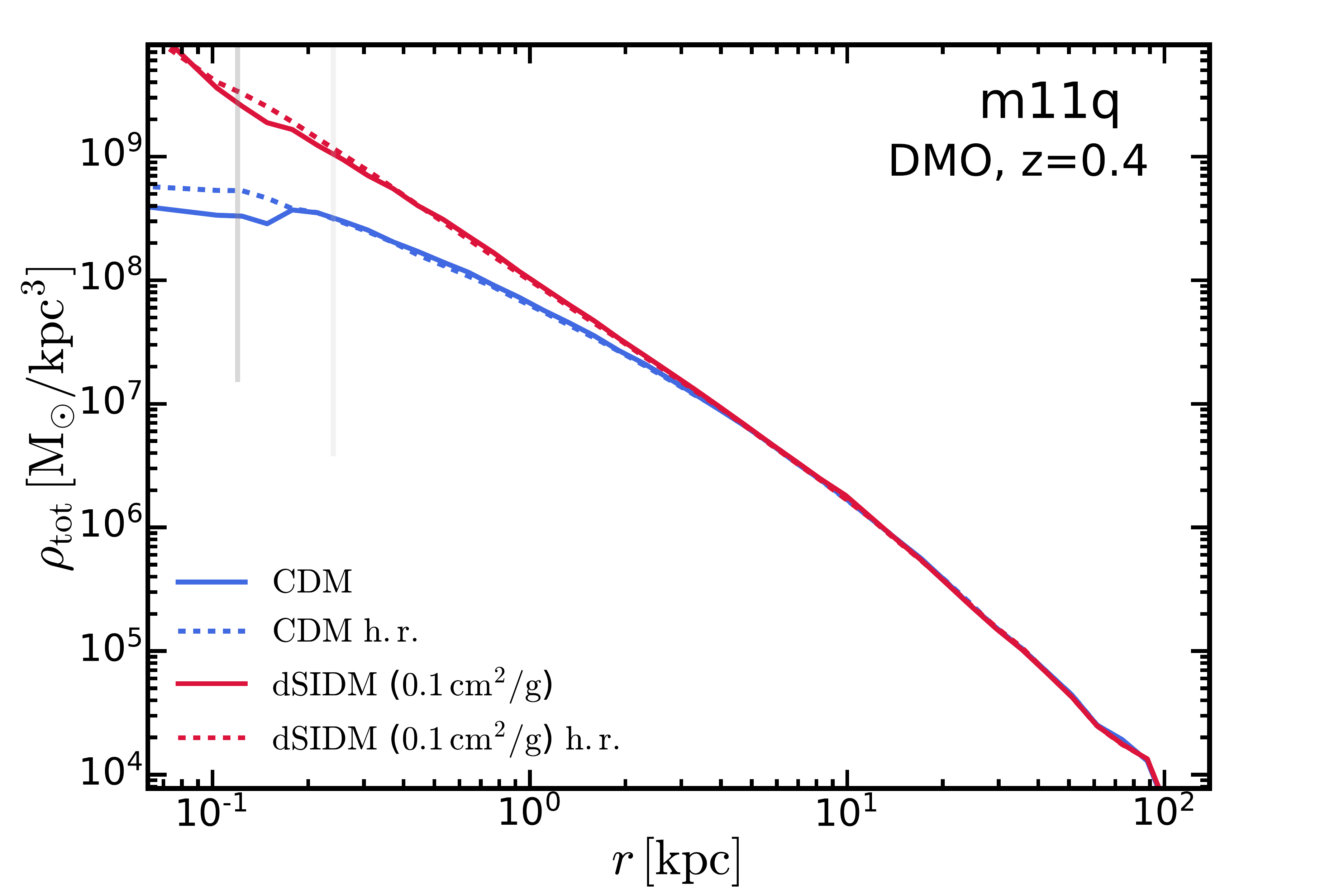}
    \caption{ \textbf{Test of convergence for the mass density profiles in CDM and dSIDM.} {\it Top panel:} Density profiles at $z=0$ of the default and low-resolution \textbf{m10q} runs with baryon. Results in different dark matter models are shown as labelled. The light (dark) vertical line indicates the convergence radius of the low-resolution (default) runs. The convergence radius is derived based on DMO runs using the \citet{Power2003} criterion \citep{Hopkins2018}. {\it Bottom panel:} Density profiles of the default and high-resolution \textbf{m11q} DMO runs. The high-resolution run stops at $z=0.4$ due to high computational cost, so we compare the profiles at that redshift. The labelling is the same as the top panel. The light (dark) vertical line indicates the convergence radius of the default (high-resolution) runs. The comparisons here demonstrate that dSIDM models have better convergence properties than CDM, whether we include baryons or not. The convergence radii quoted in the main text are conservative estimates of the true convergence radii of dSIDM runs. }
    \label{appfig:convtest}
\end{figure}

In this section, we perform convergence tests for dSIDM simulations. For the first test, we run a low-resolution version of \textbf{m10q} (with baryons; with $8$ ($2$) times poorer mass (spatial) resolution) in CDM and dSIDM with $(\sigma/m)=1\cpm$, respectively. The total mass density profiles are compared to the default runs in the top panel of Figure~\ref{appfig:convtest}. In the CDM case, the convergence radius derived based on the \citet{Power2003} criterion roughly describes where low-resolution run deviates from high-resolution one. In the dSIDM case, the convergence of the density profile extends to much smaller radii. For the second test, we run a high-resolution version of \textbf{m11q} (with $8$ ($2$) times better mass (spatial) resolution, but DMO this time) in CDM and dSIDM with $(\sigma/m)=0.1\cpm$, respectively. We stop the simulation at $z=0.4$ to save computational cost, but we verify in the default resolution simulation that the density profile at $z=0.4$ does not differ much from the $z=0$ one. Again in DMO runs of \textbf{m11q}, the dSIDM model shows much better convergence properties than CDM. In summary, the tests here demonstrate that dSIDM models in general give smaller convergence radii than CDM, and this property is not affected by whether we include baryons or not. Therefore, the convergence radii quoted in the main text are rather conservative estimates of the true convergence radii of dSIDM runs.


\bsp	
\label{lastpage}
\end{document}